\documentclass[twocolumn]{aastex62}
\usepackage{multirow}
\hypersetup{linkcolor=red,citecolor=green,filecolor=cyan,urlcolor=magenta}

\shorttitle{}
\shortauthors{}

\begin{document}

\title{Evolution of ONeMg Core in Super-AGB Stars towards Electron-Capture Supernovae: Effects of Updated Electron-Capture Rate}
\author{Shuai Zha\thanks{Email address: szha@phy.cuhk.edu.hk}}

\affiliation{Department of Physics, the Chinese
	University of Hong Kong, Hong Kong S. A. R., China}

\author{Shing-Chi Leung\thanks{Email address: shingchi.leung@ipmu.jp}}

\affiliation{Kavli Institute for the Physics and 
	Mathematics of the Universe (WPI), The University 
	of Tokyo Institutes for Advanced Study, The 
	University of Tokyo, Kashiwa, Chiba 277-8583, Japan}

\author{Toshio Suzuki  \thanks{Email address: suzuki@phys.chs.nihon-u.ac.jp}}
\altaffiliation{Visiting Researcher, National Astronomical Observatory of Japan, \\ Mitaka, Tolyo 181-8588, Japan}
\affiliation{Department of Physics, College of Humanities and Sciences, Nihon University,
	Sakurajosui 3, Setagaya-ku, Tokyo 156-8550, Japan}

\author{Ken'ichi Nomoto\thanks{Email address: nomoto@astron.s.u-tokyo.ac.jp}}

\affiliation{Kavli Institute for the Physics and 
	Mathematics of the Universe (WPI), The University 
	of Tokyo Institutes for Advanced Study, The 
	University of Tokyo, Kashiwa, Chiba 277-8583, Japan}

\accepted{by ApJ on October 4, 2019}

\begin{abstract}
Stars with $\sim 8-10~{M}_{\odot}$ evolve to form a strongly
degenerate ONeMg core. When the core mass becomes close to the
Chandrasekhar mass, the core undergoes electron captures on $^{24}$Mg
and $^{20}$Ne, which induce the electron-capture supernova (ECSN). In
order to clarify whether the ECSN leads to a collapse or thermonuclear
explosion, we calculate the evolution of an $8.4~M_\odot$ star from
the main sequence until the oxygen ignition in the ONeMg core.  We
apply the latest electron-capture rate on $^{20}$Ne including the
second forbidden transition, and investigate how the location of the
oxygen ignition (center or off-center) and the $Y_e$ distribution
depend on the input physics and the treatment of the semiconvection
and convection.  The central density when the oxygen deflagration is
initiated, $\rho_{\rm c,def}$, can be significantly higher than that
of the oxygen ignition thanks to the convection, and we estimate
$\log_{10}(\rho_{\rm c,def}/\mathrm{g~cm^{-3}})>10.10$.  We perform
two-dimensional simulations of the flame propagation to examine how
the final fate of the ONeMg core depends on the $Y_e$ distribution and
$\rho_{\rm c,def}$.  We find that the deflagration starting from
$\log_{10}(\rho_{\rm c,def}/\mathrm{g~cm^{-3}})>10.01 (< 10.01)$ leads
to a collapse (thermonuclear explosion).  Since our estimate of
$\rho_{\rm c,def}$ exceeds this critical value, the ONeMg core is
likely to collapse, although further studies of the convection and
semiconvection before the deflagration are important. 
\end{abstract}


\section{Introduction \label{sec:intro}} 
A non-rotating solar-metallicity star with the zero-age-main-sequence
mass ($M_{\rm ZAMS}$) in the range of $8-10~{M_\odot}$ forms a strongly
degenerate oxygen-neon-magnesium (ONeMg) core after the 2nd dredge up of the He
layer \citep{1984ApJ...277..791N}. Subsequently, the ONeMg core grows its
mass through the H-He double shell burning and the star evolves along
the super-asymptotic giant branch (SAGB) in the HR diagram. During the SAGB
evolution, the H-rich envelope is losing its mass by various mechanisms
such as a dust-driven wind, Mira-like pulsation,
etc. \citep[see a review by][]{2018A&ARv..26....1H}. The fate of these stars is
either the formation of an ONeMg white dwarf (WD) if almost all H-rich
envelope is lost for $M_{\rm ZAMS}<M_{\rm up,Ne}$, or the electron-capture
supernova (ECSN) if the ONeMg core mass reaches near the Chandrasekhar limit
($M_{\rm Ch}$) for $M_{\rm ZAMS}>M_{\rm up,Ne}$
\citep{1979wdvd.coll...56N,1980PASJ...32..303M,1984ApJ...277..791N,2013ApJ...772..150J,2013ApJ...771...28T,2013ARA&A..51..457N,2015MNRAS.446.2599D}. \added{Here, $M_{\rm up,Ne}$ is the upper mass limit for the star which leaves behind an ONeMg WD \citep{2013ARA&A..51..457N}.} In
the latter case, the ONeMg core undergoes various electron-capture and
URCA processes.

Initiated by heating due to electron capture on $^{20}{\rm Ne}$,
the oxygen {\sl ignition} takes place in the central region \added{(within $\sim$100 km of the center)}. Here the {\sl ignition} is defined as the stage where the nuclear
energy generation rate exceeds the thermal neutrino losses. We denote the central density at the oxygen ignition as $\rho_{\rm c,ign}$.
Subsequently, oxygen burning grows into the thermonuclear runaway (when
the timescale of temperature rise gets shorter than the dynamical
timescale), and forms an oxygen deflagration wave behind which nuclear
statistical equilibrium (NSE) is realized at temperature $T>5\times10^9$ K. We denote the central density when the oxygen deflagration starts as $\rho_{\rm c,def}$, which may be larger than $\rho_{\rm c,ign}$ if the convective energy transport after the oxygen ignition is efficient.

Further evolution of the ONeMg core depends on the competition between
the nuclear energy release by the propagating oxygen deflagration wave
and the reduction of the degeneracy pressure due to electron capture
in the NSE ash behind the deflagration wave
\citep{1991ApJ...367L..19N,1992ApJ...396..649T,2016A&A...593A..72J,2019PASA...36....6L}.

Recent multi-dimensional simulations of the oxygen-deflagration have shown that
the result of the above competition depends sensitively on the
parameterized $\rho_{\rm c,def}$.  If $\rho_{\rm c,def}$ is higher than a
certain critical density $\rho_{\rm cr}$, the core collapses to form a neutron star
(NS) because of the electron capture
\citep{1999ApJ...516..892F,2006A&A...450..345K,2017ApJ...850...43R}.
If $\rho_{\rm c,def} < \rho_{\rm cr}$, on the other hand, thermonuclear
energy release dominates to induce the partial explosion of the ONeMg core
\citep{2016A&A...593A..72J}.

For the critical density, $\log_{10}(\rho_{\rm cr}/\mathrm{g~cm^{-3}}) = {9.90 - 9.95}$ and $\log_{10}(\rho_{\rm cr}/\mathrm{g~cm^{-3}})={9.95 - 10.3}$ have been obtained by two-dimensional (2D)
\citep{2017hsn..book..483N,2019PASA...36....6L,2019arXiv190111438L}
and three-dimensional (3D) \citep{2016A&A...593A..72J} hydrodynamical
simulations, respectively.  We should note that there still exists a
big uncertainty in the treatment of the propagation of the oxygen
deflagration \citep{1992ApJ...396..649T}, as
well as the electron-capture rates of the NSE composition
\citep{2009ADNDT..95...96S}  to obtain $\rho_{\rm cr}$.

We should also note that $\rho_{\rm c,def}$ is subject to uncertainties
involved in the calculation of the final stages of the ONeMg core
evolution (see below).  $\log_{10}(\rho_{\rm c,def}/\mathrm{g~cm^{-3}})$ is currently evaluated in the
range of ${9.9-10.2}$
\citep{2015MNRAS.453.1910S,2017MNRAS.472.3390S,2019ApJ...871..153T}.

The uncertainties in the core evolution include: (1) the growth rate
of the degenerate ONeMg core mass, which gives the rate of core
contraction and compressional heating rate. This is determined by
thermal pulses of He shell burning and the 3rd dredge-up, which
require quite a lot of computational efforts.  (2) Rates of URCA processes
of $^{23}\rm Na$ and $^{25}\rm Mg$, which cool down the core.  (3)
Electron-capture rates on $^{24}{\rm Mg}$ and $^{20}{\rm Ne}$
\citep{1978ApJ...226..996I,2013ApJ...772..150J,2015MNRAS.453.1910S,2017MNRAS.472.3390S}.
(4) The initial abundances of $^{24}{\rm Mg}$ \citep{2005A&A...435..231G} and residual $^{12}{\rm C}$ \citep{2019ApJ...872..131S} in the
ONeMg core.
(5) Treatment of the criterion for the convective stability \citep{2018ApJS..234...34P}.

Most of the weak rates for these processes are theoretically
calculated with the reliable $sd$-shell model until recently
\citep{2013PhRvC..88a5806T,2014PhRvC..89d5806M} and provided with
either analytic formulae or tables. However, there is still an uncertainty in the
strength of the second-forbidden transition of $^{24}{\rm Mg}$
and $^{20}{\rm Ne}$, which affects $\rho_{\rm c,def}$ substantially
\citep{2015MNRAS.453.1910S}.

Electron-capture processes not only reduce the electron number fraction
($Y_e$) but also heat the core through the energy deposition
from $\gamma$-rays as well as distort the electron distribution function \citep[e.g., ][]{1980PASJ...32..303M}. Such heating makes the electron-capture front over-stable according to the Ledoux criterion and in the region of semiconvection
\citep{1987ApJ...318..307M}. Including semiconvection prescription
proposed in \cite{1992A&A...253..131S}, \cite{2019ApJ...871..153T}
found $\log_{10}(\rho_{\rm c,def}/\mathrm{g~cm^{-3}})\simeq{10.2}$, while without
any convection \cite{2017MNRAS.472.3390S} obtained $\log_{10}(\rho_{\rm c,ign}/{\rm g~cm^{-3}})\simeq9.95$.
Apart from altering $\rho_{\rm c,def}$, convection may enlarge the initial
size of oxygen flame and change $Y_e$ inside it, which can greatly
affect the subsequent hydrodynamical behavior
\citep{2019arXiv190111438L}.

The newest electron-capture rate of $^{20}$Ne including the second-forbidden transition
\citep{2018arXiv180508149K,Suzuki2019} can strongly affect how fast $^{20}$Ne is converted to $^{20}$F and the corresponding energy deposition. 
Such a heat source can alter the temperature profile and 
the convective structure of the core prior to the oxygen deflagration. So far there has not been much discussion on how this updated nuclear physics input affects the final fate of SAGB stars. 
Therefore, we calculated the detailed late-phase evolution 
of SAGB stars and modeled the subsequent propagation of the oxygen deflagration wave.

The structure of this paper is as follows.
In Section~\ref{sec:evol}, we present the evolutionary path of SAGB stars until the 
onset of the oxygen ignition, with the new weak rates and different convection criteria.
In Section~\ref{sec:hydro}, we use 2D hydrodynamical simulations to model the oxygen deflagration phase through the collapse or explosion. We also discuss the dependence of the outcomes on the stellar evolution and other physical inputs. 
We summarize our results in Section~\ref{sec:summary}.


\section{Evolution of SAGB Stars \label{sec:evol}}
\subsection{Methods \label{subsec:method_evol}}
We evolve a non-rotating solar-metallicity star with
$M_\mathrm{ZAMS}=8.4~{M}_\odot$ starting from the main-sequence phase, and follow the formation and growth of the degenerate ONeMg core until the ignition of oxygen burning, using Modules for
Experiments in Stellar Astrophysics \citep[MESA;][]{2011ApJS..192....3P,2013ApJS..208....4P,2015ApJS..220...15P,2018ApJS..234...34P,2019arXiv190301426P},
revision 8118. Until the formation of the ONeMg core, we use the MESA
inlist of \cite{2013ApJ...772..150J}. In short, the initial
metallicity is $Z=0.014$, the mixing-length parameter 1.73 and the
overshooting parameter $f_{\rm ov}=0.014$ at all convective boundaries
with the exception of $f_{\rm ov}=0.005$ at the base of burning
convective shells. Mass loss includes the Reimers prescription
\citep{1975MSRSL...8..369R} for the RGB phase with $\eta=0.5$ and the
Bl\"ocker prescription \citep{1995A&A...299..755B} with $\eta=0.05$
during the AGB phase. One difference is that we use the MESA nuclear reaction network \texttt{sagb\_NeNa\_MgAl.net} consisting of 22 isotopes to cover the H, He and C burning phases \citep{2015ApJ...807..184F}. We add the important nuclear reaction
$^{22}\mathrm{Ne}(\alpha,n)^{25}\mathrm{Mg}$ to produce the URCA cooling
element $^{25}\mathrm{Mg}$ \citep[][p. 203]{2012sse..book.....K}.

The modeling of the thermally pulsing AGB phase is computationally very
expensive and numerical difficulties for modeling the thermal pulse and high
temperature hydrogen ingestion make the calculation of the whole star up to the oxygen
ignition impossible with current MESA \citep{2019ApJ...872..131S}. \added{There is a sharp density and temperature gradient at the interface between the degenerate ONeMg core and the H \& He envelope, so they are in very loose contact with each other \citep[see, Figure 15 of][]{1984ApJ...277..791N}. The later evolution of the core is expected to be independent of the envelope except the mass accretion \citep{1987ApJ...322..206N,2013ApJ...771...28T}. } Therefore,
when the degenerate ONeMg core \added{of $1.28~M_\odot$} is formed, we remove the envelope with an artificial mass loss rate
($0.1-1~M_\odot~{\rm yr}^{-1}$). Nuclear burning during this phase is
switched off for numerical simplicity. The resulting ONeMg core has a
thin hot CO layer and is evolved to cool down until matter can be
accreted. We checked that the cooling time \added{(ranging from $1~\mathrm{yr}$ to $1~\mathrm{Myr}$)} does not affect the
following evolution.

We model the ONeMg core growth phase until the oxygen ignition by assuming a
constant mass accretion rate and the same accreted composition as the surface
layer. The accretion rate is set to be $10^{-6}$ or
$10^{-7}~M_{\odot}~{\rm yr}^{-1}$ to account for the uncertainties involved in the
H-He double shell burning and the associated third dredge-up of the He layer
\citep{2017PASA...34...56D}. The nuclear network further includes the URCA processes of
$^{23}$Na$\rightleftharpoons^{23}$Ne, $^{25}$Mg$\rightleftharpoons^{25}$Na and
$^{25}$Na$\rightleftharpoons^{25}$Ne, and the electron-capture chains
$^{24}$Mg$(e^-,\nu_e)^{24}$Na$(e^-,\nu_e)^{24}$Ne and
$^{20}$Ne$(e^-,\nu_e)^{20}$F$(e^-,\nu_e)^{20}$O by using the
state-of-the-art rate tables \citep[provided
by][]{2013PhRvC..88a5806T,2016ApJ...817..163S}. We consider the
rate for the second forbidden transition of
$^{20}$Ne$(e^-,\nu_e)^{20}$F \citep{Suzuki2019} as discussed in
\S\ref{ne20rate}. \deleted{The semiconvection during electron capture is
not accurately modeled with the mixing-length treatment in MESA, so we
investigate the theoretical limits by using the
Ledoux and Schwarzschild criteria.} 
\replaced{The calculation is terminated
at the oxygen ignition, when in the mass zone with the maximum nuclear energy generation rate, the nuclear energy generation rate by oxygen burning exceeds the thermal
neutrino losses.}{The calculation is terminated
when oxygen ignites in the mass zone with the maximum nuclear energy generation rate. Oxygen ignition is defined as when the nuclear energy generation rate by oxygen burning exceeds the thermal neutrino losses.}

\added{The semiconvection during electron capture is not accurately modeled with the mixing-length treatment in MESA, so we investigate the theoretical limits by using the
two extreme criteria for the convective stability \citep{1987ApJ...318..307M,2012sse..book.....K}. They are the Schwarzschild criterion:
\begin{equation}
	\nabla_{\rm rad} < \nabla_{\rm ad},
\end{equation} 
and the Ledoux criterion:
\begin{equation}
	\nabla_{\rm rad} < \nabla_{\rm ad}+({\chi_{Y_e}}/{\chi_T}) \nabla_{Y_e}.
\end{equation} 
Here, $\nabla_{\rm rad}$ and $\nabla_{\rm ad}$ are the radiative (assuming energy transport by only radiation and conduction) and adiabatic temperature gradients
\begin{equation}
	\nabla_{\rm rad(ad)} \equiv \bigg(\frac{\partial \ln T}{\partial \ln P}\bigg)_{\rm rad(ad)};
\end{equation}
$\chi_{Y_e}$ and $\chi_T$ are the derivatives related to the equation of state
\begin{equation}
	\chi_{Y_e} \equiv \bigg( \frac{\partial \ln P}{\partial \ln Y_e}\bigg)_T, \quad
	\chi_{T}  \equiv \bigg( \frac{\partial \ln P}{\partial \ln T}\bigg)_{Y_e};
\end{equation}
and
\begin{equation}
	\nabla_{Y_e} \equiv -\frac{d \ln Y_e}{d \ln P}.
\end{equation}
The semiconvective region is treated as convectively unstable (stable) when using the Schwarzschild (Ledoux) criterion, although the growth timescale of overstability needs to be taken into account \citep{1987ApJ...318..307M}.} 

\subsection{Electron-capture Rate of $^{20}$Ne \label{ne20rate}}
Here, we discuss the electron-capture rates on $^{20}$Ne used in the 
present work, especially focusing on the forbidden transition, 
$^{20}$Ne (0$_{\rm g.s.}^{+}$) $\rightarrow$ $^{20}$F (2$_{\rm g.s.}^{+}$).
Possible important roles of the forbidden transition in electron 
capture on $^{20}$Ne was pointed out in \cite{2014PhRvC..89d5806M}.
While the experimental transition rate was not well determined and 
only the lower limit of the log {\it ft} value was given for the 
second-forbidden $\beta$-decay transition $^{20}$F (2$_{\rm g.s.}^{+}$) 
$\rightarrow$ $^{20}$Ne (0$_{\rm g.s.}^{+}$), the transition was taken 
to be an allowed Gamow-Teller transition with $B$(GT) corresponding
to the lower limit value of log {\it ft} =10.5; {\it ft} =6147/$B$(GT). \added{Here, the GT transition strength $B$(GT) is defined by 
\begin{equation}
B\mbox{(GT)} = \frac{(g_A/g_V)^2}{2J_i +1} \mid \langle f || \sum_k \vec{\sigma}^{k} \vec{t}_{\pm}^{k} || i\rangle \mid^2,
\end{equation}
where $g_A$ and $g_V$ are the weak axial-vector and vector transition coupling constants, respectively;  $J_i$ is the total spin of the initial state; $\vec{\sigma}$ and $\vec{t}_{\pm}$ are the Pauli spin matrix and isospin operator, respectively; $|i(f)\rangle$ is the initial (final) state. 
The $\beta$-decay rate $\lambda$ at high density $\rho$ and high temperature $T$ can be expressed by using the $ft$ value as
\begin{equation}
\lambda = \frac{f(\rho, T, \mu)}{ft} \ln 2,
\end{equation}
where $f(\rho, T, \mu)$ with the electron chemical potential $\mu$ is the phase space factor for the transition \citep{1980ApJS...42..447F}.
As the rate $\lambda$ is proportional to the transition strength $B\mbox{(GT)}$ and the phase space factor $f(\rho, T, \mu)$, the $ft$ value is given as $ft$ =D/$B\mbox{(GT)}$ with a constant D= $\frac{2\pi^3\hbar^7  \ln 2}{g_{V}^2 m_{e}^5 c^4}$ ($m_e$ is electron mass) \citep{1994ADNDT..56..231O,2001ADNDT..79....1L}. 
}
But this prescription using a constant strength is an approximation.

Here, we treat the forbidden transition $^{20}$Ne 
(0$_{\rm g.s.}^{+}$) $\rightarrow$ $^{20}$F (2$_{\rm g.s.}^{+}$) properly, 
and evaluate the electron-capture rates by using the multipole expansion 
method \citep{1972PhRvC...6..719O,Wal2,Wal3}. An explicit formula for the electron-capture rate for finite density and 
temperature is given, for example, in \cite{Par1} and \cite{Par2}.
For a 0$^{+}$ $\rightarrow$ 2$^{+}$ transition, there are contributions from Coulomb, longitudinal and electric transverse terms induced by
weak vector current as well as axial magnetic term induced by weak 
axial-vector current with multipolarity $J = 2$. \added{Here $J$ denotes the angular momentum transferred from the initial to the final states.}
Among them, about 60$\%$ contributions come from the Coulomb and longitudinal terms. The transition strength becomes electron energy dependent in contrast to the case of the prescription assuming an allowed transition.  Note that the transition strengths or shape factors in forbidden transitions are energy dependent in general.  

\begin{figure}[t!]
	\plotone{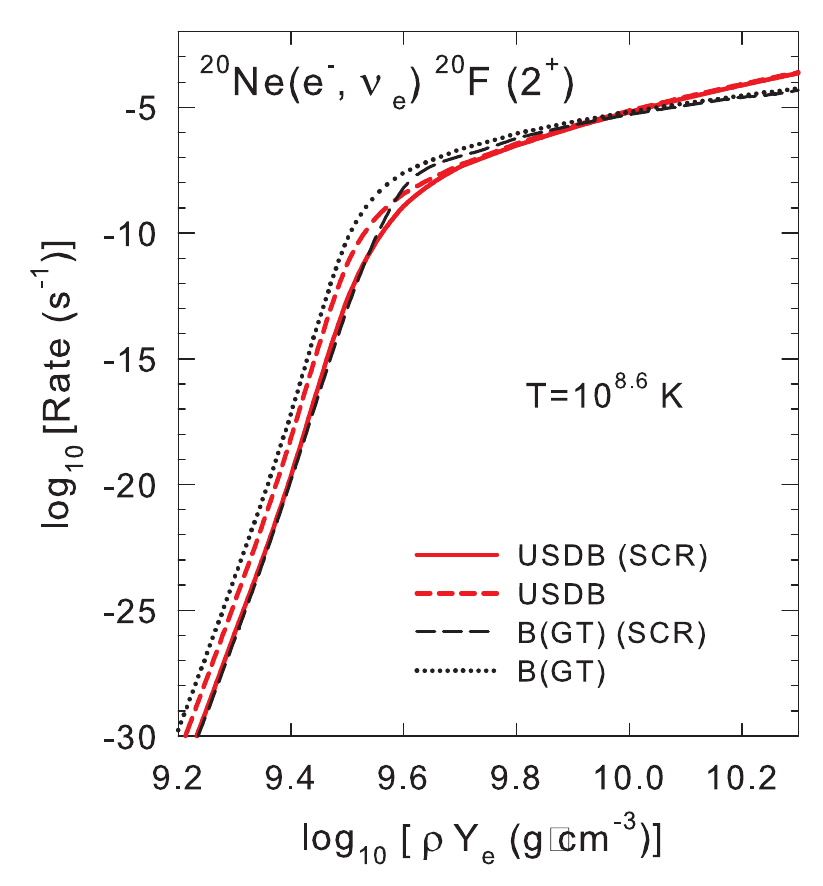}
	\caption{Calculated electron-capture rates for $^{20}$Ne 
		(0$_{\rm g.s.}^{+}$) $\rightarrow$ $^{20}$F (2$_{\rm g.s.}^{+}$) obtained 
		with the USDB Hamiltonian with and without the Coulomb (screening) 
		effects for $\log_{10}(T/{\rm K})=8.6$. \added{(SCR) in the legends denotes that the screening effects on electrons and ions are included.} 
		Calculated rates obtained as an allowed transition with a $B$(GT) 
		value corresponding to log {\it ft} =10.47 for the 
		inverse $\beta$-decay \citep{2018arXiv180508149K} are also shown. \label{fig:ecap-ne202p}}   
\end{figure}

Calculated electron-capture rates for the forbidden transition obtained 
with the USDB Hamiltonian \citep{2006PhRvC..74c4315B} with and without the Coulomb 
effects are shown in Figure \ref{fig:ecap-ne202p} for log$_{10} (T/{\rm K})$ $=8.6$.  
Screening effects on both electrons and ions are taken into account 
for the Coulomb effects \citep{2010NuPhA.848..454J,2013PhRvC..88a5806T,2016ApJ...817..163S}. Here, the quenching factor \added{$q$} for the axial-vector coupling constant 
$g_A$ is taken to be $q =0.764$ \citep{2008PhRvC..78f4302R}.
The Coulomb effects shift the electron-capture rates toward the higher density 
region due to an increase of the Q-value. 

Recently, a new measurement for the $\beta$-decay $^{20}$F 
(2$_{\rm g.s.}^{+}$) $\rightarrow$ $^{20}$Ne (0$_{\rm g.s.}^{+}$) has been 
carried out, and the rate is determined to be 
log {\it ft} = 10.47$\pm$0.11 \citep{2018arXiv180508149K}, which is very close to the lower limit value 
log {\it ft} =10.5. Calculated rates obtained as an allowed transition with a $B$(GT)
value corresponding to log {\it ft} =10.47 are also shown 
in Figure \ref{fig:ecap-ne202p}.        
The rates obtained with a constant $B$(GT) are found 
to be enhanced (reduced) compared with those with the USDB at 
log$_{10}$($\rho Y_e$/g~cm$^{-3}$) $<$ ($>$) 9.9 in case without the Coulomb effects. 
In case with the Coulomb effects, the former rates are enhanced 
compared with the latter at 9.6 $<$ log$_{10}$($\rho Y_e$/g~cm$^{-3}$) $<$ 9.9 
though the difference is smaller.    
These tendencies are due to the difference in the electron energy 
dependence of the transition strength between the two methods. The density dependence of the calculated rates with USDB by the multipole expansion method derived from energy dependent transition strength should be  considered as more reliable than that obtained assuming an allowed transition. The Coulomb effects are non-negligible and the rates with the Coulomb effects obtained with USDB are recommended to be used in astrophysical calculations.            

\begin{figure}[t!]
	\plotone{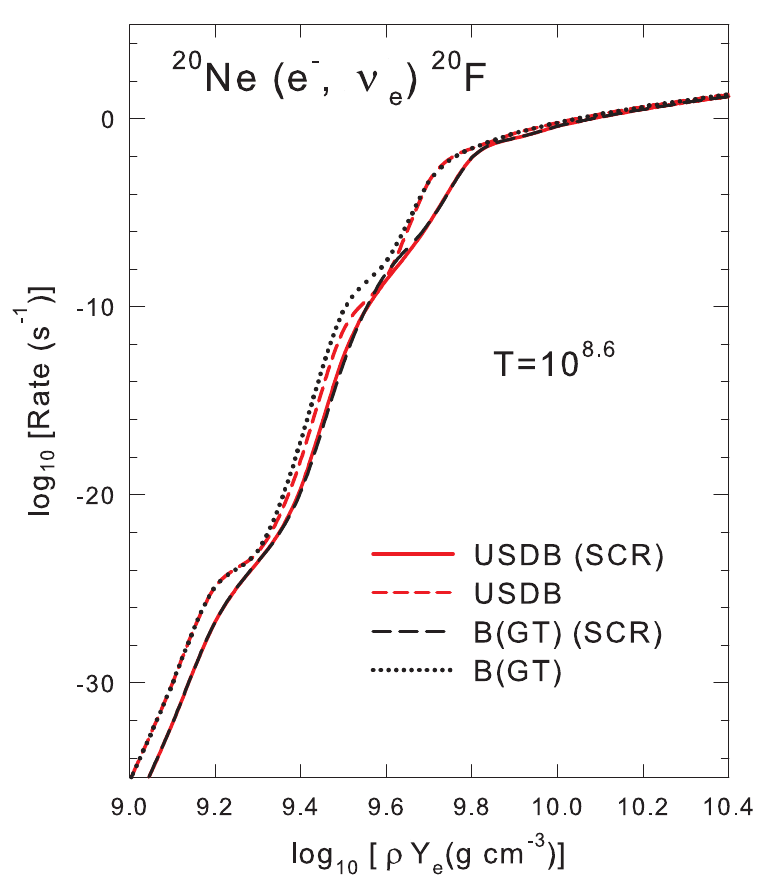}
	\caption{The same as in Figure  \ref{fig:ecap-ne202p} but for the total 
		electron-capture rates for $^{20}$Ne $\rightarrow$ $^{20}$F at log$_{10}$T(K) =8.6. \label{fig:sumecap-ne20}}
\end{figure}

Total electron-capture rates on $^{20}$Ne are shown in Figure \ref{fig:sumecap-ne20}.  
Contributions from Gamow-Teller transitions from 0$_{\rm g.s.}^{+}$ 
and 2$_{1}^{+}$ states in $^{20}$Ne to 1$^{+}$, 2$^{+}$ and 3$^{+}$ 
states in $^{20}$F obtained with USDB are included as well as the forbidden transition, 
0$_{\rm g.s.}^{+}$ $\rightarrow$ 2$_{\rm g.s.}^{+}$.

\subsection{Evolution of ONeMg core up to electron capture on $^{24}$Mg}
\begin{figure}[t!]
	\plotone{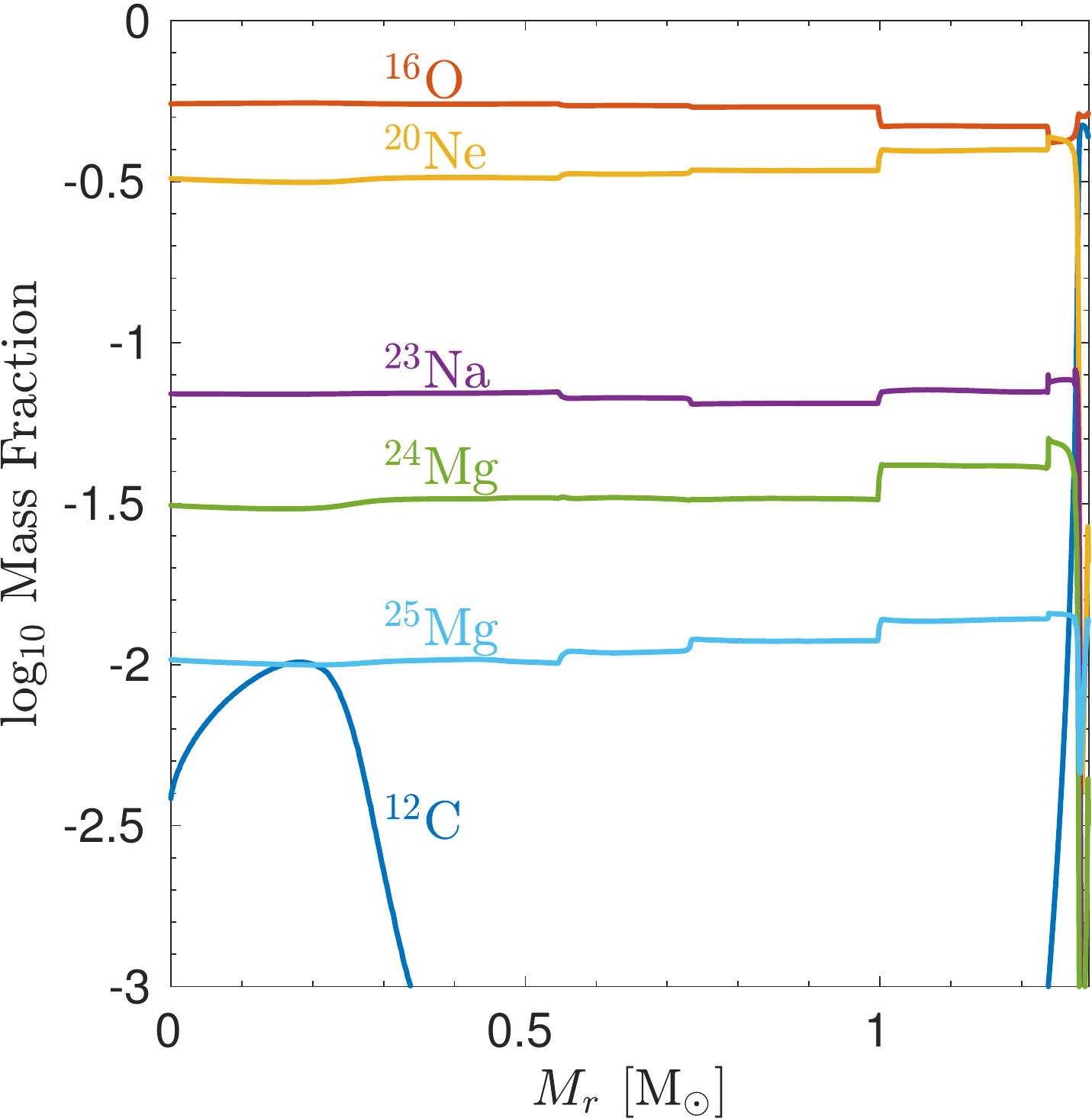}
	\caption{\replaced{Composition profile in the degenerate ONeMg core when we terminate the calculation of the whole star.}{Composition profile of the degenerate ONeMg core  of 1.28 $M_\odot$ evolved from a non-rotating solar-metallicity 8.4 $M_\odot$ star. The prescription for the evolution calculation with MESA was given in \S\ref{subsec:method_evol}.} \label{fig:core_abn}}
\end{figure}

The evolution of the $8.4~M_\odot$ star until the formation of the degenerate ONeMg core is similar to 
the lower mass models of \cite{2013ApJ...772..150J}. Carbon is ignited
slightly off-center at $M_r = 0.07~{M}_\odot$ ($M_r$ is the mass coordinate).  The carbon burning shell steadily propagates to the center, similar to the off-center carbon flame models in
\cite{2015ApJ...807..184F}. After we stop the evolution of the whole star and
remove its envelope, an $1.28~M_\odot$ core (with a
$\sim0.01~{M}_\odot$ CO layer) is left behind with the abundance
profile shown in Figure~\ref{fig:core_abn}. The abundances of key
isotopes for URCA process and electron capture are listed in
Table~\ref{tab:core_abn} in comparison with other works
\citep{2013ApJ...771...28T,2015ApJ...807..184F,2017MNRAS.472.3390S}. The
composition agrees well with \cite{2015ApJ...807..184F} except we produce $1\%$ $^{25}$Mg
with the addition of $^{22}\mathrm{Ne}(\alpha,n)^{25}\mathrm{Mg}$. Note that we also find a
residual carbon island at $M_r\lesssim0.3~{M}_\odot$, but the maximum
abundance is only $\sim1\%$ so that oxygen burning cannot be ignited
by this residual carbon burning at $\log_{10}(\rho_{\rm c}/\mathrm{g~cm^{-3}})<9.8$
\citep{2019ApJ...872..131S}.

\begin{table}[t!]
	\caption{Comparison for the key isotopic abundances of the ONeMg core in different studies. T13 stands for \cite{2013ApJ...771...28T}, F15 for \cite{2015ApJ...807..184F} and SQB17 for \cite{2017MNRAS.472.3390S}. \label{tab:core_abn}}
	\centering
	\begin{tabular}{ccccc}
		\toprule
		Isotope & This work & T13 & F15 & SQB17 \\ \hline
		$^{16}$O & 0.570 & 0.480 & 0.490 & 0.500 \\
		$^{20}$Ne & 0.311 & 0.420 & 0.400 & 0.390 \\
		$^{23}$Na & 0.069 & 0.035 & 0.060 & 0.050 \\
		$^{24}$Mg & 0.031 & 0.050 & 0.030 & 0.050 \\
		$^{25}$Mg & 0.010 & 0.015 & 0.002 & 0.010 \\
		\hline
	\end{tabular}
\end{table}

\added{
\begin{figure}
	\plotone{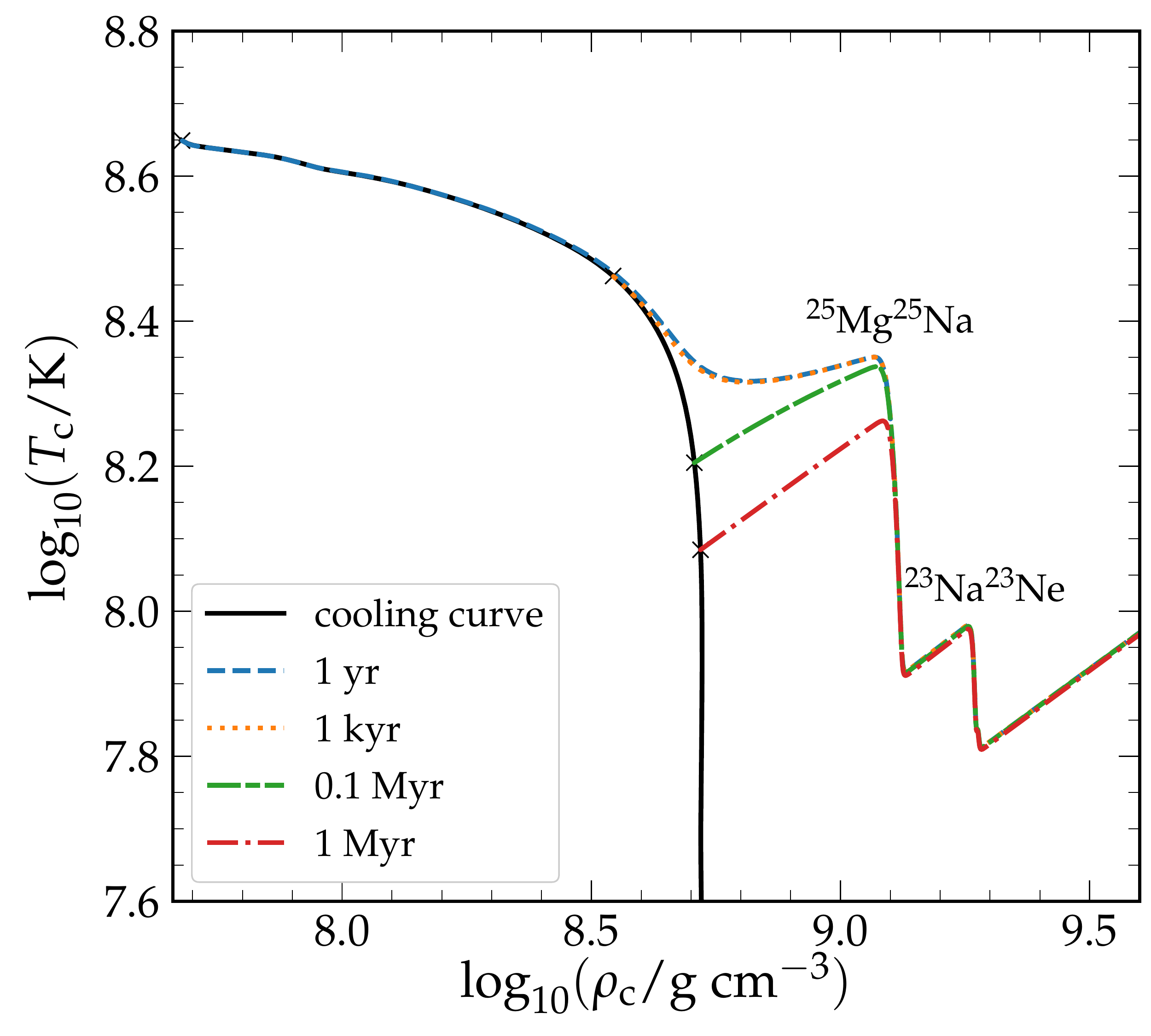}
	\caption{\added{The evolution of the accreting ONeMg core in the central-density temperature plane after a range of cooling times. The black solid line is the cooling curve for the ONeMg core after removing the envelope, and the colored lines show the evolution after a cooling time of 1 yr, 1 kyr, 0.1 Myr and 1 Myr. The black crosses mark the starting points of the accretion. The results with different cooling times converge after the URCA cooling process by the $^{25}$Mg$^{25}$Na pair at $\log_{10}(\rho_{\rm c}/{\rm g~ cm^{-3}})\simeq9.1$. \label{fig:cooling}}}
\end{figure}
	
After removing the envelope, the degenerate ONeMg core of $1.28~M_\odot$ is evolved to cool down before the accretion starts. In Figure~\ref{fig:cooling} we show the later evolution of the accreting core until $\log_{10}(\rho_{\rm c}/{\rm g~cm^{-3}})=9.6$, with the cooling time ranging from 1~yr to 1~Myr. The evolutionary paths undergo the same cooling due to the $^{25}$Mg$^{25}$Na and $^{23}$Na$^{23}$Ne  URCA pairs irrespective of the earlier cooling. In other words, after the URCA cooling starts, the evolution does not depend on the earlier history.

}
For the accretion phase, until electron capture on $^{24}$Mg takes
place, no convective instability is expected for the central region
\citep{2017MNRAS.472.3390S}. The thermal evolution of the core is
dominated by the compressional heating, thermal neutrino losses and URCA
cooling. Then, $^{24}$Mg electron captures produce a negative temperature gradient 
and a positive $Y_e$ gradient.
The energy transport and mixing in such semiconvective region is not well
constrained yet \citep{2013A&A...552A..76S}, so we use the two
extreme stability criteria, i.e.,
the Schwarzschild and Ledoux criteria, for investigating the uncertainties. In the
following, we discuss the evolution after
$^{24}$Mg$(e^-,\nu_e)^{24}$Na starts (when $\log_{10}(\rho_{\rm c}/{\rm g~cm^{-3}})\simeq9.6$) upon the usage of each criterion and set the
theoretical uncertainty on the final outcomes.

\subsection{Evolution of ONeMg Core with Ledoux Criterion}
\begin{figure}[t!]
	\plotone{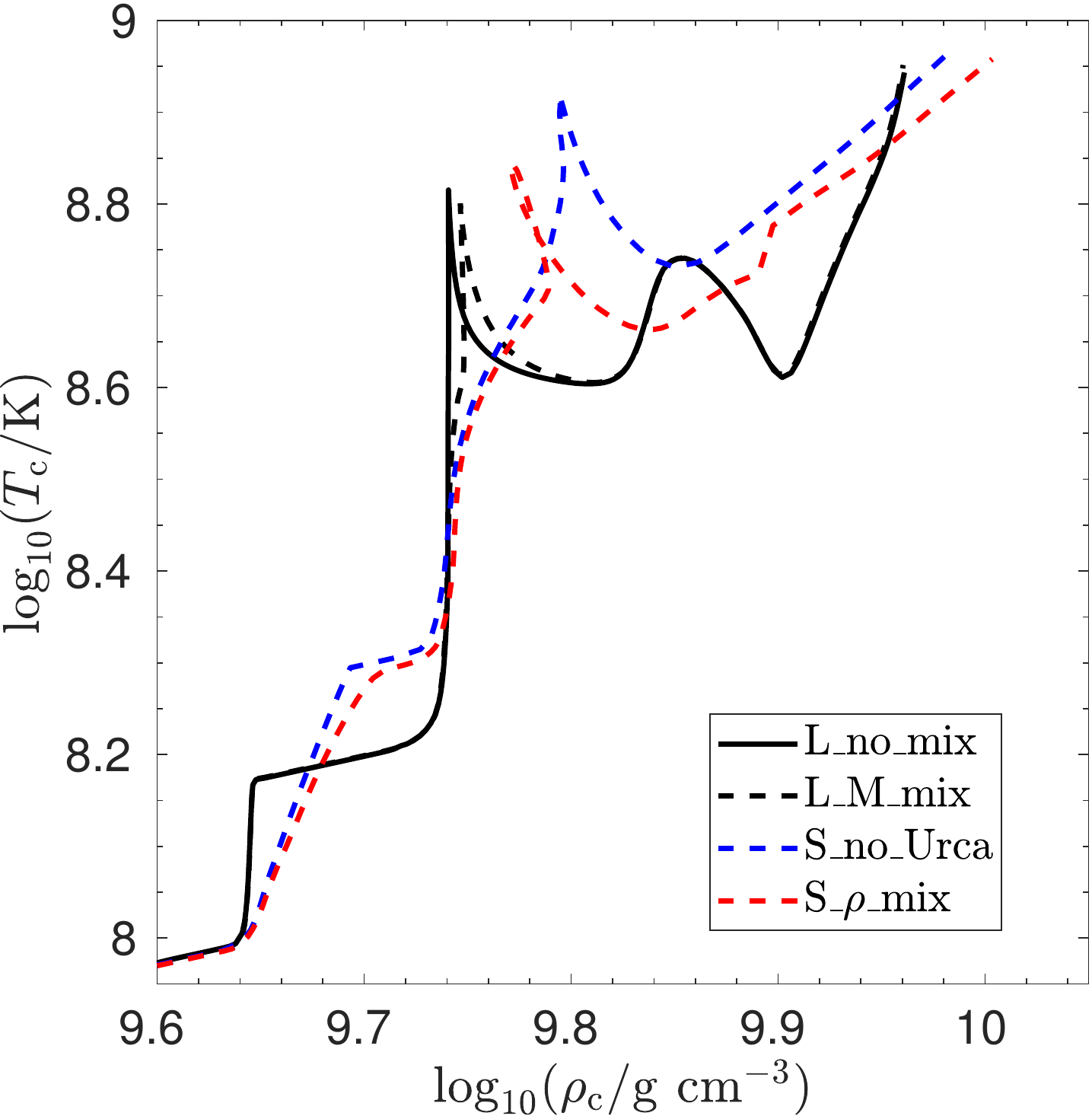}
	\caption{The evolution of the accreting ONeMg core in the central-density temperature plane for different treatments of the convection. `L' stands for the Ledoux criterion and `S' stand for the Schwarzschild criterion. The additional model nomenclature is explained in the main text.\label{fig:rho_tc}}
\end{figure}

We first focus on the model with a mass accretion rate of
$10^{-6}~{M_\odot}$~yr$^{-1}$. When using the Ledoux criterion, the $Y_{e}$
gradient is able to stabilize against convective instability during
electron capture on $^{24}$Mg. But after the onset of
$^{24}$Na$(e^-,\nu_e)^{24}$Ne, a convective core develops. Numerical
difficulty occurs when this convective core grows to
$\sim0.055~{M_\odot}$. The rapid change of the convective boundary
cannot be resolved with the Newton iteration solver in MESA
\citep{2017MNRAS.472.3390S,2019ApJ...872..131S}. Two approaches are
employed to further evolve the model. One is to mute the
mixing-length theory treatment of convection by using the MESA option
\texttt{mlt\_option='none'} (model ``L\_no\_mix"). Another is to limit
the convection inside a mass coordinate $M_{\rm mix}=0.05~{M_\odot}$
beyond which we found the numerical difficulty (model
``L\_M\_mix"). The evolution of the accreting ONeMg core in the central
density-temperature plane is shown in Figure~\ref{fig:rho_tc}. The two
approaches differ for the carbon ignition density, which is
$\sim2\%$ larger for L\_M\_mix due to the convective energy
transport. Carbon burning does not ignite oxygen burning due to its
low mass fraction in our ONeMg core model. Apart from this, the two
models show the same evolution afterwards, as the thermal neutrino cooling drags the evolution
back to a contractor (in balance between the compressional heating and thermal neutrino losses) at $\log_{10}(\rho_{\rm c}/{\rm g~cm^{-3}})\simeq9.8$
\citep{2019ApJ...872..131S}.

\begin{figure*}[t!]
	\plotone{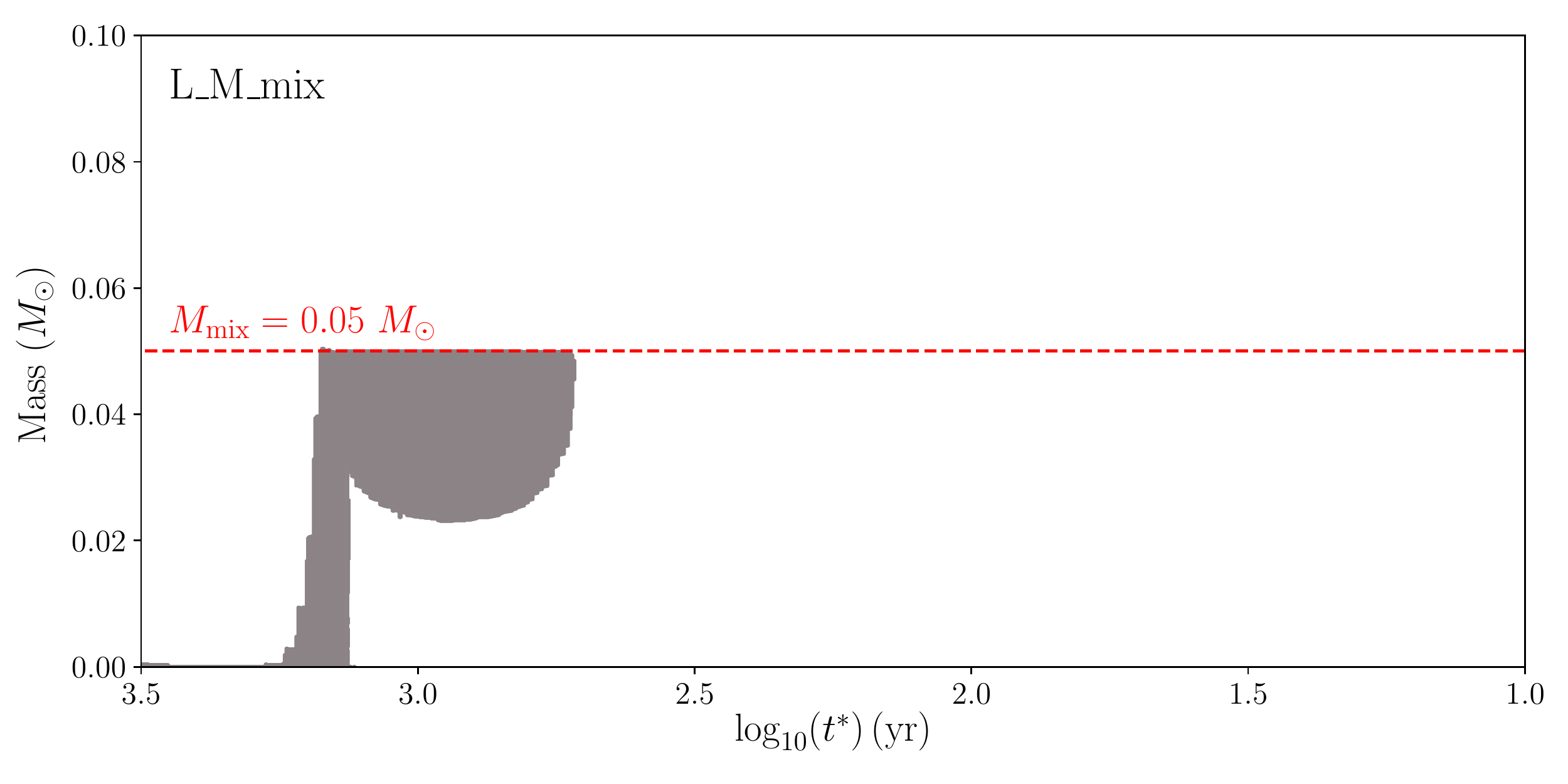}
	\caption{The Kippenhahn diagram for the accreting ONeMg core after the onset of electron capture on $^{24}$Mg. The gray shaded areas are convectively unstable. $t^*$ is the time left before the end of the calculation. The Ledoux criterion is used for the convective stability and convection is allowed only inside 0.05~$M_\odot$ to avoid numerical difficulties. A convectively unstable core is driven by $^{24}$Na$(e^-,\nu_e)^{24}$Ne, until $^{24}$Na is depleted in the central region at $\log_{10}(\rho_{\rm c}/{\rm g~cm^{-3}})\simeq9.75$. Later, the core remains convectively stable until oxygen burning is ignited. \label{fig:kip_Lmix}}
\end{figure*}

Later, the central region is heated by
the second forbidden transition of $^{20}$Ne$(e^-,\nu_e)^{20}$F at
$\log_{10}(\rho_{\rm c}/{\rm g~cm^{-3}}) \ge 9.8$ and cooled by the URCA process
$^{25}$Na$\leftrightharpoons^{25}$Ne at
$\log_{10}(\rho_{\rm c}/{\rm g~cm^{-3}}) \approx 9.85$. The second forbidden transition is
unable to ignite oxygen burning due to the slow increase of the electron-capture rate with respect
to the density. The oxygen ignition then takes place 
mildly off-center at $M_r=6\times10^{-4}~{M_\odot}$ when $\log_{10}(\rho_{\rm c}/{\rm g~cm^{-3}})\simeq9.96$. The convective structure for the
L\_M\_mix model is shown in Figure~\ref{fig:kip_Lmix}. A convectively
unstable core is driven by $^{24}$Na$(e^-,\nu_e)^{24}$Ne, until
$^{24}$Na is depleted in the central region at $\log_{10}(\rho_{\rm
	c}/{\rm g~cm^{-3}})\simeq9.75$. Heat released by electron capture on $^{20}$Ne does
not result in the convection of the core.

\subsection{Evolution of ONeMg Core with Schwarzshild Criterion}
If we use the Schwarzschild criterion, the convection already emerges
in the central region due to electron capture on $^{24}$Mg. The
convective core continues to grow and eventually reaches the layers
with ongoing URCA process of $^{23}$Na$\leftrightharpoons^{23}$Ne. Similar to \cite{2015MNRAS.447.2696D} and
\cite{2017ApJ...851..105S}, the convective URCA process heats the core
substantially in the MESA model. Due to this heating process, the core
starts to expand at $\log_{10} ({\rho_c/{\rm g~cm^{-3}}})\simeq9.7$. As the work done by
convection to transport degenerate electrons to the high density region is not
self-consistently taken into account, this heating could be
artificial. For our purpose, we get rid of the convective URCA process
by two means. One is to mute the URCA reactions when the convective
core reaches the corresponding layers (model ``S\_no\_URCA"). Another
is to limit the convective core below the URCA cooling shells (model
``S\_$\rho$\_mix"). The URCA process of
$^{25}$Na$\leftrightharpoons^{25}$Ne is muted as its threshold
density is higher than that of
$^{24}$Mg$(e^-,\nu_e)^{24}$Na. 

\begin{figure*}[t!]
	\plotone{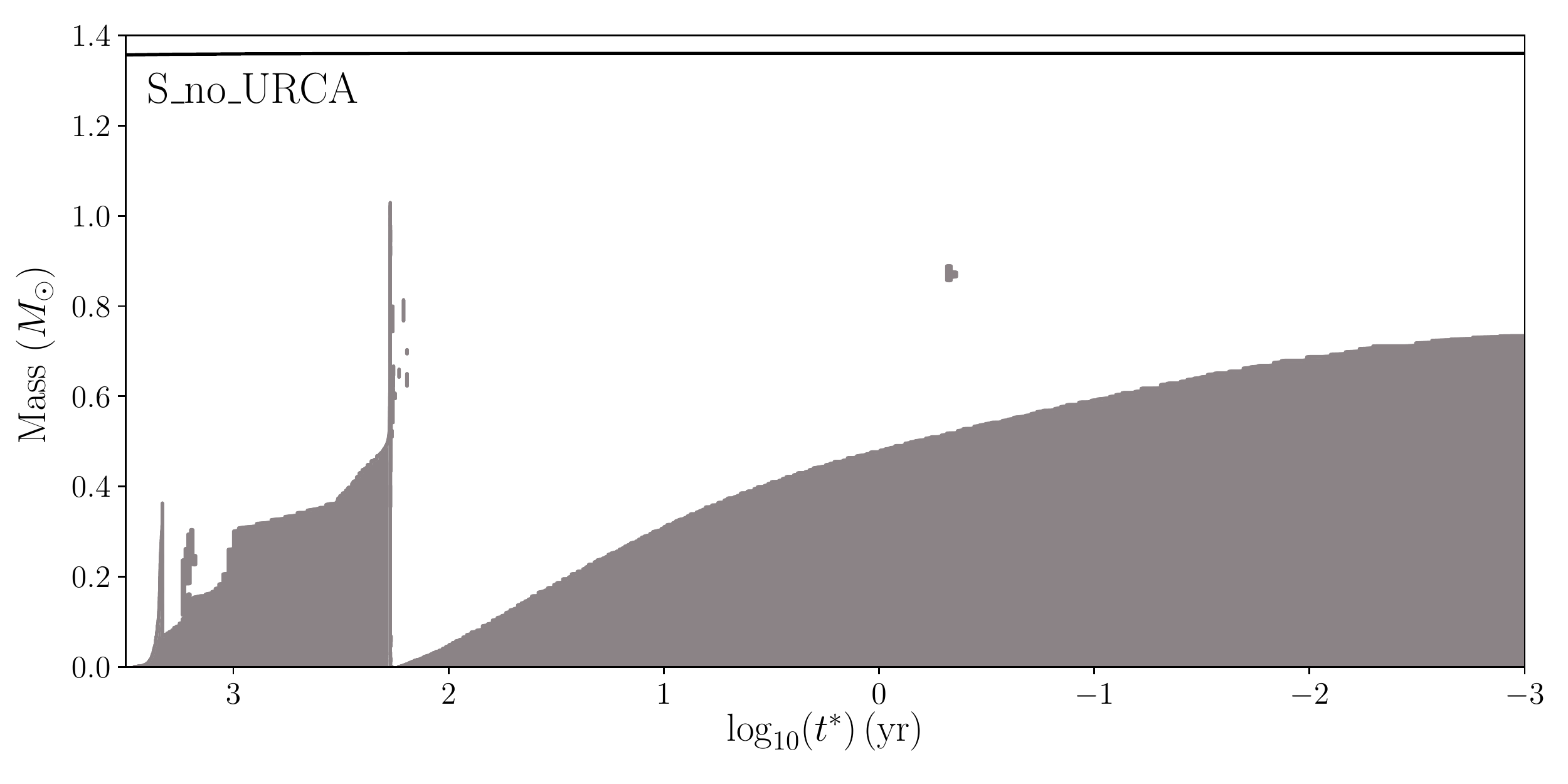}
	\caption{Same as Figure~\ref{fig:kip_Lmix}, but the Schwarzschild criterion is used for the convective stability and URCA processes are muted. When oxygen burning is ignited, the convective core grows to $\sim0.74~{M_\odot}$. \label{fig:kip_SnoURCA}}
\end{figure*}

\begin{figure*}[t!]
	\plotone{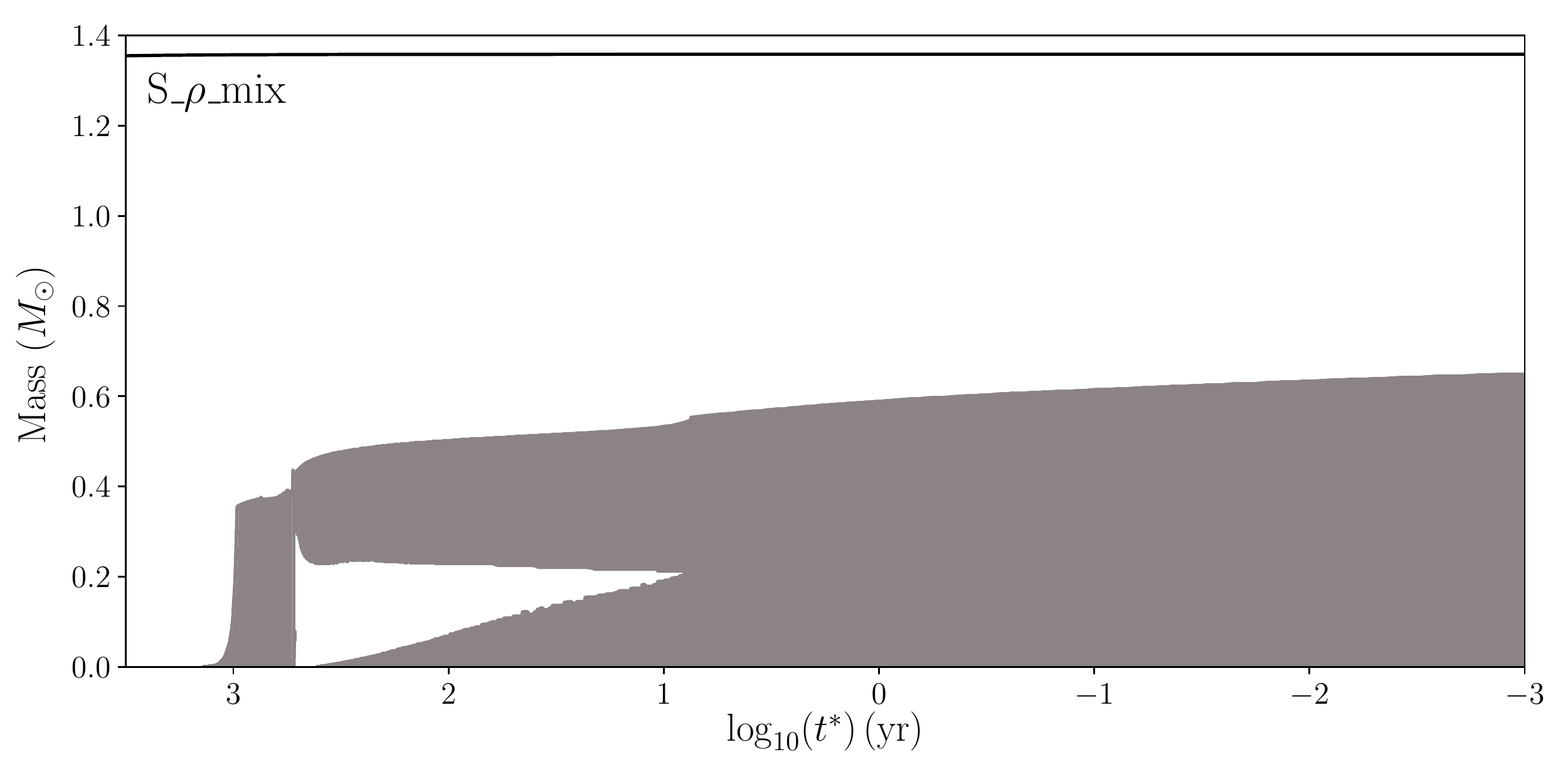}
	\caption{Same as Figure~\ref{fig:kip_Lmix}, but the Schwarzschild criterion is used for the convective stability and the convective core is limited below the URCA cooling shells. When oxygen burning is ignited, the convective core grows to $\sim0.66~{M_\odot}$. \label{fig:kip_limitRho}}
\end{figure*}

Evolution of the accreting ONeMg core in
the central density-temperature plane is also shown in
Figure~\ref{fig:rho_tc}, in comparison with the Ledoux models. In this
case, electron capture processes always make the core convectively
unstable and the temperature increases slowly but to a higher value as
more fuel is mixed into the center. Carbon burning is ignited at
$\log_{10}(\rho_{\rm c}/{\rm g~cm^{-3}})\simeq 9.8$ and cannot ignite oxygen
burning. Convective structure of these two models are shown in
Figures~\ref{fig:kip_SnoURCA} and \ref{fig:kip_limitRho}. An extended
convective core is found in both models and has a mass of
$0.74~{M_\odot}$ and $0.66~{M_\odot}$ at the oxygen ignition. 

\subsection{Oxygen Ignition}
As seen in Figure~\ref{fig:rho_tc}, the contraction of the ONeMg core
eventually leads to the oxygen ignition at $\log_{10}(T/K) \approx 9.0$.  For the Schwarzshild
criterion, due to the convective energy transport, the oxygen ignition
takes place at the center and is delayed to a higher central density than that of the Ledoux models.

\begin{figure}[t!]
	\plotone{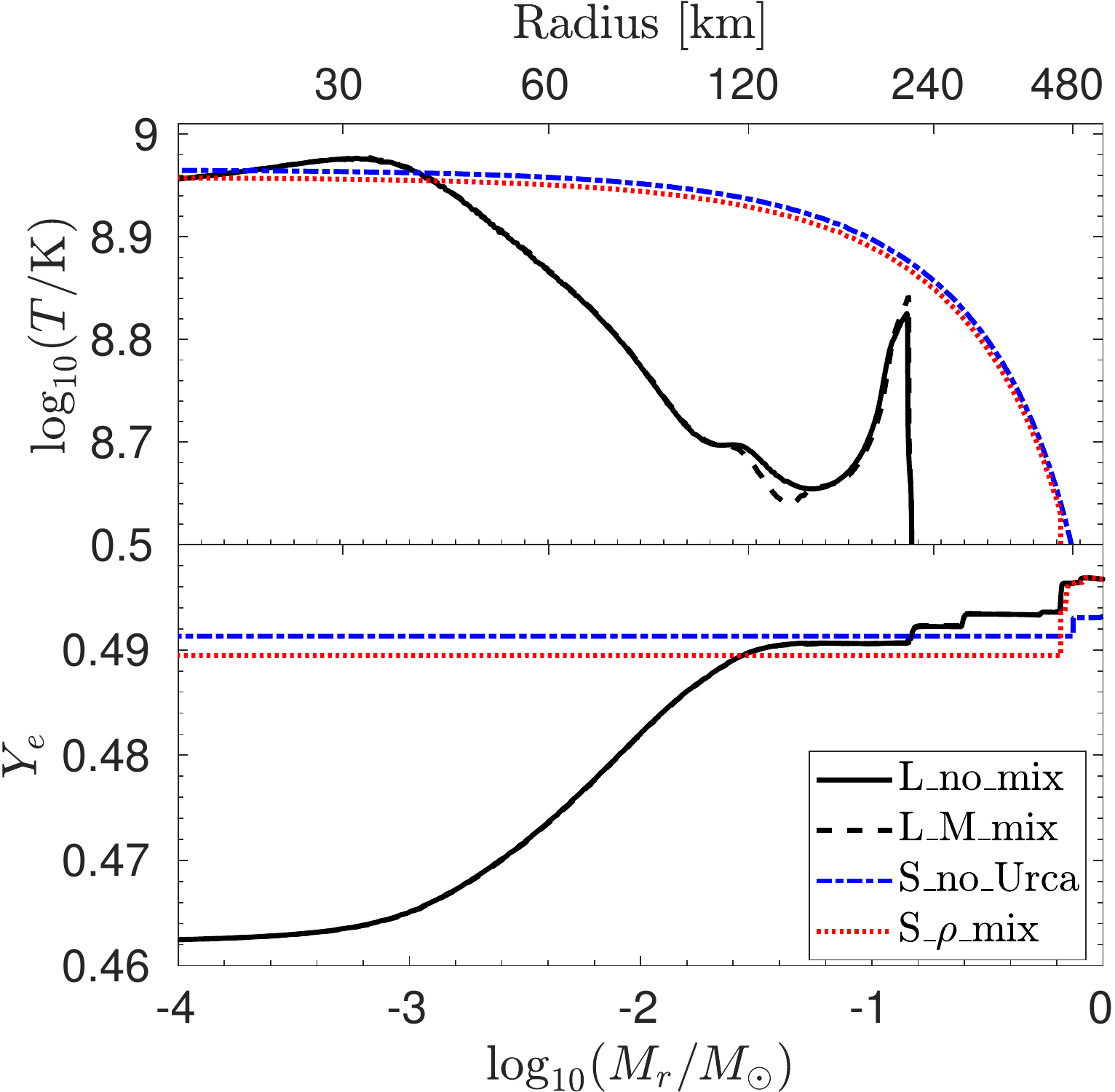}
	\caption{Temperature (top panel) and $Y_e$ (bottom panel) profiles at the oxygen ignition for the 4 models with different treatments of the convection.  The two models with the Ledoux criterion, i.e., L\_no\_mix and L\_M\_mix, overlap with each other.  \label{fig:final}}
\end{figure}

We compare the temperature and $Y_e$ profiles at the oxygen
ignition for the above 4 models in Figure~\ref{fig:final}. \added{The temperature profile is determined by heating due to the electron capture and compression, cooling due to the thermal and URCA neutrino losses, and the energy transport.} 

For the Ledoux criterion, the result
with allowing the convection \replaced{inside $M_{\rm mix}=0.05~{M_\odot}$}{at $M_r<M_{\rm mix}=0.05~M_\odot$} is identical to that
with convection suppressed.  The temperature profile is
super-adiabatic at the central region stabilized by the positive $Y_e$
gradient. \added{As the electron capture on $^{24}$Mg takes place at $M_r\sim0.1~M_\odot$, its heating results in another temperature inversion there. The temperature drops at $M_r\lesssim0.1~M_\odot$ due to the thermal neutrino losses and falls off rapidly at $M_r\gtrsim0.1~M_\odot$ due to the rapid decrease of the electron capture rate below the threshold density. There is also a sharp $Y_e$ rise at $M_r\sim0.1~M_\odot$ corresponding to the $^{24}$Mg electron capture front.} 

The two models with the Schwarzschild criterion have a similar central
temperature structure except that the convective core is more extended in
S\_$\rho$\_mix. \added{The temperature profile is adiabatic because of the efficient convective energy transport.} Both models have
a homogeneous $Y_e$ profile in the central convective region. The key parameters
for these models\replaced{, as the inputs for the subsequent hydrodynamical
simulations, are summarized in Table~\ref{tab:star_param}.}{ are summarized in Table~\ref{tab:star_param} as the inputs for the subsequent hydrodynamical
simulations.}

\begin{table*}[t!]
	\centering
	\caption{Key parameters for the profiles at the oxygen ignition for different models, with a mass accretion rate ($\dot{M}$) of $10^{-6}$ or $10^{-7}~M_{\odot}~{\rm yr}^{-1}$. $\rho_{\rm c,ign}$ is the central density at the oxygen ignition and $Y_{e, \rm ign}$ the electron fraction at the ignited mass zone. $M_{\rm conv}$ and $M_{\rm final}$ are the masses of the convective core and whole ONeMg core, respectively. $r_{\rm ign}$ and $M_{r,{\rm ign}}$ are respectively the radial position and mass coordinate of the oxygen ignited zone, and 0 indicates the central ignition.  \label{tab:star_param}}
	\begin{tabular}{|c|l|c|c|c|c|c|c|}
		\hline
		$\dot{M}~[M_\odot~{\rm yr}^{-1}]$ & Model  & $\log_{10} (\rho_{\rm c,ign}/{\rm g~cm^{-3}})$ & $Y_{e,\rm ign}$ & $M_{\rm conv}~[ M_\odot]$ & $M_{\rm final}~[M_\odot]$ & $r_{\rm ign}$~[km] & $M_{r,{\rm ign}}~[M_\odot]$
		\\ \hline
		\multirow{4}{*}{$10^{-6}$} 
		&L\_no\_mix    & 9.96  & 0.464 & ---  & 1.361 & 32 & $0.6\times10^{-3}$ \\
		&L\_M\_mix     & 9.96  & 0.464 & ---  & 1.361 & 32 & $0.6\times10^{-3}$ \\ \cline{2-8}
		&S\_no\_URCA   & 9.98  & 0.491 & 0.74 & 1.360 & 0 & 0 \\
		&S\_$\rho$\_mix& 10.00 & 0.489 & 0.66 & 1.358 & 0 & 0 \\
		\hline
		\multirow{3}{*}{$10^{-7}$} & L\_no\_mix & 9.97 & 0.466 & --- & 1.359 & 61 & $4.4\times10^{-3}$ \\ \cline{2-8}
		&S\_no\_URCA   &  9.98 & 0.491 & 0.80 & 1.358 & 0 & 0 \\
		&S\_$\rho$\_mix& 10.00 & 0.489 & 0.66 & 1.357 & 0 & 0 \\
		\hline
	\end{tabular}
\end{table*}

\subsubsection{Off-center Oxygen Ignition}
\begin{figure}[t!]
	\plotone{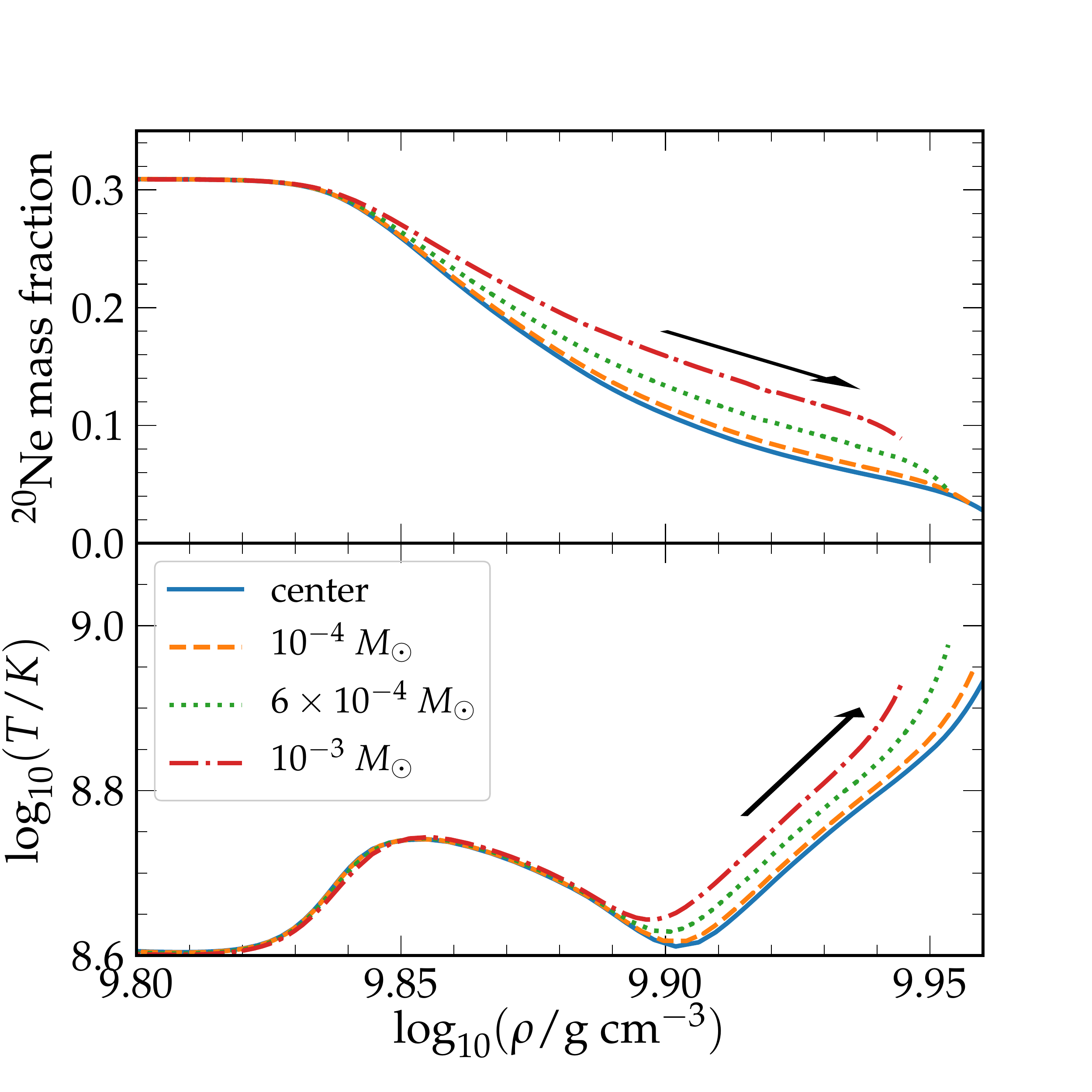}
	\caption{The evolution of the mass fraction of $^{20}$Ne \added{(the upper panel) and temperature (the lower panel)} as a function of the increasing local density for 4 mass zones for the model ``L\_no\_mix" \added{with the `Kunz' rate for  $^{12}$C$(\alpha,\gamma)^{16}$O and the core growth rate of $10^{-6}~M_\odot~\mathrm{yr}^{-1}$}. \added{Here, the density is used as a metric for time as indicated by the arrow.}. The oxygen ignition takes place at $M_r=6\times10^{-4}~{M_\odot}$. \label{fig:ne_evol}}
\end{figure}

We found that with the inclusion of the second forbidden transition for
$^{20}$Ne$(e^-,\nu_e)^{20}$F and using the Ledoux criterion, the oxygen ignition starts slightly off-center. This behavior was found
to lower the critical $\rho_{\rm c,def}$ below which the star explodes
instead of collapsing \citep{2019arXiv190111438L}. To address the reason
for this off-center ignition, we show the evolution of the mass fractions of $^{20}$Ne \added{and temperature}
as a function of the increasing local density for 4 representative mass zones in
Figure~\ref{fig:ne_evol}. The oxygen ignition takes place at
$M_r=6\times10^{-4}~{M_\odot}$. \added{During accretion, the core is compressed and the densities in the mass zones increase, so the density can be used as a metric for time as indicated by arrows in Figure~\ref{fig:ne_evol}. When reaching the same density, the outer zones has a larger $^{20}$Ne fraction than the innermost zone. This means that} electron capture on $^{20}$Ne is
slower for the outer zones than the innermost zone (`center' in
Figure~\ref{fig:ne_evol}). The higher $^{20}$Ne fraction in the outer
zone results in a larger heating effect and temperature inversion. As
a result, the oxygen ignition takes place mildly off-center.

\subsubsection{Dependence on Core Growth Rate \label{subsec:mdot}}
To test the uncertainty of progenitor properties involved in the mass growth
process, we calculate another set of models with a mass accretion rate
of $10^{-7}~M_\odot~{\rm yr}^{-1}$. The key parameters for the models at
the oxygen ignition are also listed in Table~\ref{tab:star_param}. Most of
the results show negligible differences compared to those with
$10^{-6}~M_\odot~{\rm yr}^{-1}$, except that for the case without any
convection (L\_no\_mix), the oxygen ignition takes place
further off-center at 61 km.

\subsubsection{Location of Oxygen Ignition and $^{20}$Ne Mass Fraction \label{subsec:ne20}} 

\begin{table*}[t!]
	\caption{Dependence of stellar evolution results on the $^{12}$C$(\alpha,\gamma)^{16}$O rate. \added{We used three options for the $^{12}$C$(\alpha,\gamma)^{16}$O rates in the MESA code: `Kunz' \citep{2002ApJ...567..643K} , `jina reaclib' \citep{2010ApJS..189..240C} and `CF88' \citep{1988ADNDT..40..283C}.} $X_{\rm i}$ is the initial mass fraction of the relevant element. $X_{\rm c,f}$ and $X_{\rm ign,f}$ are the final mass fractions of the relevant element in the central and oxygen ignited zones, respectively. $r_{\rm ign}$ and $M_{r,{\rm ign}}$ are the radial position and mass coordinate of the oxygen ignited zone. \label{tab:ne_fraction}}
	\centering
	\begin{tabular}{|c|c|c|c|c|c|c|}
		\hline
		Rate option & $X_{\rm i}(^{20}\rm Ne)$ & $X_{\rm i}(^{16}\rm O)$ & $X_{\rm c,f}(^{20}\rm Ne)$  & $X_{\rm ign,f}(^{20}\rm Ne)$  & $r_{\rm ign}$~[km] & $M_{r,{\rm ign}}~[M_\odot]$ \\ \hline
		Knuz 			& 	0.311 & 0.570 & 0.026 & 0.039 & 32 & $0.6\times10^{-3}$  \\ \hline
		jina reaclib	& 	0.325 & 0.549 & 0.036 & 0.036 & 0  & 0 \\ \hline 
		CF88 			& 	0.296 & 0.595 & 0.018 & 0.043 & 45 & $1.8\times10^{-3}$  \\
		\hline	
	\end{tabular}
\end{table*}

\begin{figure}[t!]
	\plotone{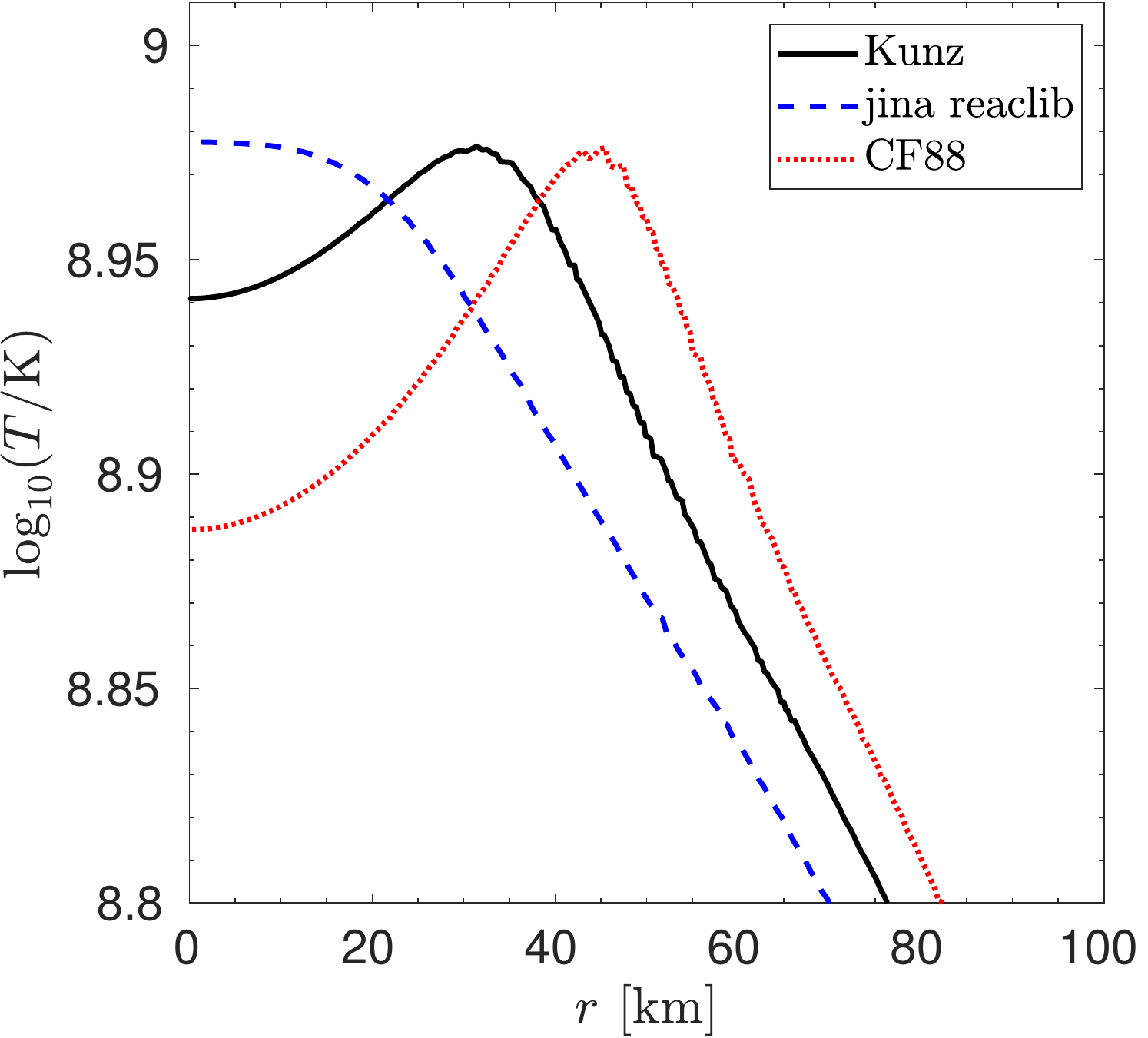}
	\caption{Comparison of the temperature profiles at the oxygen ignition for different $^{12}$C$(\alpha,\gamma)^{16}$O rates. \added{The temperature peak indicates the oxygen ignition site. For the model with the `jina reaclib' rate, the mass fraction of $^{20}$Ne is large enough to result in the central ignition. The `CF88' rate gives the lowest $^{20}$Ne fraction, and so the ignition site has the largest distance from the center.} \label{fig:n20_effect}}
\end{figure}

There is still an uncertainty for the reaction rate of $^{12}$C$(\alpha,\gamma)^{16}$O \citep{2016ApJ...817L...5A}. It is difficult to measure the rate at energies relevant for astrophysics. This rate affects the relative abundance of $^{12}$C and $^{16}$O after He burning, thus affecting the mass fraction of $^{20}$Ne $X(^{20}\rm Ne)$ after carbon burning. We check how this uncertainty affects our results by using three available rates for $^{12}$C$(\alpha,\gamma)^{16}$O in the MESA code. In the previous models, we use `Kunz' \citep{2002ApJ...567..643K}, which results in $X(^{20}\rm Ne)=0.311$. The other rates are `jina reaclib' \citep{2010ApJS..189..240C} and `CF88' \citep{1988ADNDT..40..283C}\footnote{The CF88 rate in MESA is larger than the original rate by a factor of 1.7.}, which result in $X(^{20}\rm Ne)=0.325$ and $X(^{20}\rm Ne)=0.296$, respectively. 

In Figure~\ref{fig:n20_effect} we compare the final temperature profiles for the three $^{12}$C$(\alpha,\gamma)^{16}$O rates. The convection is suppressed as in the `L\_no\_mix' model. Table~\ref{tab:ne_fraction} summarizes $X(^{20}{\rm Ne})$ and $X(^{16}{\rm O})$ before and after electron capture on $^{20}$Ne and the location of the oxygen ignition. Although the difference in $X(^{20}\rm Ne)$ is relatively small, the ignition position of oxygen burning is different.  In particular, for the largest $X(^{20}\rm Ne)$ with the `jina reaclib' rate, the oxygen ignition takes place at the center. For the smallest $X(^{20}\rm Ne)$ with the `CF88' rate, the ignition takes place further off-center at $r_{\rm ign}=$ 45~km. The difference in the ignition position affects the final outcome of the hydrodynamical phase, which is explored in Section~\ref{sec:hydro}.

\subsubsection{Effects of Residual Carbon}   
In the ONeMg core, there is a trace amount of residual carbon 
\citep{2019ApJ...872..131S}. In our case, the $\sim1\%$ carbon is not 
enough to ignite oxygen burning at low density even if no mixing is 
allowed.  However, \cite{2019ApJ...872..131S} showed that with 
$\sim3\%$ residual carbon (in a lower $M_{\rm ZAMS}$ star) and without 
mixing, oxygen burning is ignited at $\log_{10}(\rho_{\rm c,ign}/{\rm 
	g~cm^{-3}})\sim9.7$.  It is important to investigate how the convective mixing 
affects the results for this high carbon abundance as well as the 
convective URCA process associated with the carbon burning. 

\subsection{From Oxygen Ignition to Deflagration \label{subsec:runaway}}
We stop the MESA calculations at the oxygen ignition when the energy generation rate by oxygen burning exceeds thermal neutrino losses at the mass zone with the maximum energy generation rate. 
At this state, the heating timescale by the local oxygen burning is estimated as 
\begin{equation}
\tau_{\rm burn\_o}=c_p T / \varepsilon_{\rm burn\_o}
\end{equation}
where $c_p$ is the specific heat at constant pressure and $\varepsilon_{\rm burn\_o}$ the nuclear energy generation rate of oxygen burning. At the oxygen ignition, $\tau_{\rm burn\_o}$ is $\sim10^{7-8}$ s, which is still 8-9 orders of magnitude larger than the dynamical timescale ($\sim0.04$ s at $\log_{10}(\rho/{\rm g~cm^{-3}})={10}$). Thus the thermonuclear runaway of local oxygen burning does not take place yet. 

Oxygen burning forms a convectively unstable region even for the Ledoux criterion. The convective region will develop above the oxygen burning region, which is numerically difficult to calculate with the current MESA code. The further evolution is estimated as follows.

Firstly, materials in the convective region will be mixed. For the Schwarzschild models, the convective mixing from the center due to oxygen burning does not make much change of the $T$ and $Y_e$ profiles seen in Figure~\ref{fig:final}.
 
For the L\_no\_mix model, $Y_e$ in the mixed region will become much higher than 0.46. If the convective region extends to $M_r \sim 0.14~M_\odot$,
$Y_e$ becomes $\sim 0.49$ as estimated from the black solid line in Figure~\ref{fig:final}. As a result, only the very small central region of
$M_r < 6 \times 10^{-4} M_\odot$ will have $Y_e \simeq 0.46$, while the outer part
will have $Y_e \sim 0.49$. Except for the very small central region, the averaged $T$ and $Y_e$ profiles may not be so different from the Schwarzschild cases.

Secondly, for all models, the timescale of the temperature rise in the burning region will become long by the convective energy transport. Then the ONeMg core will continue to rapidly contract because of electron capture in the core whose mass is close to
the ``effective'' $M_{\rm Ch}$ with low $Y_e$.
During contraction, the evolution in $\log_{10}\rho-\log_{10}T$ of the burning shell
for all cases in Figure~\ref{fig:rho_tc} will be close to
the S\_$\rho$\_mix model (red dashed line) because of the convective energy transport. 
Eventually, the
temperature reaches $\log_{10}(T/{\rm K}) \sim 9.3$ where the
thermonuclear runaway occurs.  To estimate $\rho_{\rm c,def}$ at the
runaway, we extrapolate the evolutionary paths of the burning
shell of all cases along the red dashed line of S\_$\rho$\_mix in
Figure~\ref{fig:rho_tc} from the oxygen ignition to $\log_{10}(T/{\rm
K}) \sim 9.3$.  We obtain $\log_{10}(\rho_{\rm c,def}/{\rm g~cm^{-3}})\approx10.18$ even for the off-center ignition case. This is consistent
with $\log_{10}(\rho_{\rm c,def}/{\rm g~cm^{-3}}) \simeq 10.2$ found by
\cite{2019ApJ...871..153T}, who took into account the semiconvective mixing.  (If the super-adiabatic temperature
gradient is taken into account, the runaway density would be a little
lower.)

Because of the uncertainty in the evolution from the oxygen ignition
through the thermonuclear runaway, in Section~\ref{sec:hydro} we use the initial
models with parameterized $Y_e$ distribution and $\rho_{\rm c,def}$ 
to study the parameter dependence of the hydrodynamical behavior.
For the central density, we adopt $\log_{10}(\rho_{\rm c,def}/{\rm g~cm^{-3}}) = {9.96-10.2}$.
For the $Y_e$ distribution, we adopt three cases as will be described in \S\ref{subsec:hydro_init} and shown in Figure~\ref{fig:ye_hydro}.


\section{Hydrodynamical Simulations of Electron-Capture Supernovae}
\label{sec:hydro}

\subsection{Methods}

\begin{table*}
\begin{center}
		\caption{The initial configuration and the final fate of the representative models studied in this work. The model series names ``S$\_\rho\_$mix'' and ``L$\_$no$\_$mix''
			respectively stand for the initial models obtained from the stellar
			evolution calculations where the the Schwarzschild and Ledoux criteria
			are used for the convection criterion.  ``Ledoux mix o-burn'' stands for
			the initial models calculated with the Ledoux criterion including the
			convective shell mixing due to the off-center oxygen ignition.
			In ``Conv. (Convection)", ``S" and ``L" stand for the Schwarzschild
			and Ledoux criteria, respectively, being used in the stellar evolution
			calculations.
			In ``Mix. (Mixing)", ``Y'' means the convective mixing is included in
			setting the initial $Y_{e}$ profile in the convective zone of
			``S$\_\rho\_$mix'' and ``Ledoux mix o-burn'', while ``N'' means no
			mixing is assumed after the oxygen ignition for ``L$\_$no$\_$mix''.
			``$Y_e$" is the initial value at the center for ``S$\_\rho\_$mix'' and
			``L$\_$no$\_$mix'', while it is the value in the off-center convective
			zone for ``Ledoux mix o-burn''.
			The model name with an ending `` -LM" makes it clear that the convective
			mixing after the oxygen ignition is included in ``Ledoux mix o-burn''.
			The initial central density $\rho_{{\rm c,def}}$ is in units of g
			cm$^{-3}$. Radius $R$ and initial flame position $r_{{\rm ign}}$ are
			in units of km. Mass $M$ is in unit of $M_{\odot}$. $\dot{M}$
			is the progenitor mass accretion rate in units of $M_{\odot}$
			yr$^{-1}$.  
			``Result" stands for the final fate with ``C" being collapse and
			``E" being explosion.
			\label{table:models}}
		\begin{tabular}{|c|c|c|c|c|c|c|c|c|c|}
			\hline
			Model & $\dot{M}$ & $\log_{10} (\rho_{\rm c,def})$ & $M$ & $R$ & $r_{{\rm ign}}$ & $Y_{e}$ & Conv. & Mix. & Result \\ \hline
			\multicolumn{10}{|c|}{Ledoux mix o-burn} \\ \hline
			6-0998-049-30-LM & $10^{-6}$ &  9.98  & 1.359 & 1400 & 30 & 0.49 & L & Y & E \\
			6-0999-049-30-LM & $10^{-6}$ &  9.99  & 1.359 & 1410 & 30 & 0.49 & L & Y & C \\
			6-1000-049-30-LM & $10^{-6}$ & 10.00  & 1.359 & 1420 & 30 & 0.49 & L & Y & C \\\hline
			
			7-1000-049-60-LM & $10^{-7}$ & 10.00  & 1.357 & 1370 & 60 & 0.49 & L & Y & E \\
			7-1002-049-60-LM & $10^{-7}$ & 10.02 & 1.358 & 1350 &  60 & 0.49 & L & Y & C \\ \hline
			
			\multicolumn{10}{|c|}{S$\_\rho\_$mix} \\ \hline
			
			6-0996-049-00 & $10^{-6}$ & 9.96  & 1.359 & 1410 &  0 & 0.49 & S & Y & C \\ 
			6-0996-049-30 & $10^{-6}$ & 9.96  & 1.359 & 1410 & 30 & 0.49 & S & Y & C \\ \hline                   
			
			6-1000-049-00 & $10^{-6}$ & 10.00 & 1.360 & 1370 &  0 & 0.49 & S & Y & C \\ 
			6-1000-049-30 & $10^{-6}$ & 10.00 & 1.360 & 1370 & 30 & 0.49 & S & Y & C \\ \hline
			
			7-0997-049-00 & $10^{-7}$ & 9.97 & 1.358 & 1410 &  0 & 0.49 & S & Y & C \\   
			7-0997-049-60 & $10^{-7}$ & 9.97 & 1.358 & 1410 & 60 & 0.49 & S & Y & E \\ 
			7-0999-049-60 & $10^{-7}$ & 9.99 & 1.359 & 1360 & 60 & 0.49 & S & Y & C \\ \hline
			
			7-1000-049-00 & $10^{-7}$ & 10.00 & 1.360 & 1370 &  0 & 0.49 & S & Y & C \\ 
			7-1000-049-60 & $10^{-7}$ & 10.00 & 1.360 & 1370 &  60 & 0.49 & S & Y & C \\ \hline
			
			\multicolumn{10}{|c|}{L$\_$no$\_$mix} \\ \hline
			6-0996-046-30 & $10^{-6}$ & 9.96  & 1.357 & 1430 & 30 & 0.46 & L & N & E \\
			6-0996-046-00 & $10^{-6}$ & 9.96  & 1.357 & 1430 &  0 & 0.46 & L & N & E \\
			6-0996-046-00b\footnote{The flame size is two times larger} & $10^{-6}$ & 9.96  & 1.357 & 1430 &  0 & 0.46 & L & N & C \\ \hline
			6-1000-046-30 & $10^{-6}$ & 10.00 & 1.357 & 1400 & 30 & 0.46 & L & N & E \\ 
			6-1000-046-00 & $10^{-6}$ & 10.00 & 1.357 & 1400 &  0 & 0.46 & L & N & C \\
			6-1010-046-30 & $10^{-6}$ & 10.10 & 1.361 & 1310 & 30 & 0.46 & L & N & C \\ \hline
			7-0997-046-00 & $10^{-7}$ & 9.97 & 1.355 & 1430 &  0 & 0.46 & L & N & E \\
			
			7-0999-046-00 & $10^{-7}$ & 9.99 & 1.356 & 1380 &  0 & 0.46 & L & N & E \\
			7-1000-046-00 & $10^{-7}$ & 10.00 & 1.357 & 1400 &  0 & 0.46 & L & N & C \\ \hline
			7-0997-046-60 & $10^{-7}$ & 9.97 & 1.355 & 1430 & 60 & 0.46 & L & N & E \\
			7-1000-046-60 & $10^{-7}$ & 10.00 & 1.357 & 1400 &  60 & 0.46& L & N & E \\ 
			7-1002-046-60 & $10^{-7}$ & 10.02 & 1.357 & 1360 &  60 & 0.46& L & N & E \\
			7-1005-046-60 & $10^{-7}$ & 10.05 & 1.358 & 1330 &  60 & 0.46& L & N & C \\ \hline
		\end{tabular}
\end{center}
\end{table*}

We use a 2D hydrodynamics code primarily developed
for the supernova modeling \citep{Leung2015a}. The code has been applied
to study Type Ia supernova \citep{Leung2015b, 2017hsn..book.1275N, Leung2018_Chand,Leung2019_subChand},
accretion-induced collapse \citep{Leung2019_DMAIC, Zha2019}
and ECSN \citep{2017hsn..book..483N, 2019PASA...36....6L}.
Here we briefly review the algorithms particularly relevant to the 
modeling of ECSN. We refer the interested readers to the detailed 
implementation reported in \cite{2019arXiv190111438L}. 

The code solves the 2D Euler equations using the 
fifth-order WENO scheme for spatial discretization \citep{Barth1999} and 
five-step third-order NSSP Runge-Kutta scheme for time discretization \citep{Wang2007}. 
We use the Helmholtz equation of state \citep{Timmes1999a}.
For the propagation speed of the oxygen deflagration, we implement sub-grid scale turbulence models
introduced in \cite{Clement1993} and \cite{Niemeyer1995b},
with the turbulent flame model given in 
\cite{Pocheau1994}, \cite{Reinecke1999b,Reinecke2002a} and \cite{Schmidt2006b}.
We use the laminar flame speed as a function of the density and composition
given in \cite{1992ApJ...396..649T}.
To capture the geometry of oxygen deflagration we use the
level-set method \citep{Reinecke1999a} with reinitialization \citep{Sussman1993}. We use the 
three-step nuclear reaction to represent the energy
production by nuclear burning \citep{Townsley2007,Calder2007}.
Effects of binding energy changes, neutrino energy
losses and mass differences between electron-proton pair and
neutron are included for matter in NSE. The individual electron-capture rates
of isotopes in the NSE ash are taken from \cite{1985ApJ...293....1F,Oda1994,1999PhRvL..83.4502M,Nabi1999}.\footnote{We remark that the computation of the electron-capture rate is still uncertain because the rate table for matter in NSE with
	$Y_e \leq 0.4$ relies on multiple tables. In particular, rates for isotopes with the mass number
	$A=45-110$ are based on \cite{Nabi1999}. These rates have been calculated in
	\cite{2010NuPhA.848..454J} based on the more sophisticated large-scale shell model as in
	\cite[see e.g.][]{Langanke2000} but the actual values are unavailable yet. }

We start the hydrodynamical phase of evolution by mapping the MESA model
onto a 2D grid in cylindrical coordinates
with a uniform spatial resolution of $\Delta x \approx$ 4 km.  (Details of the initial models
are described in the \S\ref{subsec:hydro_init}.)
To trigger the initial flame, we consider a central
flame of a ``three-finger" structure \citep[see a similar illustration in][]{Reinecke1999b} and 
an off-center flame with a one-bubble 
structure \citep[also see][]{Reinecke1999b}. 
The bubble is put at $r_{\rm ign}=$ 30 or 60 km away from the center\footnote{The location of the oxygen ignition is $r_{\rm ign}=32$ (61) km from the stellar evolutionary models with $\dot{M}=10^{-6}~(10^{-7})~{M_\odot~\mathrm{yr}^{-1}}$. For the hydrodynamical simulations, the results are not sensitive to the exact values for the adopted finite grid resolution.}.
The initially ignited matter is assumed to be burnt into NSE. A typical mass of $\sim 10^{-4}$ to $10^{-3}~M_{\odot}$
for the initial ash is assumed. In all simulations, we follow the propagation
of the oxygen deflagration wave until the ONeMg core reaches
a central density of $\log_{10}(\rho_{\rm c}/{\rm g~cm^{-3}})={10.7}$ ($9.0$)
for the collapse (explosion) case. 

\subsection{Initial Models \label{subsec:hydro_init}}

\begin{figure}[t!]
	\plotone{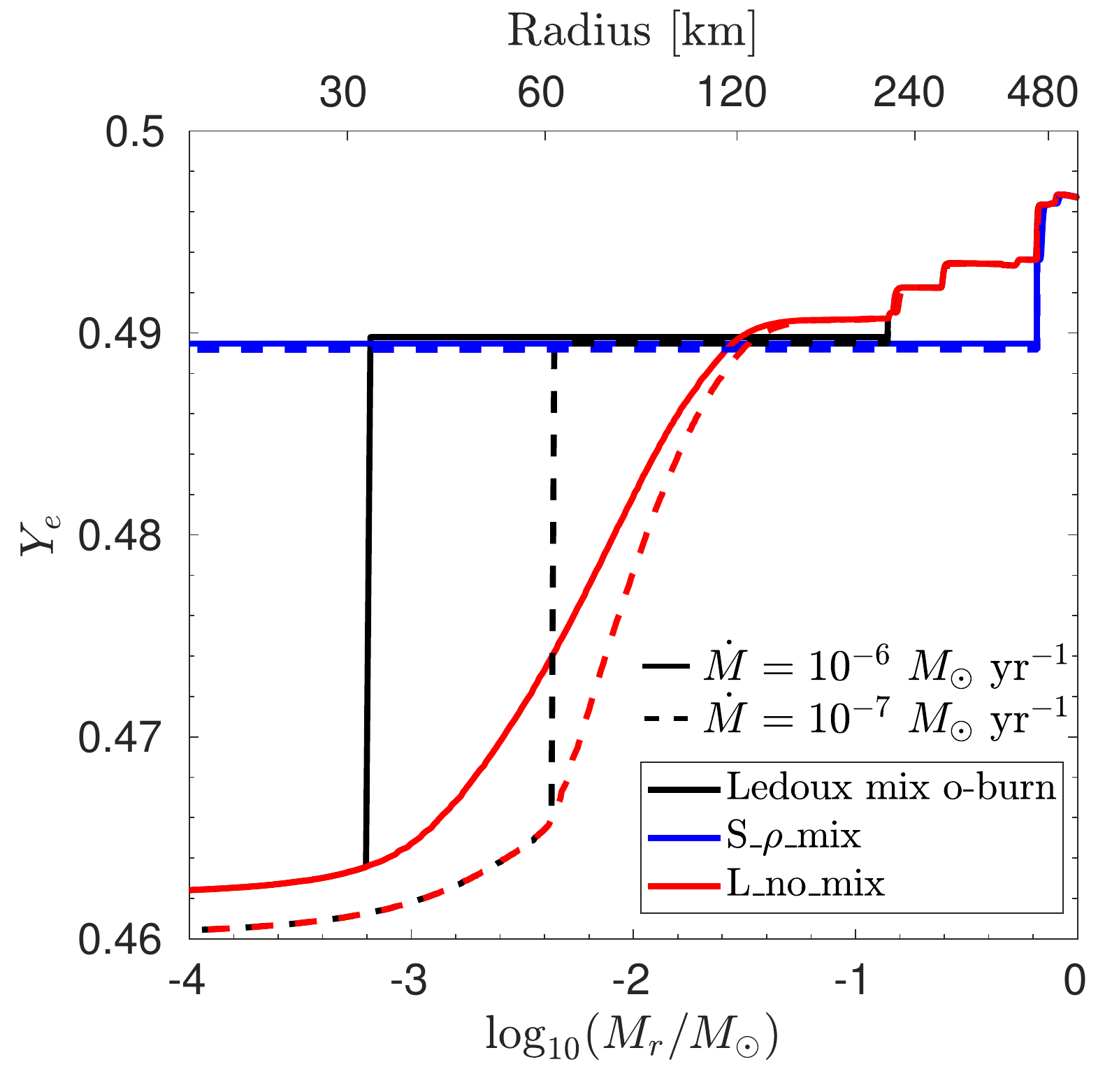}
	\caption{$Y_e$ profiles for the construction of initial models for the hydrodynamical simulations, with different assumptions of the convective mixing as described in \S\ref{subsec:hydro_init}. The solid (dashed) lines are with $\dot{M}=10^{-6}~(10^{-7})~M_\odot~{\rm yr}^{-1}$. In the Ledoux models, the position of the oxygen ignition is $r_{\rm ign}=32$ (61) km and $M_r=0.6\times10^{-3}$ ($4.4\times10^{-3}$) $M_\odot$ for $\dot{M}=10^{-6}~(10^{-7})~M_\odot~{\rm yr}^{-1}$. \label{fig:ye_hydro}}
\end{figure}

In building the initial models in the hydrostatic equilibrium at the
initiation of the deflagration, we use the stellar evolutionary models
with $\dot{M} = 10^{-6} ~M_{\odot}$ yr$^{-1}$ and $10^{-7} ~M_{\odot}$
yr$^{-1}$.  We use the $^{12}$C$(\alpha,\gamma)^{16}$O rates by `Kunz' (Table~\ref{tab:star_param}) and
also `jina\_reaclib' (Table~\ref{tab:ne_fraction}) to include the case of the central oxygen ignition.

For the $Y_e$ and temperature profiles, we take into account the
dependence on the convection criteria as shown in
Figure~\ref{fig:final} and Table~\ref{tab:star_param}. We also take into account the
convection which develops after the oxygen ignition even for the
Ledoux criterion (\S\ref{subsec:runaway}).
The convection mixes the high $Y_{e}$  matter in the outer part
of the ONeMg core with the low $Y_{{e}}$ materials at the oxygen-burning
site (center or off-center).

Therefore, for the initial $Y_e$ profile, 
we constructed the following 3 cases (1)-(3) shown in Figure
\ref{fig:ye_hydro}.  We examine the
dependence of the final fate of the ONeMg core on these initial
$Y_{e}$ distributions in \S\ref{subsec:hydro_mix} - \ref{subsec:hydro_nomix}. More details on the
configuration are described in each subsection.

\noindent
(1) ``Ledoux mix o-burn": L$\_$no$\_$mix + mixed region above the oxygen ignited shell (\S\ref{subsec:hydro_mix}). 

\noindent
(2) ``S$\_\rho\_$mix": Schwarzshild criterion with almost full mixing. This also accounts for the convective mixing after the central ignition due to the usage of the `jina\_reaclib' rate for the Ledoux criterion (\S\ref{subsec:hydro_cmix}).

\noindent
(3) ``L$\_$no$\_$mix": Ledoux criterion with no mixing (\S\ref{subsec:hydro_nomix}).

In these models, $\rho_{\rm c,def}$ is a model parameter. As discussed in
\S\ref{subsec:runaway}, the convective energy transport above the oxygen ignited shell can significantly delay the thermonuclear runaway,
thus increasing $\rho_{\rm c,def}$. However, the exact details of the convective energy transport and mixing remain unknown due to numerical difficulties with MESA.
Therefore, the exact $\rho_{\rm c,def}$ when the deflagration starts
and its position are not well determined.  As estimated in \S\ref{subsec:runaway},
$\log_{10}(\rho_{\rm c,def}/{\rm g~cm^{-3}})$ can be as high as ${10.18}$.
Here we take $\log_{10}(\rho_{\rm c,def}/{\rm g~cm^{-3}})$ ranging
from ${9.96}$ to ${10.2}$.

We do not directly map the MESA density profile because we find that the
discretization produces global motion of the ONeMg core, which may affect the initial propagation of the flame and the final fate. Instead, we recalculate the hydrostatic equilibrium explicitly for a central density $\rho_{\rm c,def}$, with $Y_e$ and temperature as a function of $M_r$.

In Table~\ref{table:models} we tabulate the parameters and the
outcomes of the hydrodynamical simulations for the models studied in this
work. We name the models as follows. In 6-0996-046-30, 7-1002-049-60,
and 6-0998-049-30-LM,, for example,

\noindent
(a) ``6'' and ``7'' stand for the ``6"-series and ``7"-series
progenitors evolved with $\dot{M} = 10^{-6}~M_{\odot}$~yr$^{-1}$ and
$10^{-7}~M_{\odot}$~yr$^{-1}$, respectively.

\noindent
(b) ``0996'', ``0998" and ``1002'' stand for $\log_{10}(\rho_{\rm c,def}/{\rm g~cm^{-3}})$ = 9.96, 9.98
and ${10.02}$, respectively.

\noindent
(c) ``046'' and ``049'' stand for $Y_e = 0.46$ and 0.49,
respectively,  at the center of
case (3) models (L$\_$no$\_$mix) and case (2) models (S$\_\rho\_$mix).  $Y_e$ of case (1) models (Ledoux mix o-burn) is
shown as ``049'' (see (e) below).

\noindent
(d) ``30'' and ``60'' stand for the initial flame at a distance of 30
and 60 km from the center, respectively.

\noindent
(e) Models with an ending ``-LM'' represent those of case (1) above,
i.e., L\_no\_mix + mixed region above the oxygen-ignited shell. In these models, ``049" stands for $Y_e=0.49$ in the oxygen-burning mixed shell.

\subsection{Off-Center Runaway with Mixing (Ledoux mix o-burn ``LM'' Models) \label{subsec:hydro_mix}}

These models come from the evolution of the ONeMg core using the Ledoux criterion
with the off-center convective mixing (see Ledoux mix o-burn in Figure~\ref{fig:ye_hydro}).
When the off-center oxygen burning is ignited, the generated energy drives the convection from the burning location
to the outer part.

To construct the initial models, we first use the $Y_{e}$ and
temperature profiles obtained from Section~\ref{sec:evol}. Then we
estimate the convective mixing which produces approximately $Y_{e} =$ 0.49.
Within the region of $M_r < M_{r,{\rm ign}}$, no mixing is assumed and
the $Y_{e}$ profile is directly taken from the stellar
evolutionary model.
Thus we set the following $Y_{e}$ distribution as seen in Figure~\ref{fig:ye_hydro}. At $M_r < M_{r, \mathrm{ign}}$, 
$Y_{e} = $ 0.46 - 0.47, while at
$M_{r, \mathrm{ign}} <  M_r < 0.14 ~M_\odot$, $Y_{e} = $ 0.49.
At $M_r > 0.14 ~M_\odot$ $Y_{e}$ follows the stellar evolutionary
model again.

We do not change the temperature
since the matter is extremely degenerate such that the role of 
temperature is unimportant compared to $Y_{\rm e}$. 

In \S\ref{subsub:rhoc}, we run the models with $\rho_{\rm c,def}$ as a parameter to see how the final fate depends on it. 

\subsubsection{$\rho_{\rm c,def}$-dependence \label{subsub:rhoc}}
\begin{figure*}[t!]
	\begin{center}
		\includegraphics*[width=8cm, height=6cm]{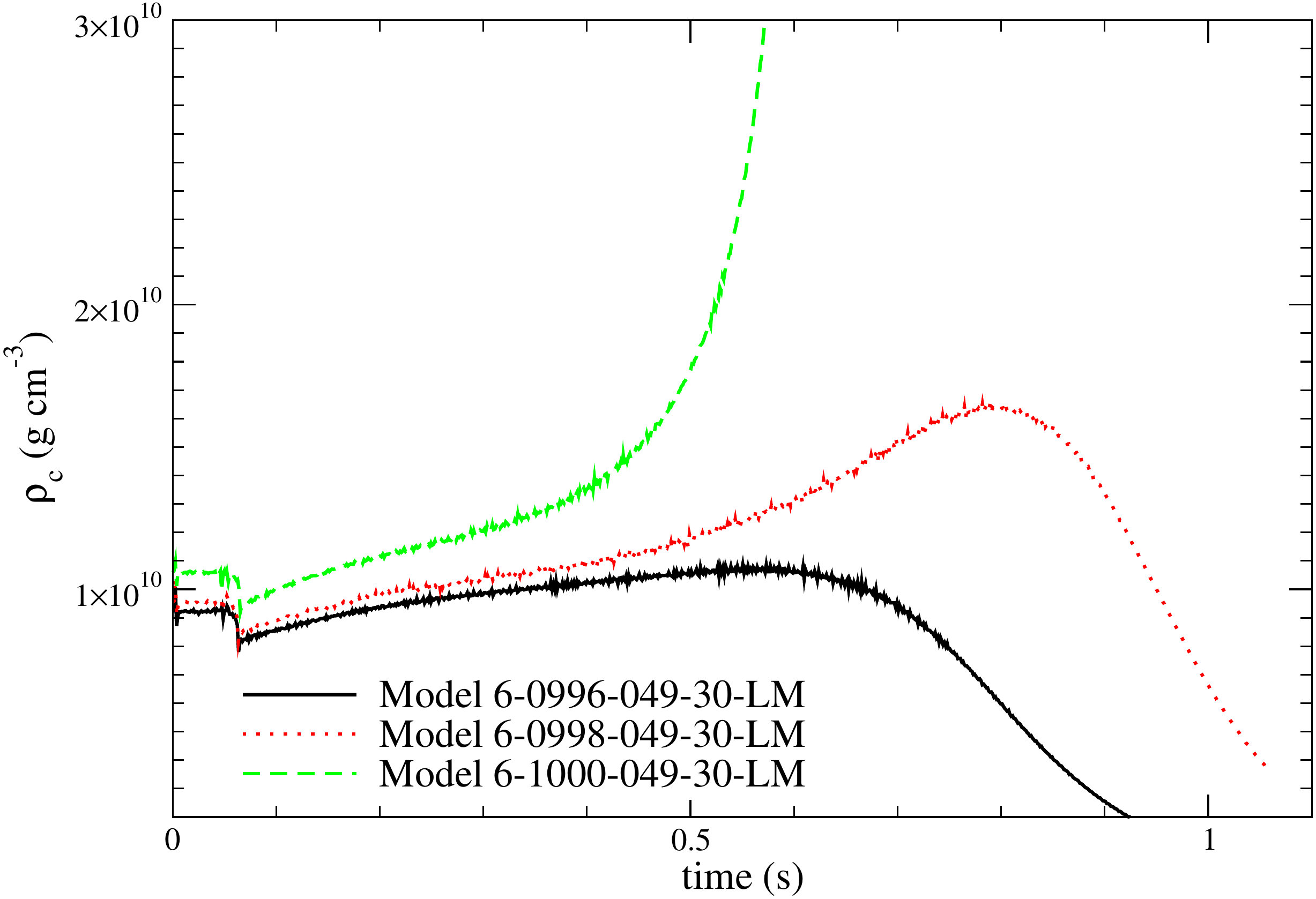}
		\includegraphics*[width=8cm, height=6cm]{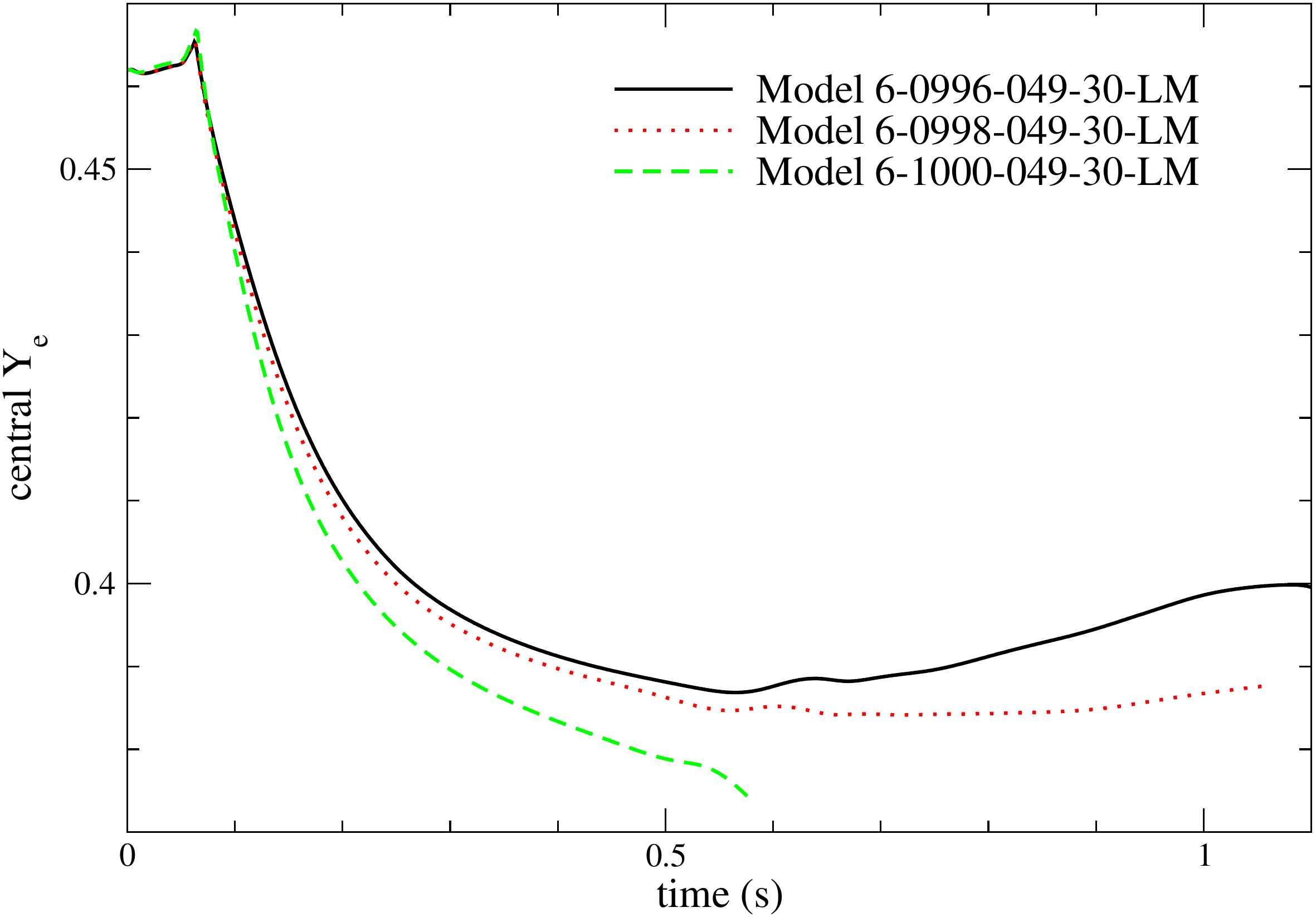}
		\caption{\added{$\rho_{{\rm c,def}}$-dependence of ``Ledoux mix o-burn" models for $\dot{M}=10^{-6}~M_\odot~\mathrm{yr}^{-1}$. }(left panel) The central density evolution of Models 
			6-0996-049-30-LM (black solid line), 6-0998-049-30-LM (red dotted line) 
			and 6-1000-049-30-LM (green dashed line). \added{The time lapse of $\sim0.1$ s is the time for the flame to arrive at the center to trigger the initial expansion. The collapsing model (green dashed line) shows a monotonic increase of the central density after the early expansion, while the other two exploding models show a turning point after which the star expands due to the energy input by oxygen deflagration.}
			(right panel) Similar to the left panel but for the central $Y_{e}$.}
	\end{center}
	\label{fig:rhoc_L_plot}
\end{figure*}

\begin{figure*}[t!]
	\begin{center}
		\includegraphics*[width=8cm, height=6cm]{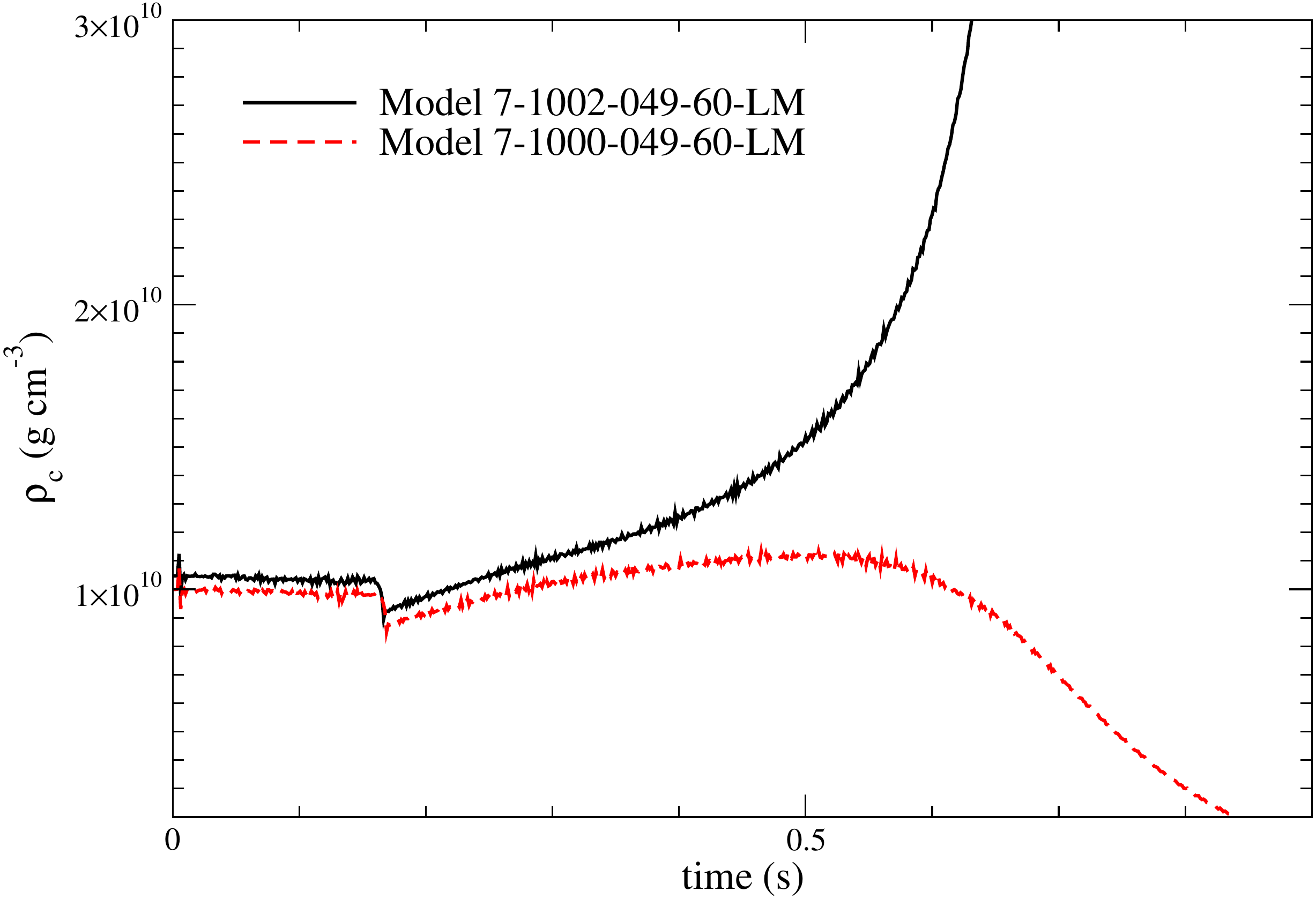}
		\includegraphics*[width=8cm, height=6cm]{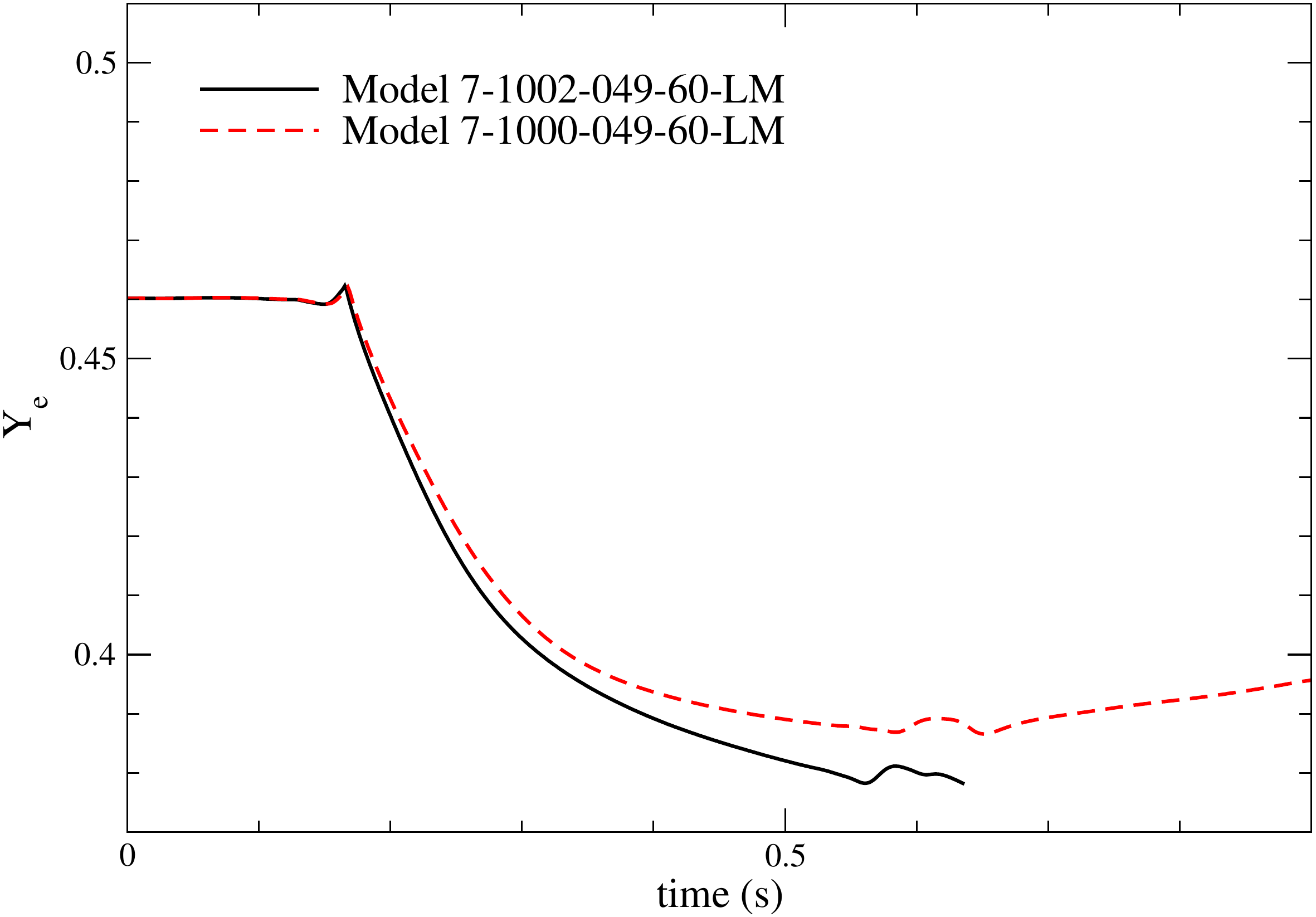}
		\caption{\added{$\rho_{{\rm c,def}}$-dependence of ``Ledoux mix o-burn" models for $\dot{M}=10^{-7}~M_\odot~\mathrm{yr}^{-1}$. }(left panel) The central density evolution of Models 
			7-1002-049-60-LM (black solid line) and 7-1000-049-60-LM (red dashed line). \added{Similar to Figure~\ref{fig:rhoc_L_plot}, the time lapse of $\sim0.2$ s is the time for the flame to arrive at the center to trigger the initial expansion. The collapsing model (black solid line) shows a monotonic increase of the central density after the early expansion, while the exploding model (red dashed line) shows a turning point after which the star expands due to the energy input by oxygen deflagration.}
			(right panel) Similar to the left panel but for the central $Y_{e}$.}
	\end{center}
	\label{fig:rhoc_L_plot2}
\end{figure*}

In Figure \ref{fig:rhoc_L_plot} we plot the central density 
and central $Y_{e}$ against time for Models 6-0996-049-30-LM, 6-0998-049-30-LM and 6-1000-049-30-LM
in the left and right panels respectively. 

There is a time lapse $\sim 0.1$ s which is the 
time for the flame to arrive at the center to trigger the early expansion. 
The model which collapses shows a monotonic increase in the central density
after the early expansion. For $\log_{10}(\rho_{\rm c,def}/{\rm g~cm^{-3}}) \gtrsim 10.00$, 
the ONeMg core collapses to form a NS.
Models which eventually explode
show a turning point in the central density evolution. 
This is the moment when the energy input by the oxygen deflagration
dominates the dynamical process in the star. We remark that for 
the model close to the bifurcation point, i.e., Model 6-0998-049-30,
the central density at the turning point is as high as $\log_{10}(\rho_{\rm c}/{\rm g~cm^{-3}})=10.18$.

We also study the hydrodynamical outcomes for the model set with 
$\dot{M} = 10^{-7} ~M_{\odot}$~yr$^{-1}$. 
In Figure \ref{fig:rhoc_L_plot2} we plot the central density and $Y_e$ evolution for Models 7-1000-046-60-LM and 7-1002-049-60-LM.
Compared with the ``6"-series, these two models have a farther off-center ignition at 60 km, which requires a longer time ($\sim 0.2$ s) for the flame 
to reach the center. It thus provides more time for the 
flame to develop in its size and surface area, which may balance
the contraction after electron capture occurs in the center.

When the flame arrives at the center, the heated core again rapidly 
expands by $\sim 20 \%$. Then the rapid electron capture in the 
NSE ash induces the first contraction.
For $\log_{10}(\rho_{\rm c,def}/{\rm g~cm^{-3}})=10.02$, the core continues to collapse. For $\log_{10}(\rho_{\rm c,def}/{\rm g~cm^{-3}})=10.00$, the expansion starts at $t \approx 0.6$ s. The electron capture fails to trigger sufficiently strong contraction before the 
flame can release the necessary energy to make the star explode. 

\subsection{Centered Runaway with Mixing (S$\_\rho\_$mix Models) \label{subsec:hydro_cmix}}

When we apply the Schwarzschild criterion, convection can develop in the core before the oxygen ignition.  The convective flow transports heat away from
the center, which is the first place expected for the nuclear runaway, and uniformly mix the material as seen in Figure~\ref{fig:final}. 
To construct the models in Figure~\ref{fig:ye_hydro}, we adopt the $Y_e$ profile of S$\_\rho\_$mix in Figure~\ref{fig:final}. 
Major differences in 
the initial models from those in \S\ref{subsec:hydro_mix} are the flat
$Y_{{e}}$ distribution in the core and the centered flame.

We also notice that this scenario is also possible for the Ledoux
criterion. As described in Table~\ref{tab:ne_fraction}, the exact abundance
of $^{12}$C and $^{20}$Ne depends on the nuclear reaction 
rate. When we use the updated reaction rate `jina\_reaclib'
the higher $X(^{20}\mathrm{Ne})$ leads to the oxygen ignition at the center. And even with the Ledoux criterion, a convective core driven by the oxygen burning develops from the center afterwards. 

\subsubsection{$\rho_{\rm c,def}$-dependence}

\begin{figure*}[t!]
	\begin{center}
		\includegraphics*[width=8cm, height=6cm]{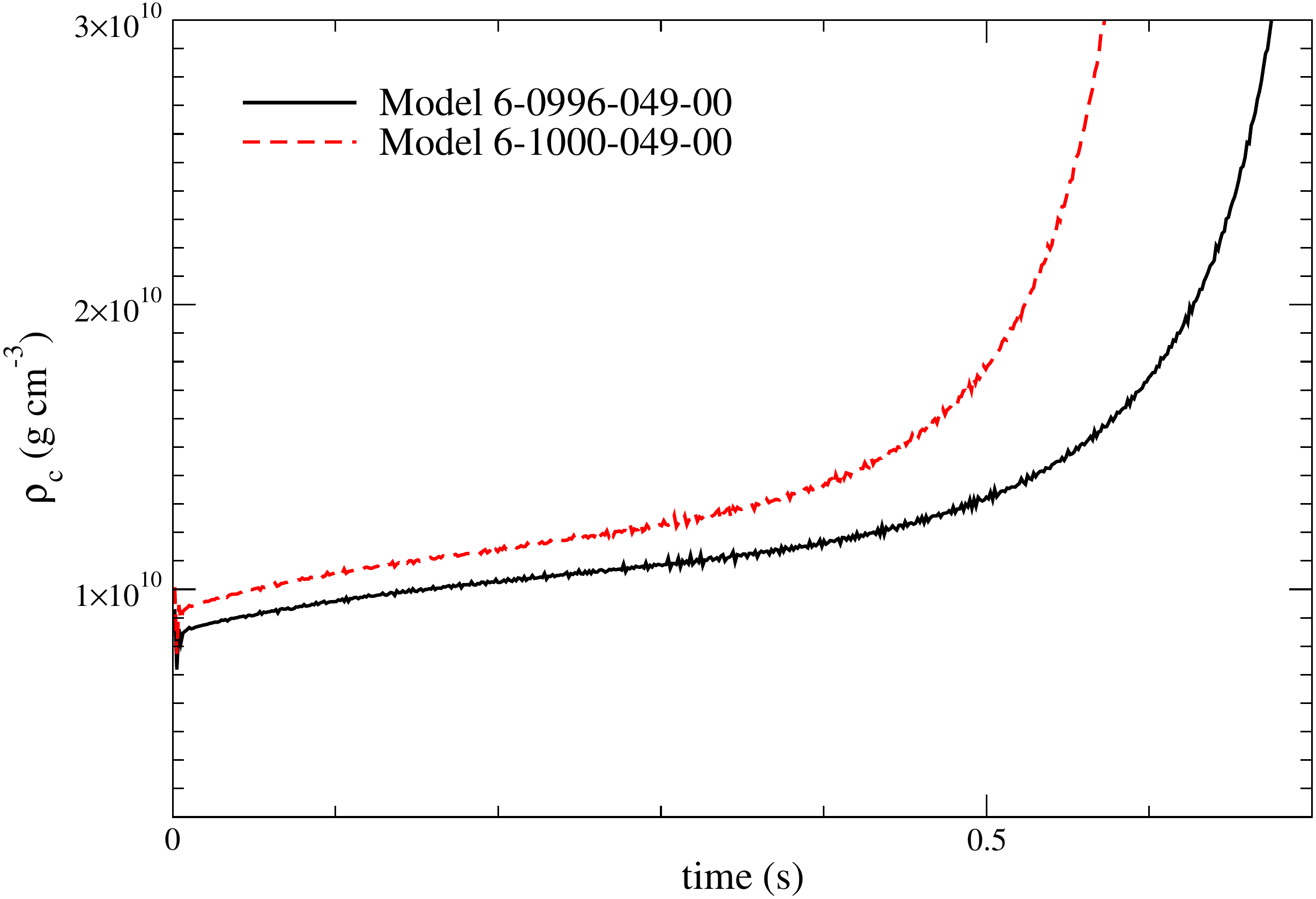}
		\includegraphics*[width=8cm, height=6cm]{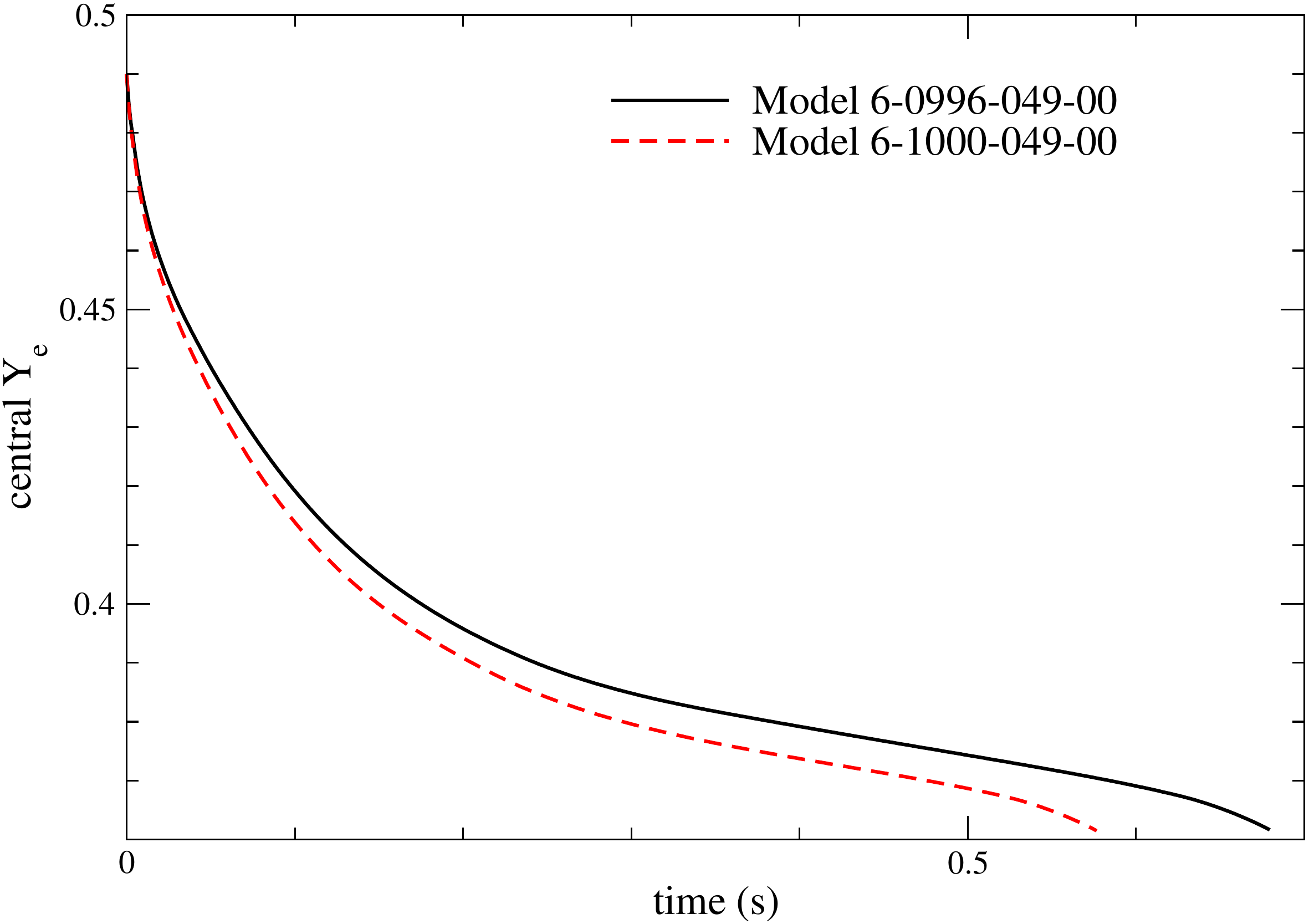}
		\caption{\added{$\rho_{{\rm c,def}}$-dependence of ``S\_$\rho$\_mix" models for $\dot{M}=10^{-6}~M_\odot~\mathrm{yr}^{-1}$. }(left panel) The central density evolution of Models 
			6-0996-049-00 (black solid line) and
			6-1000-049-00 (red dashed line).
			Model 6-1000-049-00 corresponds to the 
			stellar evolutionary model S$\_\rho\_$mix. 
			(right panel) Similar to the left panel but for the central $Y_{{\rm e}}$.}
	\end{center}
	\label{fig:rhoc_rhoc_S_plot}
\end{figure*}

\begin{figure*}[t!]
	\begin{center}
		\includegraphics*[width=8cm, height=6cm]{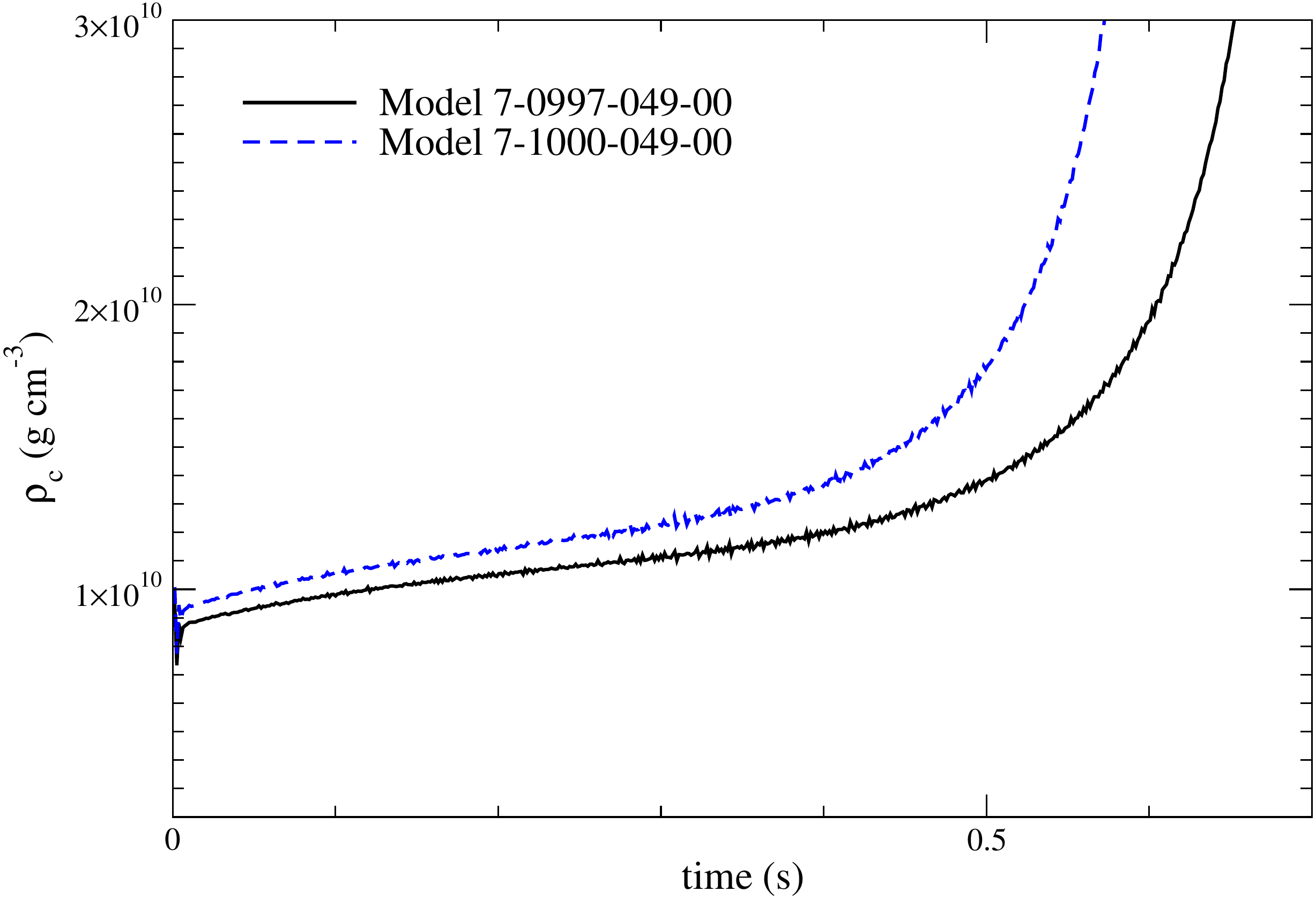}
		\includegraphics*[width=8cm, height=6cm]{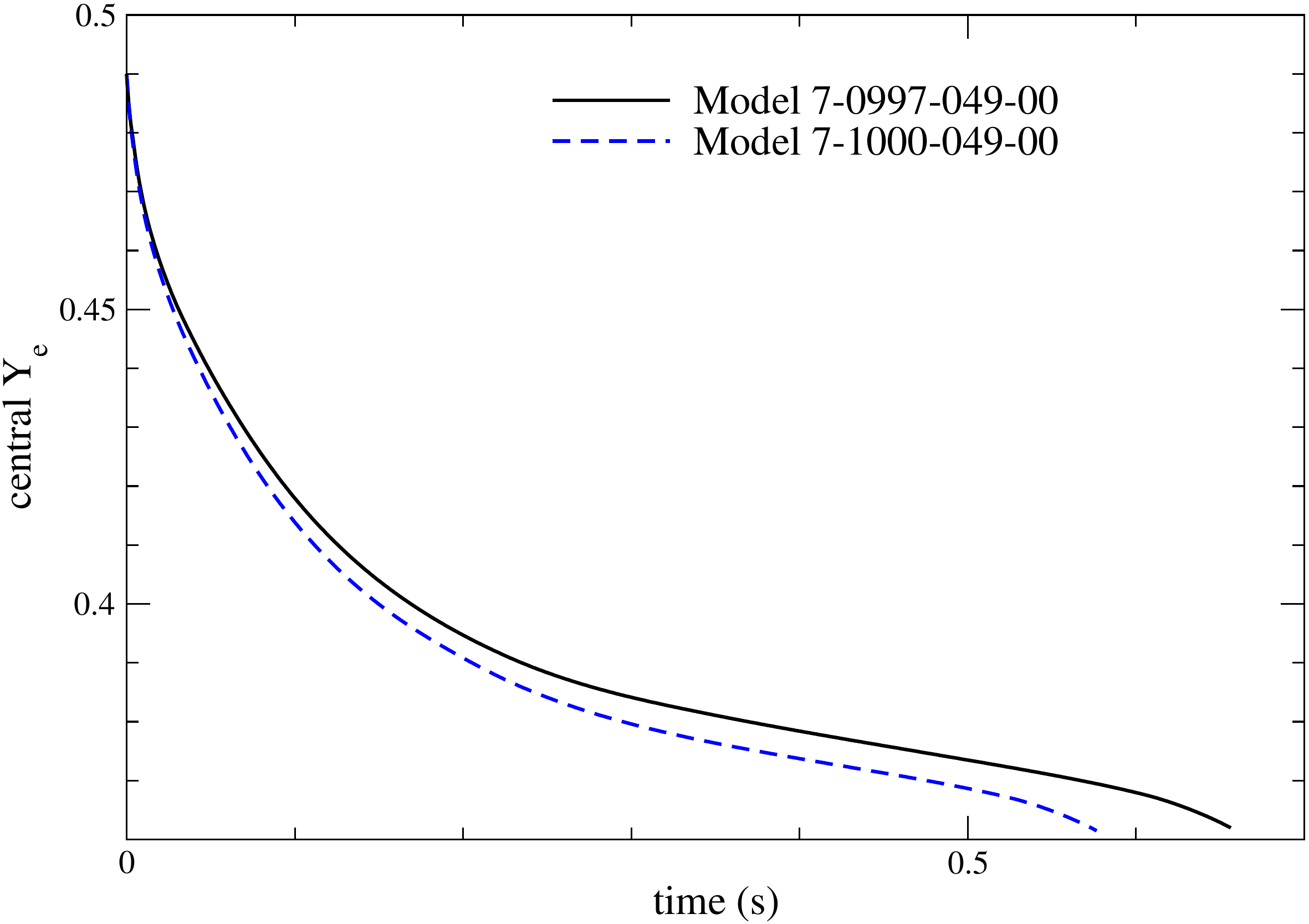}
		\caption{\added{$\rho_{{\rm c,def}}$-dependence of ``S\_$\rho$\_mix" models for $\dot{M}=10^{-7}~M_\odot~\mathrm{yr}^{-1}$. }(left panel) The central density evolution of Models 
			7-0997-049-00 (black solid line) and 
			7-1000-049-00 (blue dashed line).
			Model 7-1000-049-00 corresponds to the 
			stellar evolutionary model S$\_\rho\_$mix. (right panel) Similar to the left panel but for the central $Y_{{\rm e}}$.}
	\end{center}
	\label{fig:rhoc_rhoc_S_plot2}
\end{figure*}

As discussed in \S\ref{subsec:runaway}, the core continues to contract to a higher
$\rho_{\rm c,def}$ until the thermonuclear runaway starts.
Since the exact $\rho_{\rm c,def}$ depends on the efficiency of the
convective energy transport, we examine how the outcome of the
deflagration depends on $\rho_{\rm c,def}$.

In the left panel of Figure~\ref{fig:rhoc_rhoc_S_plot} we show the central density evolution of two models with $\log_{10}(\rho_{\rm c,def}/{\rm g~cm^{-3}})=9.96$ and 10.00. 
Both models directly collapse. 
The minimum $\log_{10}(\rho_{\rm c,def}/{\rm g~cm^{-3}})$ for the ONeMg core to collapse 
is 9.96, which is even lower than 10.00 of the MESA model with the Schwarzschild criterion (S\_$\rho$\_mix). 
We also plot the central $Y_{e}$ evolution in the right panel, which smoothly decreases without any bump. 

The ``7"-series models with different $\rho_{\rm c,def}$ (7-0997-049-00 and 7-1000-049-00) are plotted in the Figure \ref{fig:rhoc_rhoc_S_plot2}. 
The high $Y_{e}$ (=0.49) again allows the ONeMg core to collapse 
at $\log_{10}(\rho_{\rm c,def}/{\rm g~cm^{-3}})=9.97$, which is lower than $\log_{10}(\rho_{\rm c,ign}/{\rm g~cm^{-3}})=$10.00 of stellar evolutionary model (S\_$\rho$\_mix). 

\subsubsection{$r_{\rm ign}$-dependence \label{subsec:Sch_off}}

\begin{figure*}[t!]
	\begin{center}
		\includegraphics*[width=8cm, height=6cm]{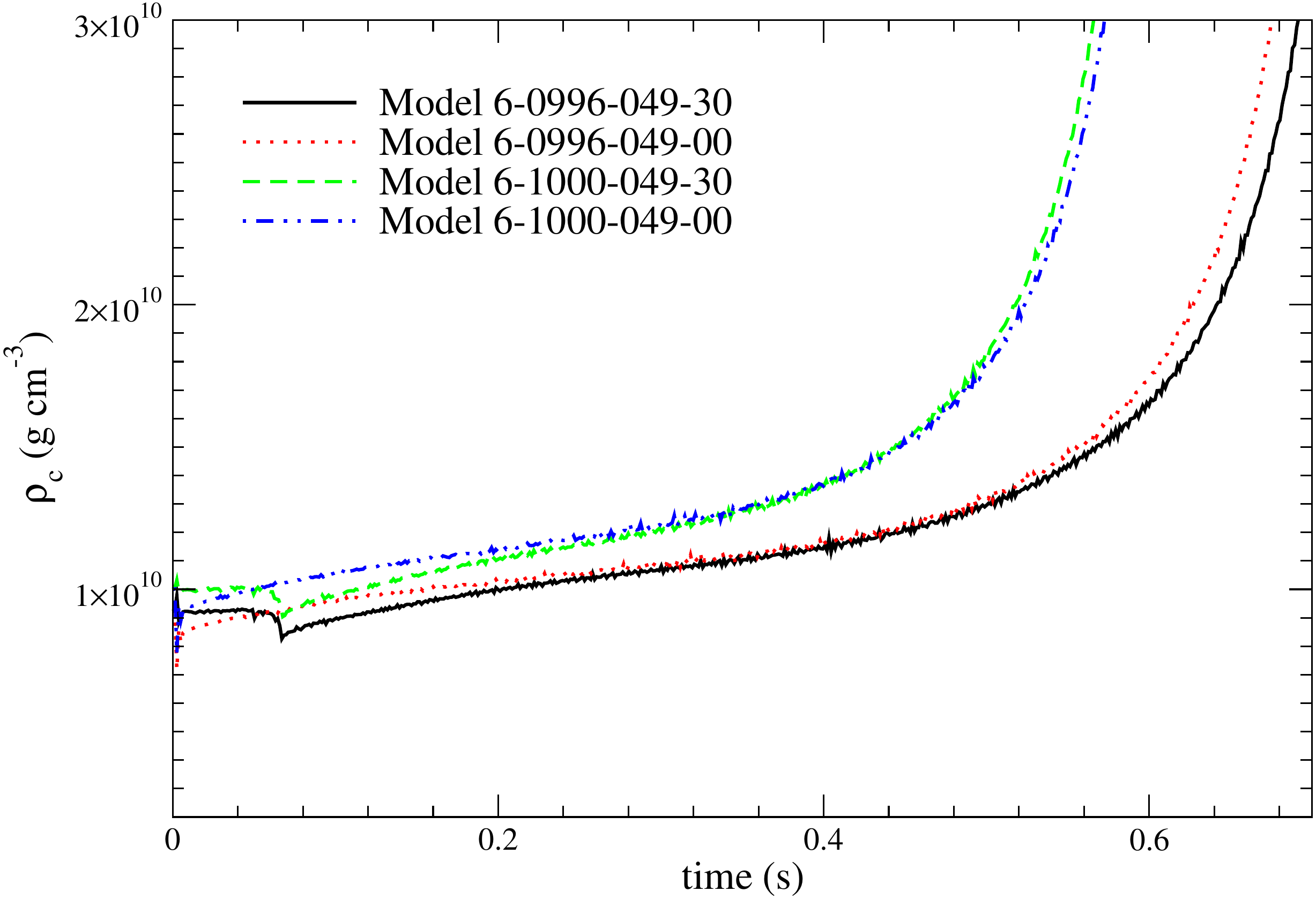}
		\includegraphics*[width=8cm, height=6cm]{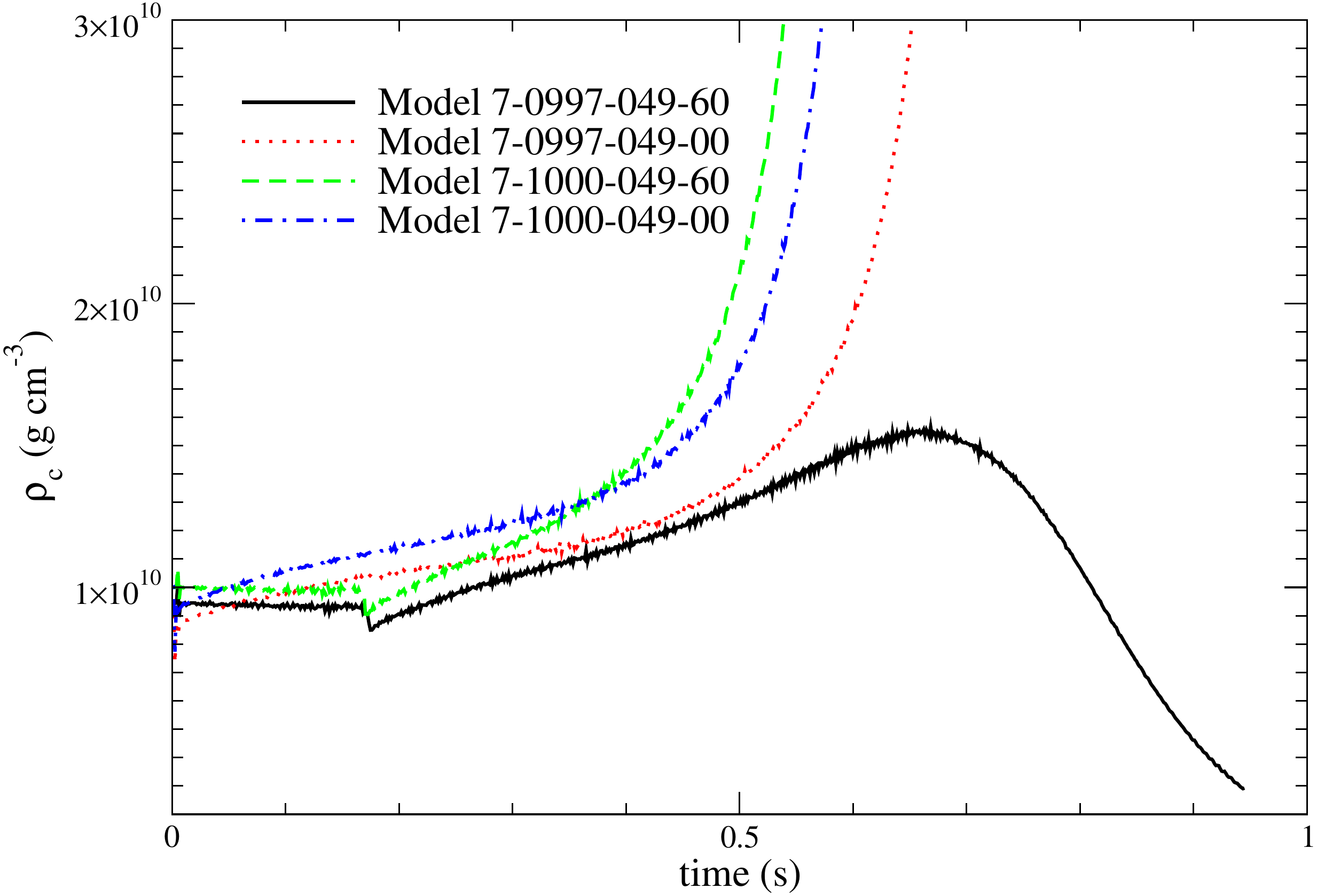}
		\caption{\added{$\rho_{{\rm c,def}}$ and $r_{\rm ign}$-dependence of ``S\_$\rho$\_mix" models for $\dot{M}=10^{-6}$ and $10^{-7}~M_\odot~\mathrm{yr}^{-1}$. }(left panel) The central density evolution of Models 
			6-0996-049-30 (black solid line), 6-0996-049-00 (red dotted line), 
			6-1000-049-30 (green dashed line),
			6-1000-049-00 (blue dot-dashed line).
			(right panel) Similar to the left panel but for the 
			Models 7-0997-049-60 (black solid line), 7-0997-049-00 (red dotted line), 
			7-1000-049-60 (green dashed line),
			7-1000-049-00 (blue dot-dashed line).}
	\end{center}
	\label{fig:rhoc_flamepos_plot}
\end{figure*}

For models with the central oxygen ignition, the center is the most likely position
for the oxygen deflagration to start because the convection developed from the center will smooth out any temperature inversion in the star. 
However, when the convective flow is strong, 
the potential fluid parcel which will undergo the nuclear runaway
may be carried away by the flow before the runaway is triggered. 
As a result, an off-center flame can be developed. 
Therefore, the exact $r_{\rm ign}$ could be 
non-zero and depends on the detailed characteristic of the convective flow. 
Here, we study the uncertainties in this parameter.

We simulate the propagation of the oxygen deflagration with
different initial flame structures. In Figure \ref{fig:rhoc_flamepos_plot}
we compare the evolution of Models 6-0996-049-30, 6-0996-049-00,
6-1000-049-30 and 6-1000-049-00. They are two sets of models with 
the centered (-00) and off-center flame (-30). The two $\rho_{\rm c,def}$ 
correspond to $\rho_{\rm c,ign}$, i.e., the lowest cases of $\rho_{\rm c,def}$ obtained from the Ledoux (0996) and Schwarzschild (1000) criteria, respectively.

Figure~\ref{fig:rhoc_flamepos_plot} shows that all four models directly collapse. As discussed in the previous section,
the position of the initial flame affects the early evolution of the central
density. Models with a centered flame show a rapid drop in $\rho_c$ at the beginning, but then 
the following electron capture makes the core contract again and 
$\rho_c$ increase until the simulations end. Models with 
an off-centered flame show no change in $\rho_c$ until the 
flame arrives at the center at $t \approx 0.08$ s. After the rapid drop by $\sim 10 \%$, 
$\rho_c$ increases again until the end of simulations. 
Therefore, for models with $r_{\rm ign}=$ 30 km, 
the position of the  initial flame is less important for the final fate of the ONeMg core.

In the right panel of Figure~\ref{fig:rhoc_flamepos_plot} we plot the central density evolution for Models 7-0997-049-60, 7-0997-049-00,
7-1000-049-60 and 7-1000-049-00. The two models with a higher $\rho_{\rm c,def}$  collapse.
However, different from the models with $r_{\rm ign}=$ 30 km, the lower $\rho_{\rm c,def}$
model with a centered flame collapses while that with an off-center flame ($r_{\rm ign}=$ 60 km)
explodes. 

\subsection{Off-center Runaway without Mixing (L$\_$no$\_$mix Models) \label{subsec:hydro_nomix}}

Here we examine the model developed from the 
model L$\_$no$\_$mix. This is another limiting case in our model survey,
where we assume no convective mixing appears despite that the oxygen burning creates
a convectively unstable region even with the Ledoux criterion.
As a result, the oxygen-ignited site becomes the site for the nuclear runaway. 

\subsubsection{$\rho_{\rm c,def}$-dependence}

We examine the dependence of the evolution of the ONeMg core on the
initial central density $\rho_{\rm c,def}$.  Even neglecting the convective energy
transport, the timescale of the temperature rise due to early phase of the
oxygen burning is still longer than the timescale of core contraction due
to electron capture.  Therefore $\rho_{\rm c,def}$ can become somewhat
higher than $\rho_{\rm c,ign}$.

\begin{figure*}[t!]
	\begin{center}
		\includegraphics*[width=8cm, height=6cm]{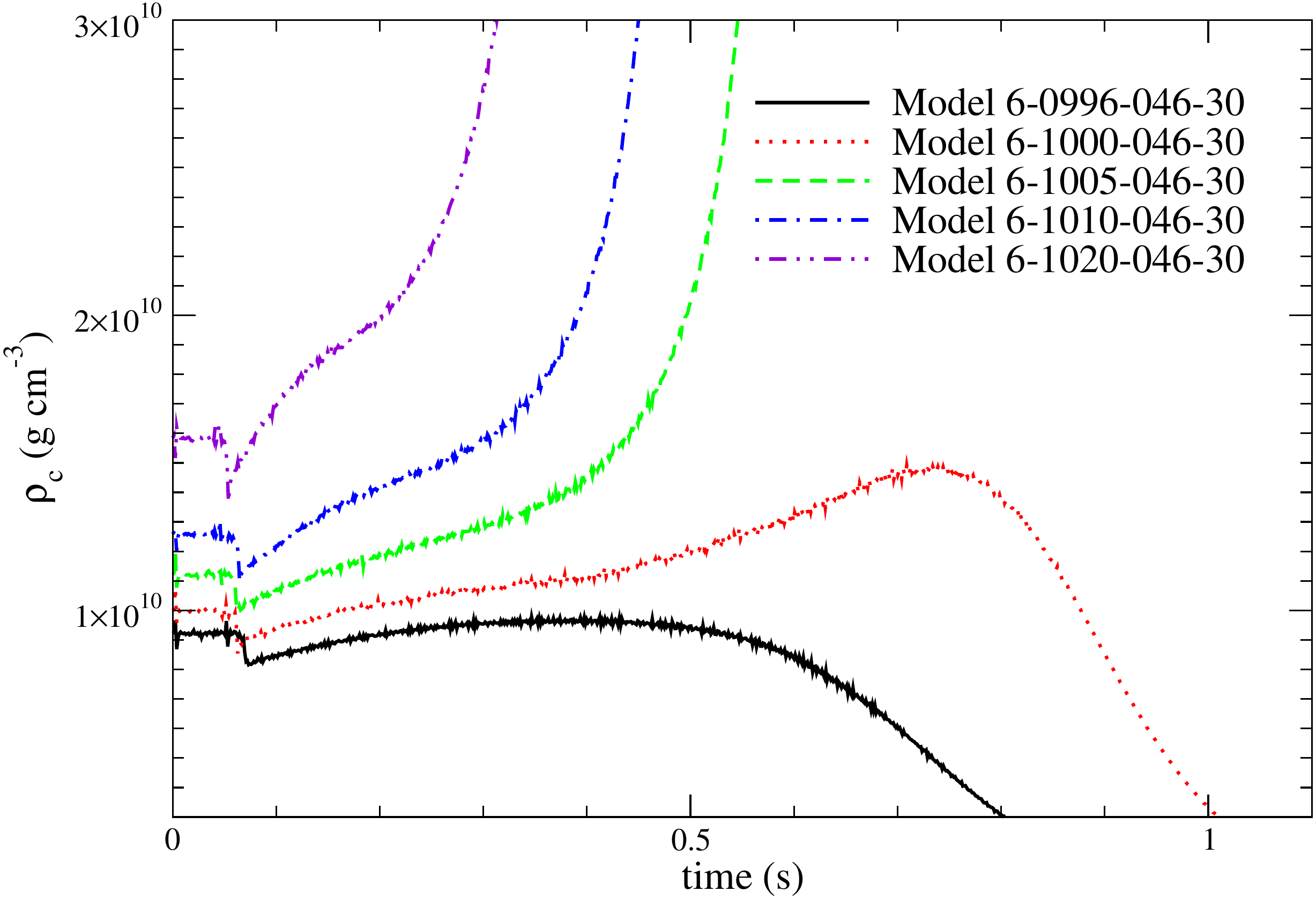}
		\includegraphics*[width=8cm, height=6cm]{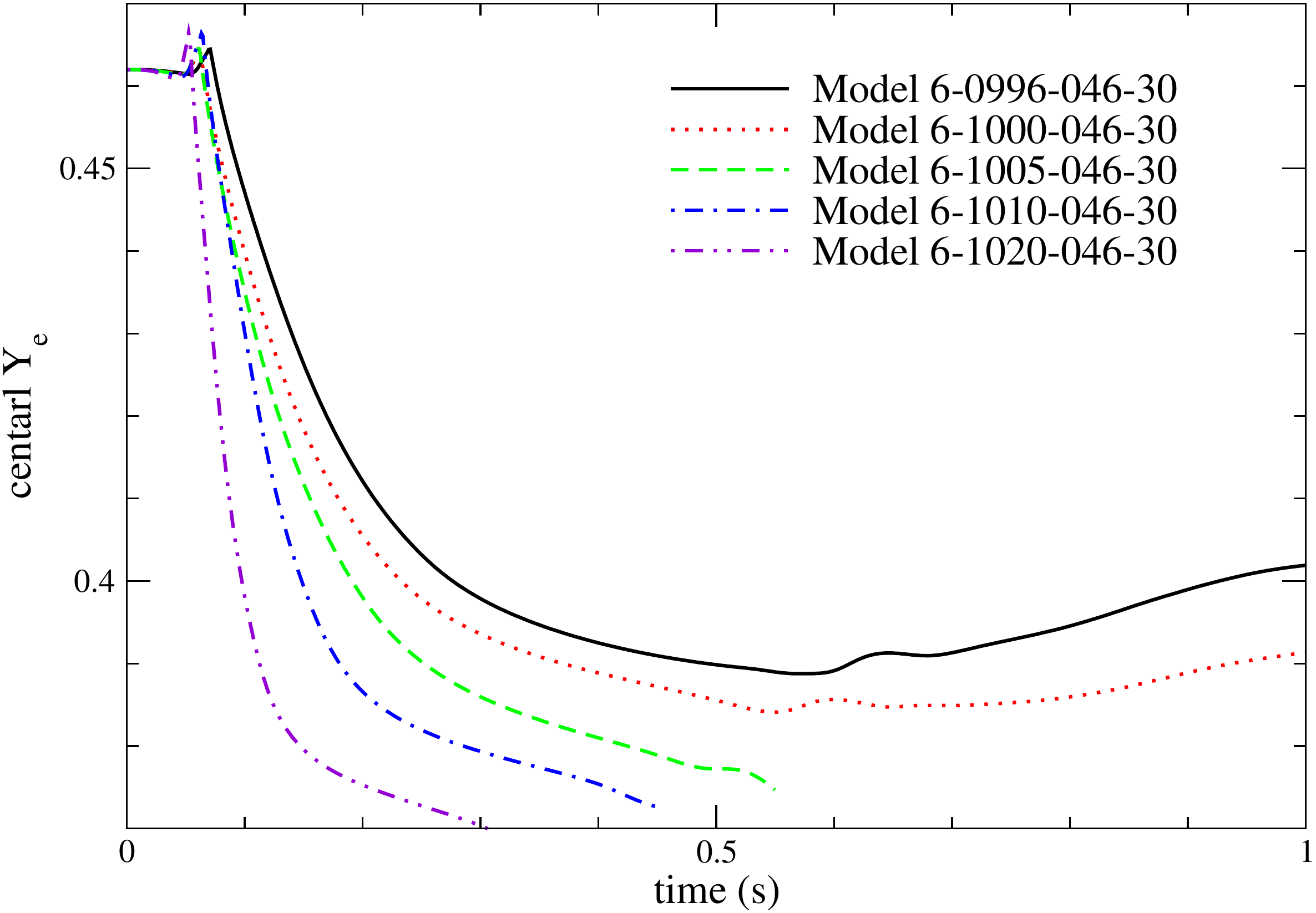}
		\caption{\added{$\rho_{{\rm c,def}}$-dependence of ``L\_no\_mix" models for $\dot{M}=10^{-6}~M_\odot~\mathrm{yr}^{-1}$. }(left panel) The central density evolution of Models 
			6-0996-046-30 (black solid line), 6-1000-046-30 (red dotted line), 
			6-1005-046-30 (green dashed line),
			6-1010-046-30 (blue dot-dashed line), 6-1020-046-30 (purple dot-dot-dashed line).
			Notice that Model 6-0996-046-30 corresponds to the 
			MESA model L$\_$no$\_$mix without contraction after the oxygen ignition, i.e., $\rho_{\rm c,def} = \rho_{\rm c,ign}$.
			(right panel) Similar to the left panel but for the central $Y_{{\rm e}}$.}
	\end{center}
	\label{fig:rhoc_1e-6_plot}
\end{figure*}

\begin{figure*}[t!]
	\begin{center}
		\includegraphics*[width=8cm, height=6cm]{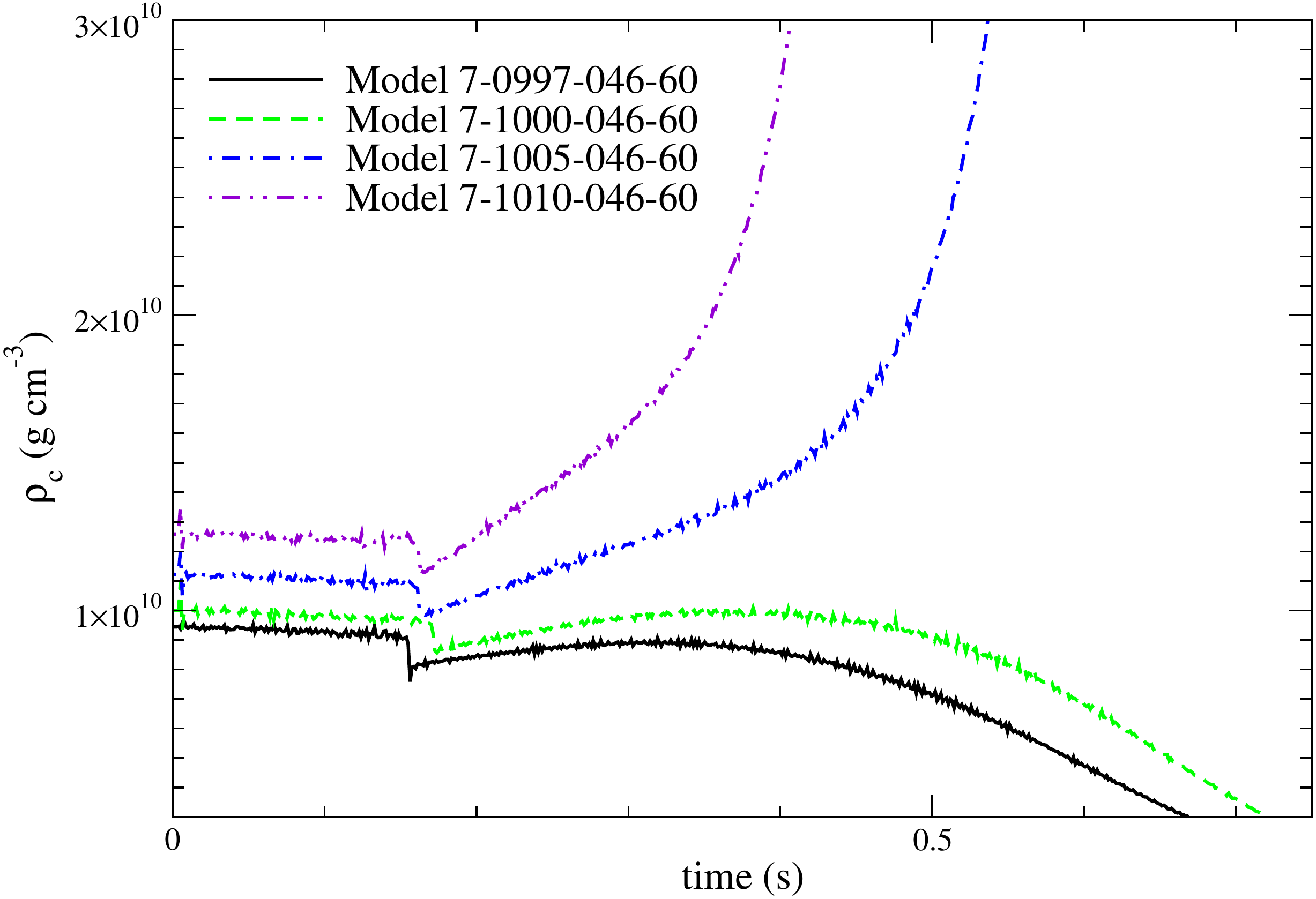}
		\includegraphics*[width=8cm, height=6cm]{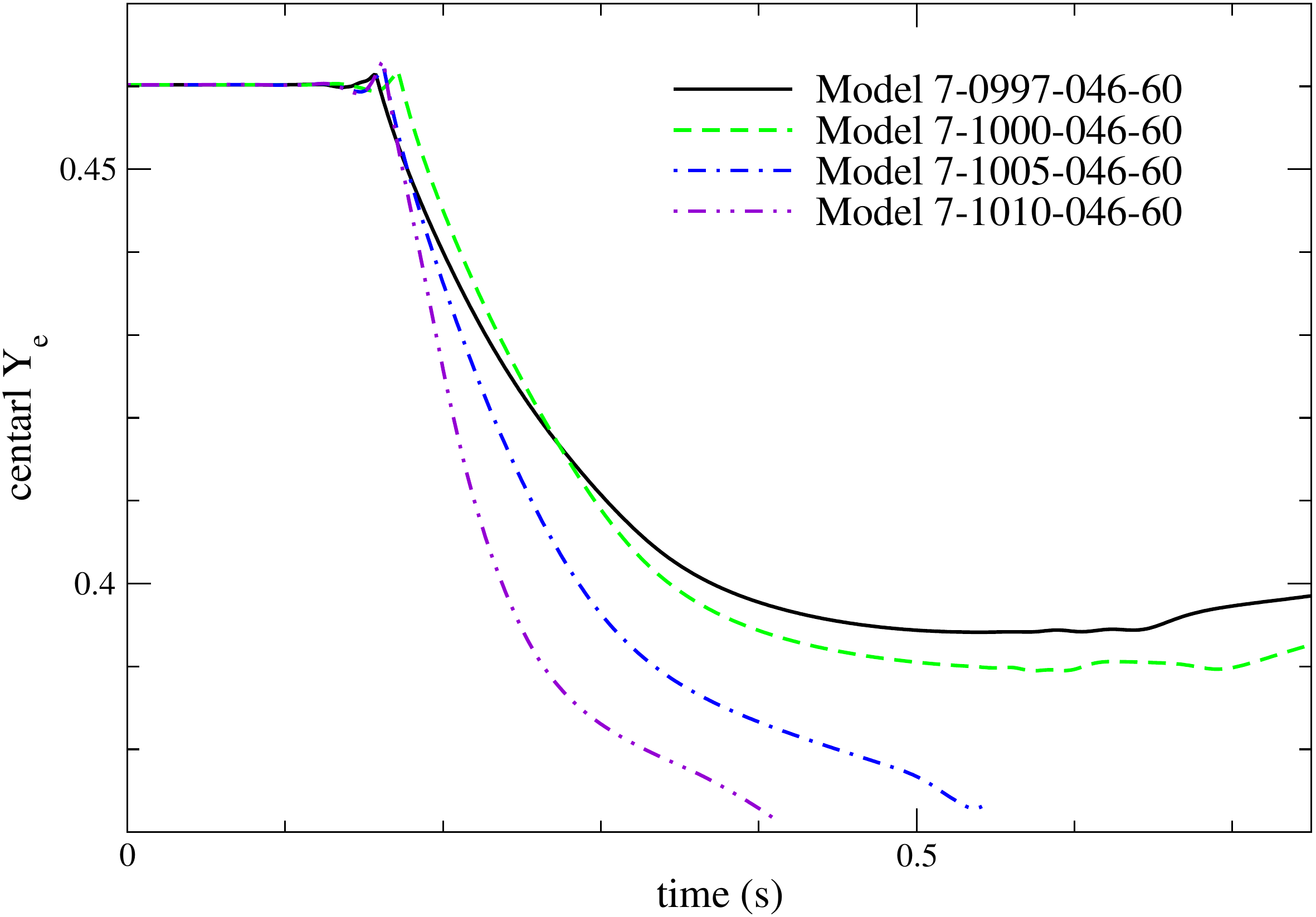}
		\caption{\added{$\rho_{{\rm c,def}}$-dependence of ``L\_no\_mix" models for $\dot{M}=10^{-7}~M_\odot~\mathrm{yr}^{-1}$. }(left panel) The central density evolution of Models 
			7-0997-046-60 (black solid line), 
			7-1000-046-60 (green dashed line),
			7-1005-046-60 (blue dot-dashed line), 7-1010-046-60 (purple dot-dot-dashed line).
			Notice that Model 7-0997-046-60 corresponds to the 
			oxygen-ignition model L$\_$no$\_$mix without contraction after the oxygen ignition, i.e., $\rho_{\rm c,def} = \rho_{\rm c,ign}$.
			(right panel) Similar to the left panel but for the central $Y_{{\rm e}}$.}
	\end{center}
	\label{fig:rhoc_1e-7_plot}
\end{figure*}

In the left panel of Figure \ref{fig:rhoc_1e-6_plot} we plot 
the central density evolution for 5 models from 6-0996-046-30 to 6-1020-046-30,
where the flame position and the temperature and $Y_e$ profiles against $M_r$ 
are the same. $Y_e$ at the center is as low as 0.46. 
Model 6-0996-046-30 corresponds to the model L\_no\_mix without contraction after the oxygen ignition, i.e., $\rho_{\rm c,def} = \rho_{\rm c,ign}$.
The central density remains unchanged again for the first 0.1~s. 
Then, the central density shows a sudden drop as the burnt matter 
in the center expands. After that, the central density increases again. 
For models with a higher $\rho_{\rm c,def}$, the contraction is faster.
Models with  $\log_{10}(\rho_{\rm c,def}/{\rm g~cm^{-3}}) \geq {10.05}$
collapse into NSs. 

In the right panel of Figure \ref{fig:rhoc_1e-6_plot}, we show the corresponding central $Y_e$ evolution. 
Before the flame reaches the center, $Y_e$ remains unchanged.
However, once the material is burnt into NSE, $Y_e$ quickly 
drops from its original value to $\sim 0.38 - 0.40$ 
within 0.1 - 0.2 s. For a higher $\rho_{\rm c,def}$, the electron capture 
takes place faster. For models which explode, the central $Y_e$
increases mildly when the central matter is
mixed with the outer high $Y_e$ material, 
until it reaches an asymptotic value. 

In the left panel of Figure~\ref{fig:rhoc_1e-7_plot}
we plot the central density evolution for 4 models 
from 7-0997-046-60 to 7-1010-046-60.
Model 7-0997-046-60 corresponds to the model L\_no\_mix without contraction after the oxygen ignition, i.e., $\rho_{\rm c,def} = \rho_{\rm c,ign}$.
For models with $\log_{10}(\rho_{\rm c,def}/{\rm g~cm^{-3}}) \geq {10.05}$ , 
they collapse into NSs. 
It takes a longer time of $\sim 0.15$ s
for the flame to reach the center.
Then the early expansion and the subsequent contraction due to electron capture take place.
The contraction is weaker than the case for $\dot{M} = 10^{-6}~M_\odot~\mathrm{yr}^{-1}$ 
because the flame has more time to propagation before central 
electron capture induces the rapid contraction. 
A maximum $\log_{10}(\rho_{\rm c}/{\rm g~cm^{-3}})$ of $\sim 10.00$ is found
for the turning point of the exploding models. 
At $t \sim 0.5$ s, the core begins its expansion. 

In the right panel, we plot the corresponding central $Y_e$ evolution.
The qualitative feature of the $Y_e$ evolution is 
similar to the ``6"-series models. Models which 
explode reach a minimum $Y_e$ of $\sim 0.39$.
For those which collapse, $Y_e$ continues to drop
before the simulations stop. 
From the two set of models, it suffices to see that, despite
the initial configurations are different, the models
still show a strong sensitivity on $\rho_{\rm c,def}$.
In particular, the exact value of $\log_{10}(\rho_{\rm c,def}/{\rm g~cm^{-3}})$ is
important because the minor change from ${10.0}$ to ${10.05}$ is sufficient to change the core from explosion to
collapse. 

\subsubsection{$r_{\rm ign}$-dependence}

Depending on the nuclear reaction network, as discussed in \S\ref{subsec:ne20}, the initial runaway position can change from off-center 
($\sim 30 - 60$ km) to the center. Here we briefly examine how the 
models vary by considering the different possible position of the initial flame. 

\begin{figure*}[t!]
	\begin{center}
		\includegraphics*[width=8cm, height=6cm]{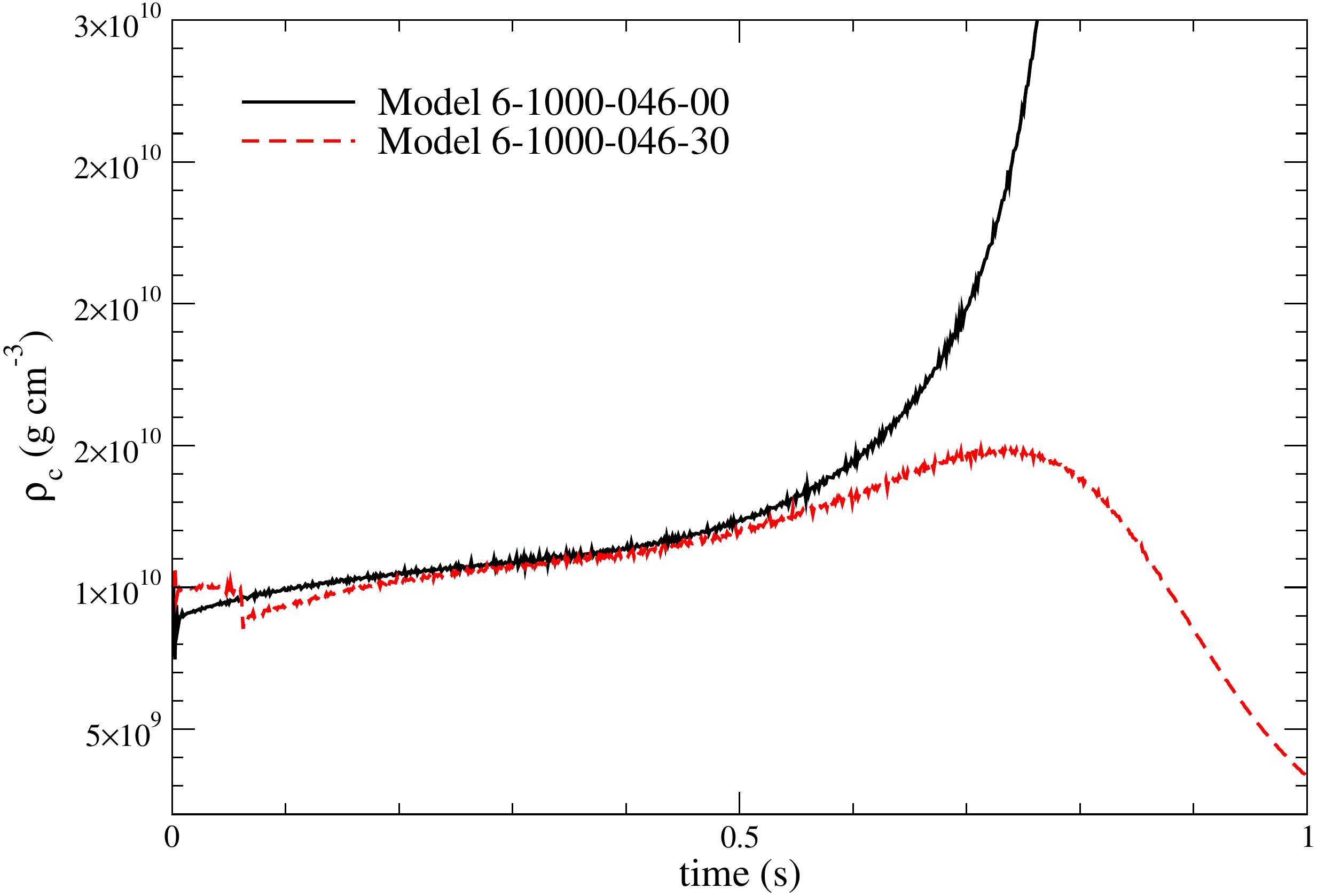}
		\includegraphics*[width=8cm, height=6cm]{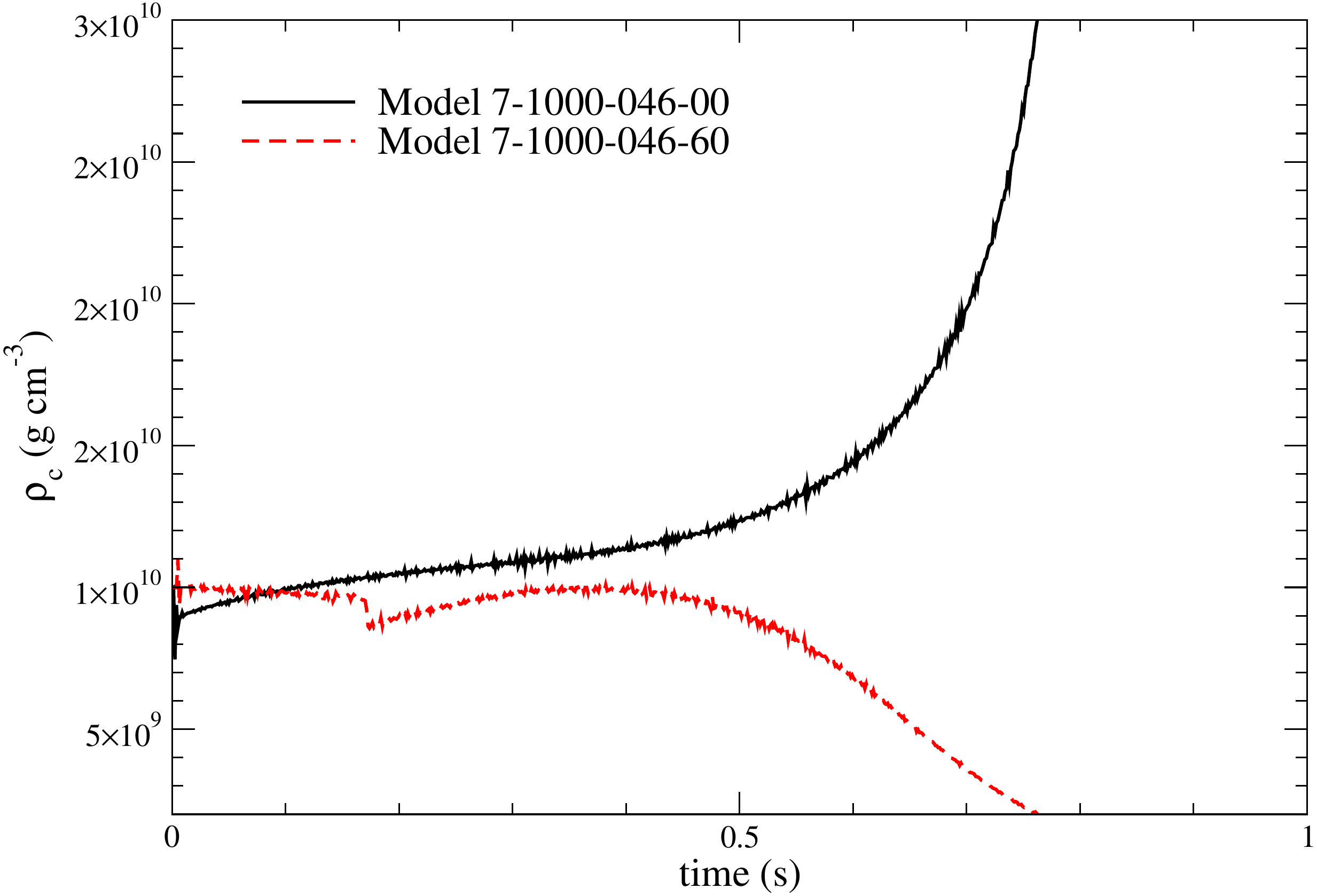}
		\caption{\added{$r_{{\rm ign}}$-dependence of ``L\_no\_mix" models for $\dot{M}=10^{-6}$ and $10^{-7}~M_\odot~\mathrm{yr}^{-1}$. }(left panel) The central density evolution of Models 
			6-1000-046-00 (black solid line) and 6-1000-046-30 (red dotted line). 
			(right panel) Similar to the left panel but for the 
			Models 7-1000-046-00 (black solid line) and 7-1000-046-60 (red dotted line). 
			}
	\end{center}
	\label{fig:rhoc_flamepos_L_plot}
\end{figure*}

In Figure \ref{fig:rhoc_flamepos_L_plot} we plot the central density
evolution for Models 6-1000-046-30 and 6-1000-046-00 in the left panel
and Models 7-1000-046-60 and 7-1000-046-00 in the right panel. 
Here we see the contrasting final fates when the flame starts
at the center or off-center. For the centered flame in both 
cases, a direct collapse of ONeMg core is observed. On the 
other hand, an off-center flame leads to explosion.
Furthermore, in the exploding case, owing to the 
high initial central density, during the contraction phase
the central density can reach as high as $\log_{10}(\rho_{\rm c}/{\rm g~cm^{-3}})\sim 10.18$.
Such a high central density allows $Y_e$ to reach as low as $\sim 0.38$.
The low $Y_e$ allows formation of extremely neutron-rich isotopes,
which may provide characteristic abundances if they are
later ejected from the core during explosion.

\added{
\subsection{Flame Structure}

\begin{figure*}
		\centering
		\includegraphics[width=0.495\textwidth]{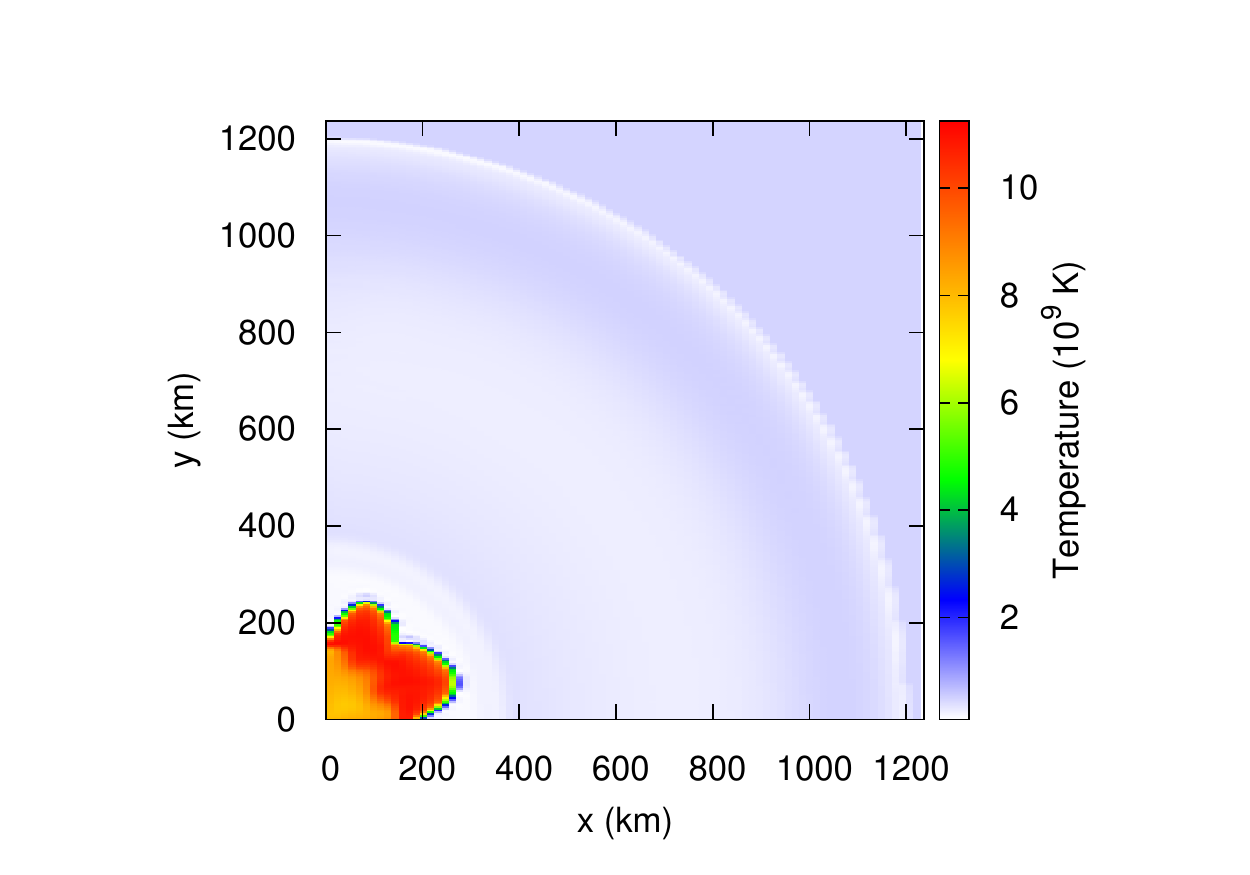}
		\includegraphics[width=0.495\textwidth]{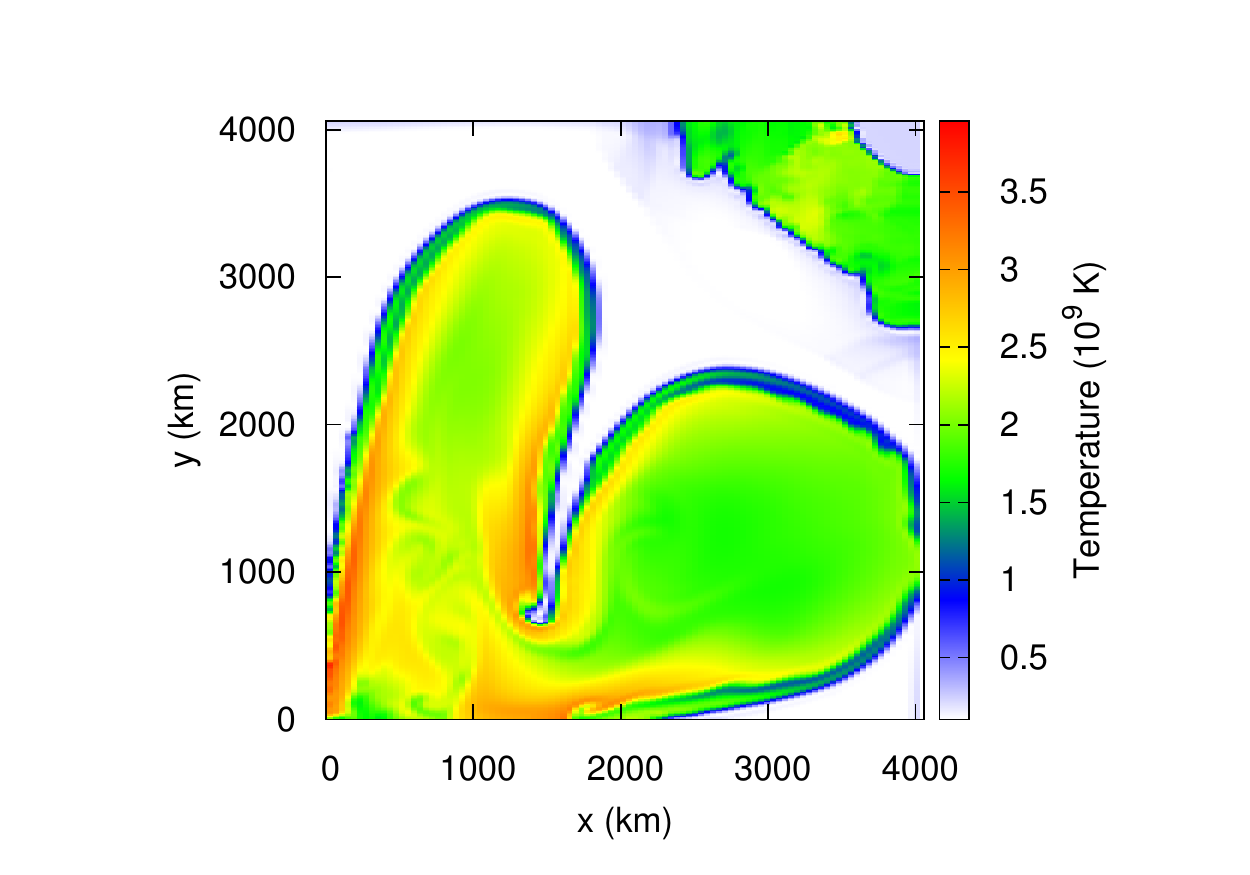}
		\caption{\added{(left panel) The temperature color map of the collapsing model 
			6-1000-046-00 at the end of simulation ($\sim 0.6$ s) where the central
			density reaches $10^{10.5}$ g cm$^{-3}$. 
			(right panel) Similar to the left panel but for the exploding model 
			6-1000-046-30 at $\sim 1.1$ s after the nuclear runaway starts. Here the central density is $\sim10^{9.5}$ g cm$^{-3}$. }\label{fig:flame_plot}}
\end{figure*}

In Figure~\ref{fig:flame_plot} we plot the temperature color map for
the collapsing (exploding) model 6-1000-046-00 (6-1000-046-30) at
time $\sim 0.6$ s ($\sim 1.1$ s) after the nuclear runaway has started. The other collapsing or exploding models share the similar properties for the flame structure.

In the collapsing model, the continuing contraction of matter prevents the 
burnt matter from reaching low density regions. The burnt matter is confined
within the radius of $\sim 200$ km. Outside the flame, most matter remains unperturbed
with a low temperature below $10^9$ K.
The flame appears to be spherical to a good approximation and the central region has in general a low temperature ($\sim 8 \times 10^9$ K).
At higher densities, the growth of hydrodynamical instabilities tends to be suppressed because the nuclear energy release relative to the internal energy is small.

For the exploding model, the structure of the oxygen deflagration is similar to the carbon deflagration \citep{Leung2018_Chand}. The flame is much more extended to a size of $\sim 3500$ km.
One major difference between the oxygen deflagration and carbon deflagration is that
the asphericity and hydrodynamical instabilities tend to be suppressed. 
This is because the nuclear energy release relative to the internal energy is smaller and the effect of electron capture is larger, which leads to the smaller buoyancy force in the oxygen flame than carbon.}

\subsection{Summary of Parameter Dependence}

\begin{figure*}[t!]
	\centering
	\includegraphics[width=1.0\textwidth]{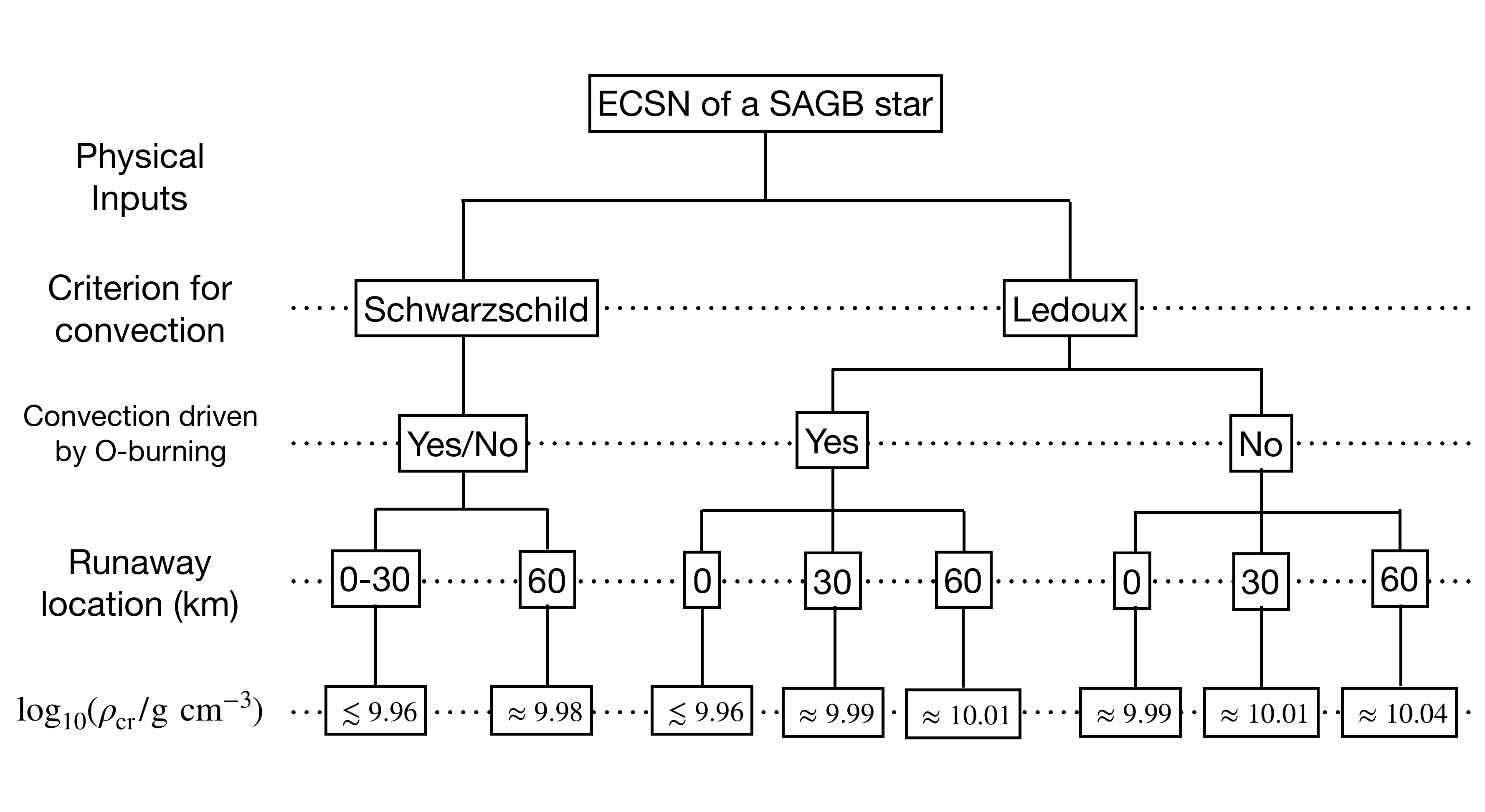}
	\caption{Summary of the critical density for the explosion-collapse bifurcation, $\rho_{\rm cr}$, from the 2D hydrodynamical simulations with different physical assumptions (see also Table~\ref{table:models} and Figures \ref{fig:trans} and \ref{fig:trans2}). The physical assumptions include the criterion for the convective stability (Schwarzschild or Ledoux), the convection driven by the oxygen burning, and the thermonuclear runaway location. Here the runaway location depends on the accretion rate (\S\ref{subsec:mdot}), $^{12}$C($\alpha,\gamma$)$^{16}$O reaction rate (\S\ref{subsec:ne20}), and the convective flow in the flame (\S\ref{subsec:Sch_off}). \label{fig:flow}}
\end{figure*}

\begin{figure*}[t!]
	\begin{center}
		\plotone{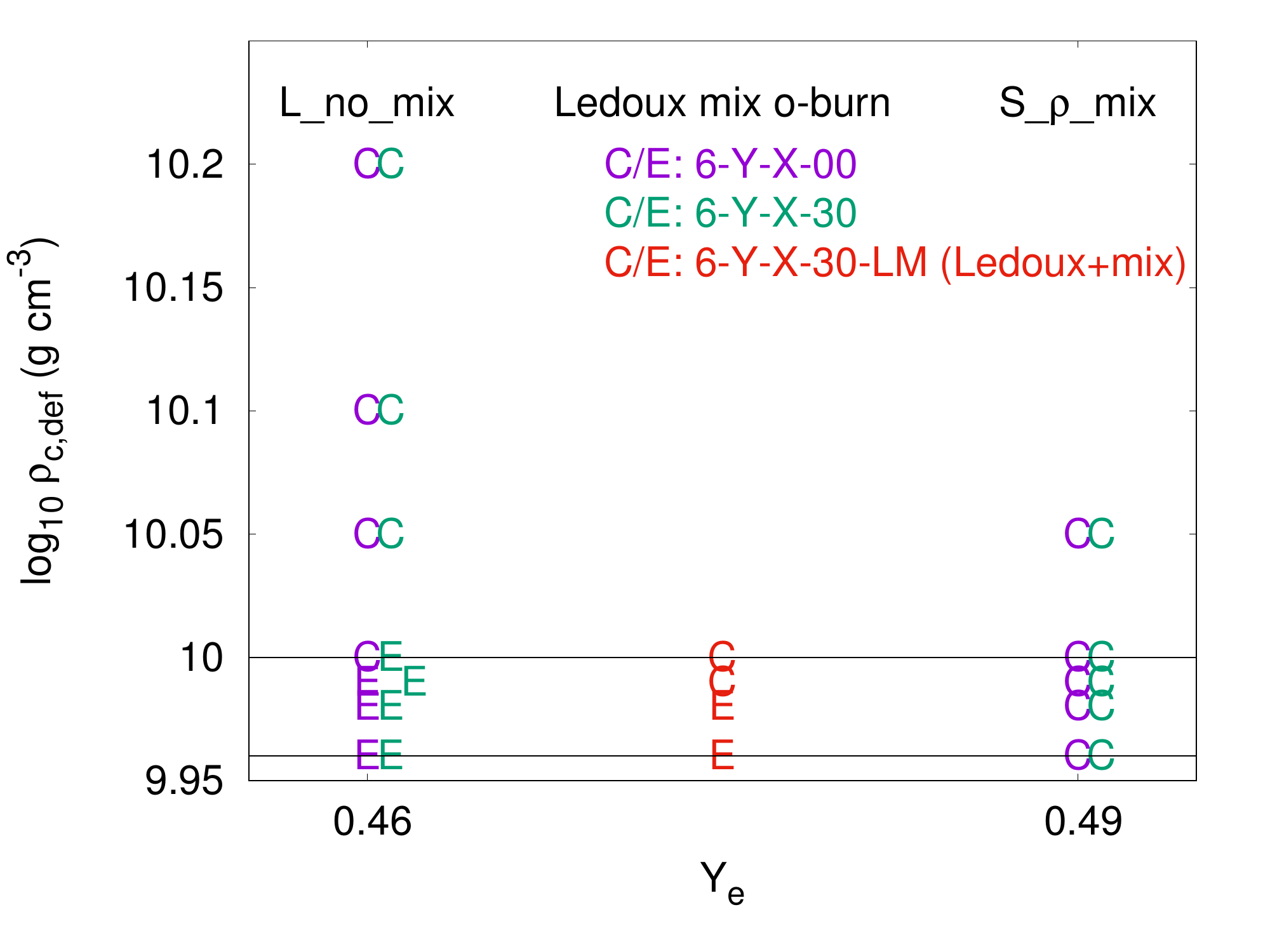}
		\caption{The explosion-collapse bifurcation diagram as a 
			function of $\rho_{\rm c,def}$
			and the initial $Y_e$ distribution. $Y_e=0.46$ and 0.49 are the central $Y_e$ of
			case (3) modesl (L$\_$no$\_$mix) and case (2) models (S$\_\rho\_$mix), respectively.
			Between these cases, case (1) models (Ledoux mix o-burn) are shown.
			The data is taken from models with $\dot{M} = 10^{-6}$ $M_{\odot}$~yr$^{-1}$.
			``E" and ``C" stand for ``Explosion'' and ``Collapse'', respectively. }
	\end{center}
	\label{fig:trans}
\end{figure*}

\begin{figure*}[t!]
	\begin{center}
		\plotone{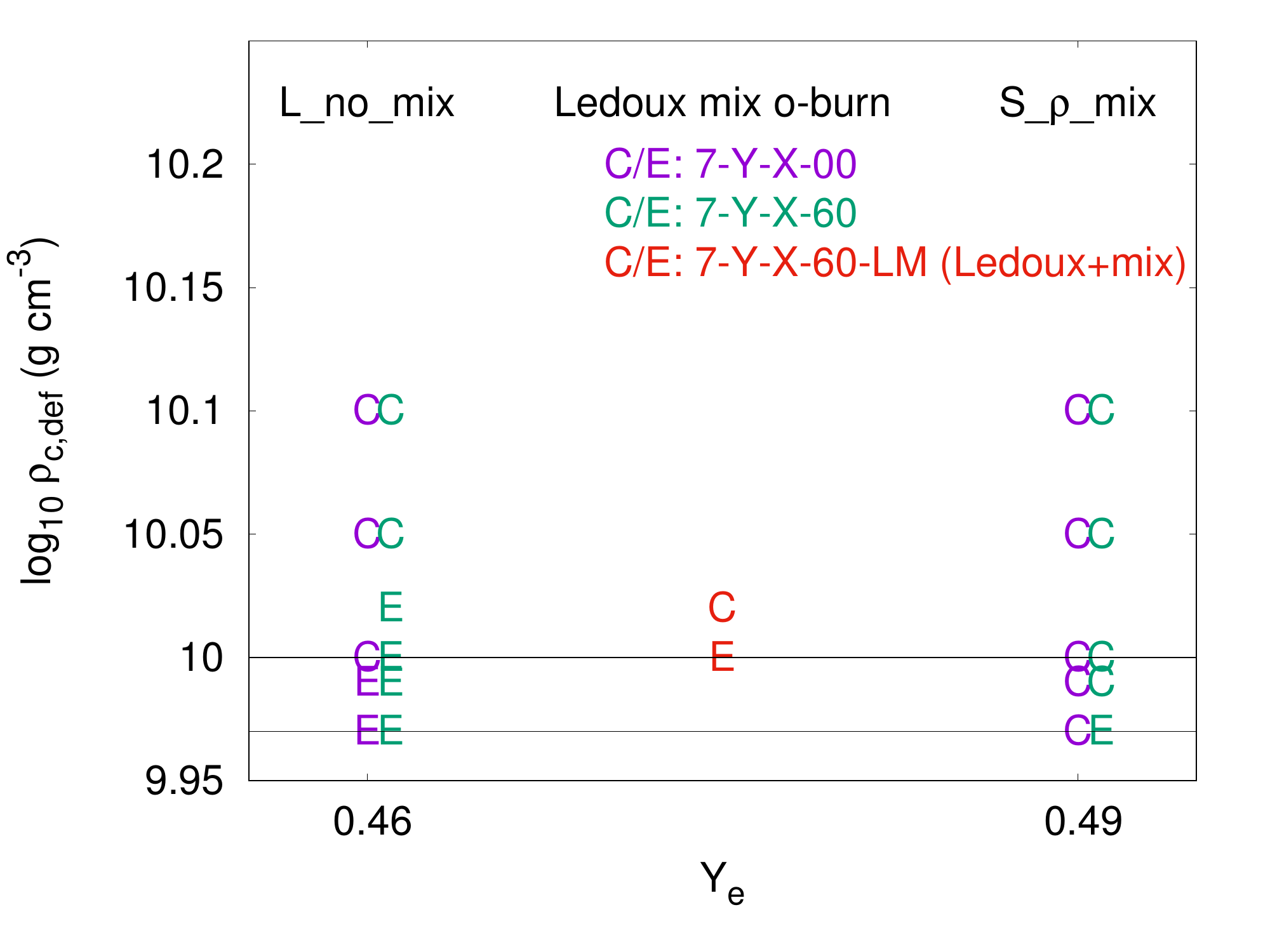}
		\caption{Similar to Figure \ref{fig:trans} but the data
			are taken from models with $\dot{M} = 10^{-7}$ $M_{\odot}$~yr$^{-1}$. }
	\end{center}
	\label{fig:trans2}
\end{figure*}

In \S\ref{subsec:hydro_mix}-\ref{subsec:hydro_nomix}, we have performed the hydrodynamical
simulations by adopting the three $Y_e$ distributions at the oxygen
ignition: (1) Ledoux mix o-burn (\S\ref{subsec:hydro_mix}), (2)
S$\_\rho\_$mix (\S\ref{subsec:hydro_cmix}), and (3) L$\_$no$\_$mix
(\S\ref{subsec:hydro_nomix}).  Such differences stem from the
different treatment of convection.  Development of the convection after
the oxygen ignition controls the further evolution and determines
$\rho_{\rm c,def}$ and the flame location ($r_{\rm ign}$) at the initiation of the
oxygen deflagration. \added{In Figure~\ref{fig:flow} we summarize the dependence of $\rho_{\rm cr}$ (the critical density for explosion-collapse bifurcation) on the physical assumptions, complementary to Table~\ref{table:models}.} Because of the numerical difficulty to follow the
evolution after the oxygen ignition, we have treated $\rho_{\rm c,def}$
and $r_{\rm ign}$ as parameters.

The outcomes of our 2D hydrodynamical simulations depend on the above parameters are summarized in
Figures~\ref{fig:trans} and \ref{fig:trans2} for $\dot{M} =
10^{-6}~M_{\odot}~{\rm yr}^{-1}$ and $10^{-7}~M_{\odot}~{\rm
  yr}^{-1}$, respectively.  Here, the final fate of the ONeMg core is
designated as either C (collapse) or E (explosion). 

In the abscissa, 3 cases of the initial $Y_e$ distributions are
shown as $Y_e = 0.46$ for case (3), $Y_e = 0.49$ for case (2), and
case (1) (-LM) in between.
In the ordinate, $\rho_{\rm c,def}$ is shown.

For case (1) (-LM), if $\log_{10}(\rho_{\rm c,def}/{\rm g~cm^{-3}}) \geq $ 9.99 and 10.02, the
final fate is the collapse for $r_{\rm ign} =$ 30 km and 60 km, respectively.  On the contrary, if
$\log_{10}(\rho_{\rm c,def}/{\rm g~cm^{-3}}) \leq {9.98}$ and 10.00, respectively,
the final fate would be the explosion.  In other words, \added{the critical density for the collapse to occur is} $\log_{10}(\rho_{\rm cr}/{\rm g~cm^{-3}}) \approx$ 9.985 and 10.01 for $r_{\rm ign} =$ 30 km and 60 km, respectively.

For case (2) ($Y_e = 0.49$) with the Schwarzshild criterion, most
models collapse, and even \added{the critical density is smaller than the central density at the oxygen ignition, i.e., }
$\log_{10}(\rho_{\rm cr}/{\rm g~cm^{-3}}) < \log_{10}(\rho_{\rm c,ign}/{\rm g~cm^{-3}}) \approx 9.97$.

For case (3) with no mixing, $\log_{10}(\rho_{\rm cr}/{\rm g~cm^{-3}}) \simeq 10.05$.
If $\log_{10}(\rho_{\rm c,def}/{\rm g~cm^{-3}}) \gtrsim 10.15$ as we estimated
for the evolution after the oxygen ignition (\S\ref{subsec:runaway}), $\rho_{\rm cr} > \rho_{\rm c,def}$, i.e.,
the collapse is the most likely outcome.

\added{
\subsection{Astrophysical Significance}
In the previous sections, we have shown that the collapse is the more likely outcome of an ECSN. This collapse would be similar to the ONeMg core of the $8.8~M_\odot$ star \citep{1984ApJ...277..791N,2006A&A...450..345K,2011ApJ...726L..15W}. If this similarity is the case, the neutrino heating mechanism works to induce a low-energy explosion thanks to the steep density gradient in the outermost layers. The very extended H-He super-AGB envelope would be easily ejected and a NS would be formed. If the mass ejection from the core is negligible, the baryonic mass of the NS is in the range of $1.357 - 1.361~M_\odot$ (Table~\ref{tab:star_param}). If the ejected mass is $1.14-1.39\times10^{-2}~M_\odot$ \citep[like in][]{2006A&A...450..345K,2011ApJ...726L..15W}, then it would be $1.343-1.350~M_\odot$. This is smaller than $\sim1.36~M_\odot$ of the $8.8~M_\odot$ model \citep{2006A&A...450..345K}, because of the lower $Y_e$ in the ONeMg core.

The super-AGB progenitors are different from more massive stars, undergoing
a large amount of mass loss. Such mass loss would lead to a small mass of the H-He envelope and a large amount of circumstellar material (CSM), possibly with lots of carbon dust.
The supernova properties depend on the masses of the H-He envelope and CSM, and the optical light curve could be similar to the Fast-evolving Blue Optical Transient \citep{2019ApJ...881...35T} or the Crab supernova \citep{1982Natur.299..803N,2013ApJ...771L..12T,2014A&A...569A..57M}. It might be Type II-L like supernovae because of the low envelope mass. If CSM is very dusty, it would be bright in infrared, like the eSPecially Red Intermediate-luminosity Transient Events \citep{2017yCat..18390088K}. The above properties are somewhat different from typical Type II-P Fe-core-collapse supernovae.

On the other hand, if the ONeMg core (partially) explodes as a result of the oxygen deflagration, it could be a weak thermonuclear explosion within a H-He envelope, leaving a WD behind. The oxygen deflagration cannot burn all materials in the star, and it results in a partial disruption of the ONeMg core \citep{2016A&A...593A..72J}. The turbulent mixing by the flame allows the ejecta to consist of both Fe-peak elements and the ONe-rich fuel. Ejecta can be rich in neutron-rich isotopes, e.g., $^{48}$Ca, $^{50}$Ti, $^{54}$Cr, and $^{60}$Fe \citep{2019A&A...622A..74J}. The light curve is dimmer than a normal Type II supernova, due to a smaller mass of $^{56}$Ni synthesized. It could be called as a Type I.5ax supernova (like Type Iax).

}

\section{Summary \label{sec:summary}}
We have calculated the evolution of the 8.4 $M_\odot$ star from the 
main sequence until the oxygen ignition in the degenerate ONeMg core, 
where the nuclear energy generation rate exceeds the thermal neutrino 
loss rate and a convective region develops.  We have applied 
the latest weak rates \citep{Suzuki2019}, including the second 
forbidden transition for the electron capture on $^{20}$Ne 
\citep{2018arXiv180508149K}.  \added{The electron-degenerate ONeMg core evolves through complicated processes of the mass accretion, election capture heating, URCA cooling, and the $Y_e$ change due to weak interactions.  The convective and semiconvective regions are formed.  Because of uncertainties in the semiconvective mixing, we have applied both the Ledoux and Schwarzshild criteria for the convective stability. Our findings of the ONeMg core evolution are summarized as follows. 
	
(1) If we apply the Ledoux criterion and assume no mixing, we have 
found the following evolution.  The second forbidden transition is so 
slow that it does not ignite oxygen burning at the related threshold 	
density, but decreases the central $Y_e$ to $\sim0.46$ during the core 	
contraction.  The oxygen ignition takes place when the central density 
reaches $\log_{10}(\rho_{\rm c,ign}/{\rm g~cm^{-3}})=9.96-9.97$.  The 
location of the oxygen ignition, i.e., center or off-center ($r_{\rm ign} \sim30 - 60~{\rm km}$), depends on the 
$^{12}$C$(\alpha,\gamma)^{16}$O reaction rate because it affects the 
$^{20}$Ne mass fraction by a few percent in the ONeMg core. 
	
(2) If we apply the Schwarzshild criterion, the convective core heated 
up by electron capture on $^{20}$Ne can grow to half of the mass of 
the ONeMg core. The oxygen ignition takes place at the center.  The 
convective energy transport delays the oxygen ignition until 
$\log_{10}(\rho_{\rm c,ign}/{\rm g~cm^{-3}}) \sim 10.0$ is reached, and 
the convective mixing makes $Y_e$ in the convective region as high as 
0.49. 
	
(3) Even with the Ledoux criterion, the oxygen ignition (at the center 
or off-center) creates the convectively unstable region and the 
convective mixing forms an extended region with $Y_e \sim 0.49$ above 
the oxygen ignited shell.  The convective energy transport would slow 
down the temperature increase, and so the thermonuclear runaway to 
form a deflagration wave is estimated to occur when the central density 
$\log_{10}(\rho_{\rm c,def}/{\rm g~cm^{-3}})$ exceeds ${10.10}$. 
(This estimate is consistent with the result by 
\cite{2019ApJ...871..153T}, who obtained $\log_{10}(\rho_{\rm 
c,def}/{\rm g~cm^{-3}})\approx 10.2$ with a semiconvection coefficient 
of \cite{1992A&A...253..131S}.) 

Then, to examine the final fate of the ONeMg core, we have performed 2D hydrodynamical simulations of the 
propagation of the oxygen deflagration wave for the three cases of 
the $Y_e$ distribution, three locations of the oxygen ignition, and 
various $\rho_{\rm c,def}$.  We have found that the deflagration 
starting from $\log_{10}(\rho_{\rm c,def}/\mathrm{g~cm^{-3}})>10.01 (< 10.01)$ 
leads to a collapse (a thermonuclear explosion).  Since our 
estimate of $\rho_{\rm c,def}$ well exceeds this critical value, the 
ONeMg core is likely to collapse irrespective of the central $Y_e$ and 
ignition position (Figures~\ref{fig:trans} and \ref{fig:trans2}). 

Our work has shown that the degenerate ONeMg core evolved in a 
SAGB star can collapse to form a relatively low mass NS. 
However, future work needs to confirm whether such a high $\rho_{\rm c,def}$ is 
reached by calculating the evolution of the core with semiconvection 
and with the full convection from the oxygen ignition 
through the initiation of the deflagration by improving the stellar evolution modeling. }


\section{Acknowledgment}
The results in this paper have been presented at the Lorentz Center
workshop on \href{https://www.lorentzcenter.nl/lc/web/2019/1122/info.php3?wsid=1122&venue=Snellius}{``Electron-Capture-Initiated Stellar Collapse''}.
We would like to thank the Lorentz Center and the workshop
participants for stimulated discussion.
Near the end of the workshop, \cite{2019arXiv190509407K} was submitted.
We will make comparisons in a separate paper. 
This work was supported by World Premier 
International Research Center Initiative 
(WPI), MEXT, Japan and JSPS KAKENHI Grant Numbers 
\deleted{JP26400222, JP16H02168,} JP17K05382, JP19K03855, and a grant
from the Research Grant Council of Hong Kong (Project
14300317).
\deleted{We acknowledge the support by the Endowed Research
Unit (Dark Side of the Universe) by Hamamatsu Photonics K.K.} We acknowledge the support of the CUHK Central High Performance Computing Cluster, on which the stellar evolution simulation in this work has been performed.
We thank F.-X. Timmes for the open-source subroutines 
of the helmholtz equation of state and torch nuclear
reaction network. We also thank MESA developers for making their code open-source. 

\software{MESA \citep{2011ApJS..192....3P,2013ApJS..208....4P,2015ApJS..220...15P,2018ApJS..234...34P,2019arXiv190301426P}; NuGridPy (available from \url{https://nugrid.github.io/NuGridPy/})}

\bibliographystyle{aasjournal}
\bibliography{the_bib}

\end{document}